\documentclass[prc,aps,twocolumn,floats,floatfix,showpacs,nofootinbib]{revtex4-2}

\usepackage{amsmath}
\usepackage{amssymb}
\usepackage{amsfonts}
\usepackage{graphicx}
\usepackage{color}  
\usepackage[english]{babel}
\usepackage{orcidlink}
\usepackage[normalem]{ulem}
\usepackage{soul}

\allowdisplaybreaks

\begin{document}

\title{Triaxiality in Mo and Ru nuclei}

\author{D. Bonatsos\orcidlink{0000-0003-1728-0910}}
\address{Institute of Nuclear and Particle Physics, N.C.S.R. Demokritos, GR-15310 Aghia Paraskevi, Attiki, Greece}

\author{G. Thiamov\'{a}}
\address{Universite Grenoble 1, CNRS, LPSC, Institut Polytechnique de  Grenoble, IN2P3, F-38026 Grenoble, France}

\author{M. Bender\orcidlink{0000-0001-8707-3410}}
\address{Universit{\'e} Claude Bernard Lyon 1, CNRS/IN2P3, IP2I Lyon, UMR 5822, F-69100 Villeurbanne, France}

\author{I. Tk\'{a}\v{c}ov\'{a}\orcidlink{0000-0002-4166-8053} and P. Alexa\orcidlink{0000-0003-0335-3189}}
\address{Department of Physics, Faculty of Electrical Engineering and Computer Science, V\v{S}B -- Technical University of Ostrava,
17. listopadu 2172/15, CZ-708 00 Ostrava, Czech Republic}

\author{D. Petrellis\orcidlink{0000-0002-8609-7522}}
\address{Nuclear Physics Institute, Czech Academy of Sciences, CZ-250 68 \v{R}e\v{z} near Prague, Czech Republic,\\
School of Physics, Aristotle University of Thessaloniki, GR-54124 Thessaloniki, Greece}

\author{P. Vesel\'{y}\orcidlink{0000-0003-3901-3186}}
\address{Nuclear Physics Institute, Czech Academy of Sciences, CZ-250 68 \v{R}e\v{z} near Prague, Czech Republic}

\date{18 May 2026}

\begin{abstract}
In this paper we focus on the evolution of triaxiality from $\gamma$-soft toward rigid triaxial shapes in Mo and Ru chains above $N=60$. Spectral criteria for the 
second-order phase transition are suggested and examined. Both chains are investigated within the phenomenological Algebraic Collective Model and in the framework 
of microscopic Hartree-Fock-Bogoliubov calculations with six representative parametrizations of the effective Skyrme interaction. The Algebraic Collective Model yields
energy surfaces that are softer towards prolate shapes, whereas the Skyrme results are all softer towards oblate shapes.
\end{abstract} 


\maketitle

\section{Introduction}
\label{Intro}

Triaxiality in even-even nuclei is recently attracting renewed attention \cite{Grosse2022}. Studies of relevant data indicate that some degree of triaxiality, 
either static or dynamic, can be expected to be found in any nucleus. This has implications for many observables and phenomena for which this degree of freedom 
has been neglected in the past. For example, 
according to Ref.~\cite{Grosse2022}, the triaxiality parameter $\gamma$ exhibits a variance of around 8$^{\rm o}$, while acquiring values of 10$^{\rm o}$ or higher
 \cite{Bonatsos2017b}. In a recent review \cite{Rouoof2024} regarding the Triaxial Projected Shell Model (TPSM), 30 nuclei with $\gamma =15^{\rm o}$-45$^{\rm o}$ are
 listed, while the prevalence of triaxial shapes within Monte Carlo Shell Model calculations in nuclei where shell evolution and high-$j$ orbitals play
a role has been attributed to the tensor force as one possible, but not unique microscopic contributor \cite{Tsunoda2021,Otsuka2023}. 

Triaxiality has been also studied in the framework of algebraic models, in particular the Interacting Boson Model (IBM) and the proxy-SU(3) model.
Triaxial shapes can be included by using three-body terms in the IBM-1 Hamiltonian \cite{VanIsacker1981,Heyde1984}.
In addition, triaxial shapes can be described within the SU(3)$^*$ limit of IBM-2, in which 
valence protons are supposed to be particles described by the SU(3) limit of IBM, corresponding to prolate shapes \cite{Zamfir1991,McCutchan2007}, while neutrons are supposed to be holes described by the $\overline {\rm SU(3)}$ limit of IBM, corresponding to oblate shapes, or \textit {vice versa} \cite{Dieperink1982,Walet1987}. 
$^{104}_{44}$Ru$_{60}$ has been considered as the textbook example of SU(3)$^*$ \cite{Dieperink1982}.
Within the proxy-SU(3) model, only particles and no holes are used. From Table~I of Ref.~\cite{Bonatsos2017b} one sees that maximal triaxiality occur for 16-20, 42-48, 74-80, 116-124 nucleons. 

Within the more fundamental macroscopic-microscopic approach, finite-range droplet model or finite-range liquid-drop model
\cite{Moeller2006,Moeller2008}, the reduction in the potential energy of the ground state due to triaxiality has been calculated.
In the Mo isotopes, substantial reductions have been obtained for $N=54$-74 with a maximum at $N=66$, while
in the Ru isotopes for $N=58$-76 with the maximum at $N=64$. 

Within microscopic nuclear models, triaxiality emerges for many nuclei and phenomena.
Self-consistent mean-field models systematically 
predict triaxial shapes for nuclei in several specific regions of the chart of nuclei, either for the ground state or excited shape isomers. 
Cranked mean-field models predict that, irrespective of the ground-state shape, states in collective rotational bands built on a mean-field minimum always become 
triaxial with increasing spin \cite{Nilsson95a} because of centrifugal and Coriolis forces that align the nucleonic single-particle states with the axis of rotation.
Also, when projecting mean-field states on angular momentum $L$, the state that yields the lowest projected energy at any $L$ is virtually always triaxially deformed
\cite{Hayashi84a}, which is a feature of the TPSM \cite{Rouoof2024} and also systematically found for the energy surfaces in the symmetry-restored 
Generator Coordinate Method (GCM) \cite{Robledo19a} when considering the entire $\beta$-$\gamma$ plane, or the Monte-Carlo shell model \cite{Tsunoda2021,Otsuka2023}.
When applying the GCM, or a collective Hamiltonian constructed from a Gaussian overlap approximation to the GCM \cite{Delaroche10a}, shape fluctuations also inevitably 
lead to an average triaxiality parameter $\gamma$ of all collective wave functions that is non-zero when limiting the averaging to one sextant of the $\beta$-$\gamma$ plane
\cite{Delaroche10a,Bally22a}. Within such models, it is therefore not straigthforward to distinguish static from dynamical triaxiality.

Self-consistent mean-field methods based on an energy density 
functional (EDF) that models the effective in-medium interaction
\cite{Bender}, as well as their extensions to microscopic collective 
Hamiltonians and the symmetry-restored GCM,
have been frequently used to analyze signatures of triaxial shapes of neutron-rich 
isotopes in the Mo-Ru region, in some cases directly accompanying 
the analysis of new experimental data. 

Results for deformation energy surfaces of even-even isotopes obtained 
with various Skyrme EDFs are reported in Refs.~\cite{Bonche85a,Aysto90a,Zhang2015b}.
The recent BSkG$x$ parametrizations, which aim at a simultaneous description
of nuclear structure and astrophysical phenomena, have been found
to describe quite well many of the available data for the masses
and charge radii of nuclei in this mass region \cite{Hukkanen23b,Hukkanen24a,Maass25x}, 
including those for odd and odd-odd nuclides 
\cite{Hukkanen23b,Hukkanen24a,Maass25x,Hukkanen23a,Stryjczyk25a}. 
Various parametrizations of the Gogny interaction were employed in either
calculations at the mean-field level \cite{Deloncle00a,Hilaire07a,Amadee,Rodriguez10a},
with a microscopic collective Hamiltonian \cite{Deloncle00a,Doherty17a,Hilaire07a,Amadee,Delaroche10a},
with an IBM Hamiltonian mapped from mean-field  \cite{Nomura16a},
and also in the framework of the symmetry-restored GCM, in this context also
called symmetry conserving configuration mixing method \cite{Garrett22a,Ralet2017,Garrett2020}.

In addition, results from mean-field calculations with various forms of 
relativistic EDFs can be found in Refs.~\cite{Abusara2017a,Shi18a,ElBassem2024}. 
All of these calculations agree that mid-shell Ru isotopes with $N \simeq 66$ 
adopt triaxial shapes, as do in many cases also the adjacent Mo isotopes, but
results differ in details about the boundaries of the region for which triaxially 
deformed ground states are found, the precise deformation 
of the ground state, and the stiffness of the energy surface around the minimum.
In all cases, however, the minimum is quite shallow compared to the axial 
saddle points, and the energy surface therefore quite soft in $\gamma$
direction.

On the experimental side, model-independent values for $\gamma$ obtained as Kumar-Cline shape invariant from multiple Coulomb excitation data
are available for about 25 nuclei \cite{Kumar72a,Cline86a,Srebrny06a,BSkG1}. 
For other nuclei, the identification of triaxiality remains empirical, and is usually based on collective models that consider 
triaxial deformation as an explicit degree of freedom, albeit usually in a way that prohibits truly axial solutions. Within the Davydov model \cite{Davydov1958,Davydov1959}, an empirical value for $\gamma$ can be obtained directly from the energy ratio $R= {E(2_2^+)/ E(2_1^+)}$ 
or from the ratio of reduced transition probabilities $R_2= {B(E2; 2_2^+ \to 2_1^+) / B(E2; 2_2^+ \to 0_1^+)}$ \cite{Casten2000}.  
The odd-even staggering within the $\gamma$-bands has also long been considered as an indicator of triaxiality \cite{Zamfir1991,McCutchan2007}. It should be noticed that  although very few nuclei exhibit odd-even staggering corresponding to rigid triaxial shapes, many nuclei exhibit odd-even staggering corresponding to soft triaxiality. 
 
The present paper provides a comprehensive analysis of the transition from $\gamma$-soft to $\gamma$-rigid triaxial shapes in Mo and Ru isotopes, focusing on the 
neutron mid-shell at $N=66$. The work is novel in its approach, combining the phenomenological Algebraic Collective Model (ACM) and the microscopic Skyrme-Hartree-Fock-Bogoliubov (HFB) method to
offer complementary insights into the structural evolution of these isotopes. It highlights the role of shell effects,
energy surfaces, and collective properties in driving triaxiality, while also supporting the proxy-SU(3) symmetry predictions. Additionally, it identifies gaps in
experimental data, such as $B(E2)$ transition rates and Kumar-Cline invariants, and emphasizes the need for further experimental investigations to deepen our understanding
of triaxiality in this region. In Sec.~\ref{sig} we investigate
possible spectral signatures of $\gamma$-soft and $\gamma$-rigid nuclei. Section~\ref{EmpTheor} focuses on both empirical systematics and theoretical predictions of triaxial
nuclei with an emphasis on the most pronounced island of triaxiality in Mo and Ru isotopes. In Sec.~\ref{ACMsec} we briefly introduce the Algebraic Collective Model (ACM), which is then applied to Mo and Ru isotopes. In Sec.~\ref{SkyrmeInt} we turn to microscopic calculations based on the Skyrme functionals.

\section{Rigid triaxiality versus $\gamma$-softness}\label{sig}

As mentioned in the Introduction within the $\gamma$-rigid Davydov model \cite{Davydov1958,Davydov1959} one can obtain $\gamma$ directly from the energy ratio 
\begin{equation}
\label{R}
R= {E(2_2^+) \over E(2_1^+)} 
\end{equation}
through the expression \cite{Casten2000}
\begin{equation} \label{g}
\gamma = {1\over 3} \sin^{-1} \left( {3\over R+1} \sqrt{R\over 2}  \right). 
\end{equation}
One can also obtain $\gamma$ numerically from the ratio of transition rates (branching ratio) 
\begin{equation} \label{R2}
R_2= {B(E2; 2_2^+ \to 2_1^+)  \over B(E2; 2_2^+ \to 0_1^+)} 
\end{equation}
through the expression \cite{Casten2000}
\begin{equation}\label{R2gamma}
R_2 = {20\over 7} { {\sin^2 3\gamma  \over 9-8 \sin^2 3\gamma} \over   1-{3-2\sin^2 3\gamma \over \sqrt{9-8\sin^2 3\gamma} }  } . 
\end{equation} 
In algebraic models, the connection between the $(\lambda,\mu)$ Elliott labels of the SU(3) irrep and the collective variable $\gamma$ is provided through a mapping 
between the invariant quantities of the collective model of Bohr and the rigid triaxial rotor, resulting in the relation \cite{Castanos1988,Draayer1989}
\begin{equation} \label{mu}
\gamma = \arctan \left( {\sqrt{3} (\mu+1) \over 2\lambda+\mu+3}  \right), 
\end{equation}
from which it becomes clear that irreps with $\lambda > \mu$ correspond to prolate ($0 <\gamma <30^{\rm o}$) shapes,  irreps with $\lambda <\mu$ correspond to oblate 
($30^{\rm o} <\gamma <60^{\rm o}$) shapes, while irreps with $\lambda=\mu$ give maximal triaxiality with $\gamma = 30^{\rm o}$. 

Odd-even staggering as an indicator of triaxiality in $\gamma$ bands \cite{Zamfir1991,McCutchan2007} can be studied using the quantity 
\begin{equation} \label{SL}
S(L) =  { E(L^+_\gamma) + E[(L-2)^+_\gamma] - 2  E[(L-1)^+_\gamma] \over E(2_1^+)}, 
\end{equation}
which measures the displacement of the $(L-1)$ level in relation to the average of its neighbors, $L$ and $(L-2)$, normalized to the energy of the first excited state of the nucleus.

Within the $\gamma$-rigid Davydov model one can see that $S(L)$ exhibits minima at odd values of the angular momentum $L$, irrespectively of the value of $\gamma$. 
In contrast, within $\gamma$-soft models, ranging from the vibrator with U(5) symmetry to the O(6) limit of the interacting boson model, $S(L)$ exhibits minima at even values of the angular momentum $L$ (see, for example, Fig. 1 of Ref. \cite{Bonatsos2025}). In addition to the opposite sign of staggering, it is seen that the size of the odd-even staggering in the rigid triaxial cases is an order of magnitude higher than the staggering in the $\gamma$-soft cases.

Another quantity exhibiting different behavior in the $\gamma$-rigid and $\gamma$-soft cases is the ratio \cite{Casten2020,Bonatsos2021}
\begin{equation} \label{RL}
R(L) = { E(L^+_\gamma)-E(L^+_g) \over E(2^+_\gamma)-E(2^+_g) }, 
\end{equation}
based on the energy differences between the $\gamma$-band and the ground state band ($g$) at the same angular momentum $L$. 
In the rigid triaxial case the ratios $R(L)$ increase rapidly and gradually bending upwards, while in the $\gamma$-soft cases the ratios $R(L)$ increase much more slowly and linearly
(see, for example, Figs.~2 and 3 of Ref.~\cite{Bonatsos2025}).

\section{Empirical systematics and theoretical predictions}\label{EmpTheor}

\begin{figure*}[t!]

    {\includegraphics[width=60mm]{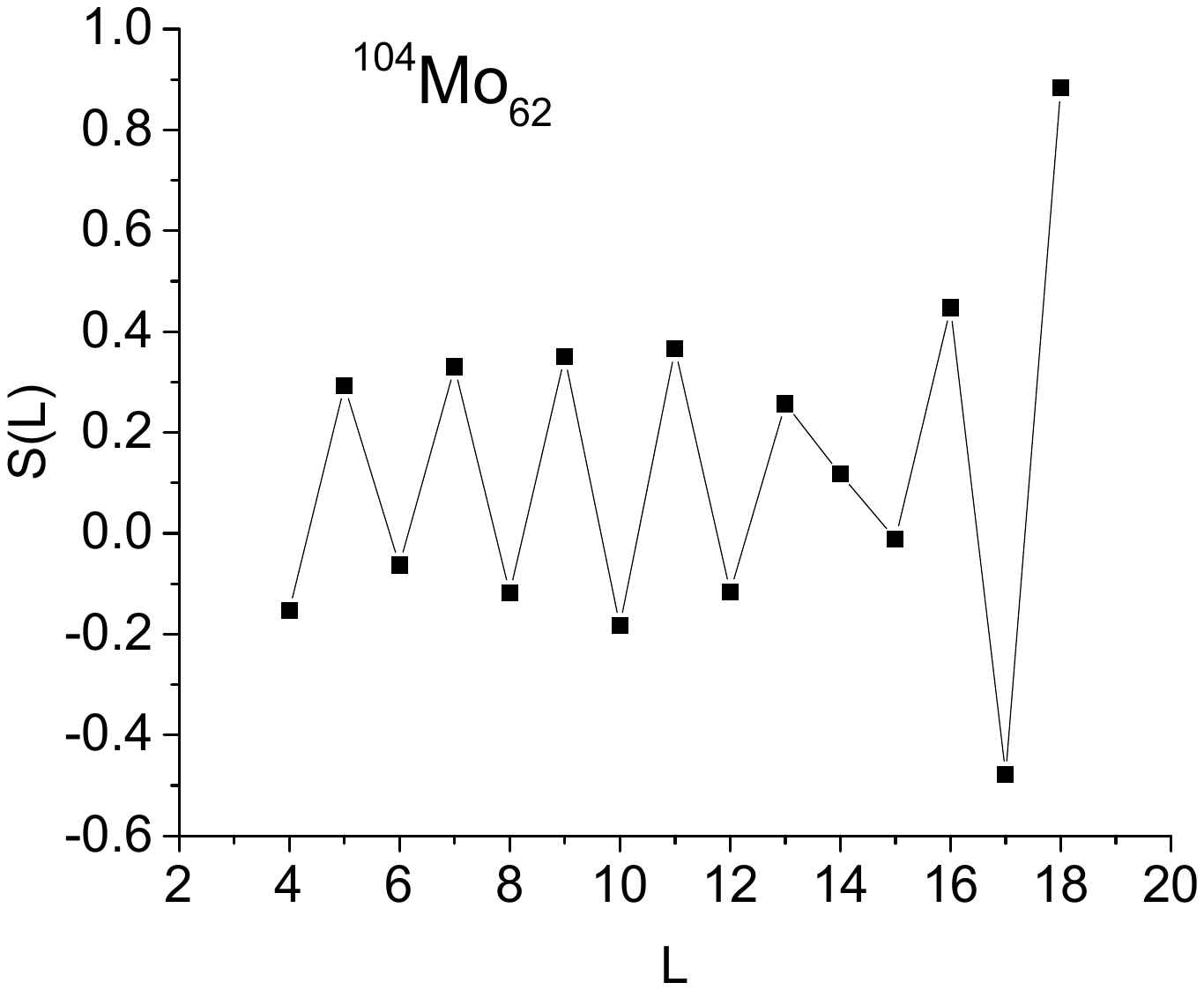}\hspace{5mm}
    \includegraphics[width=60mm]{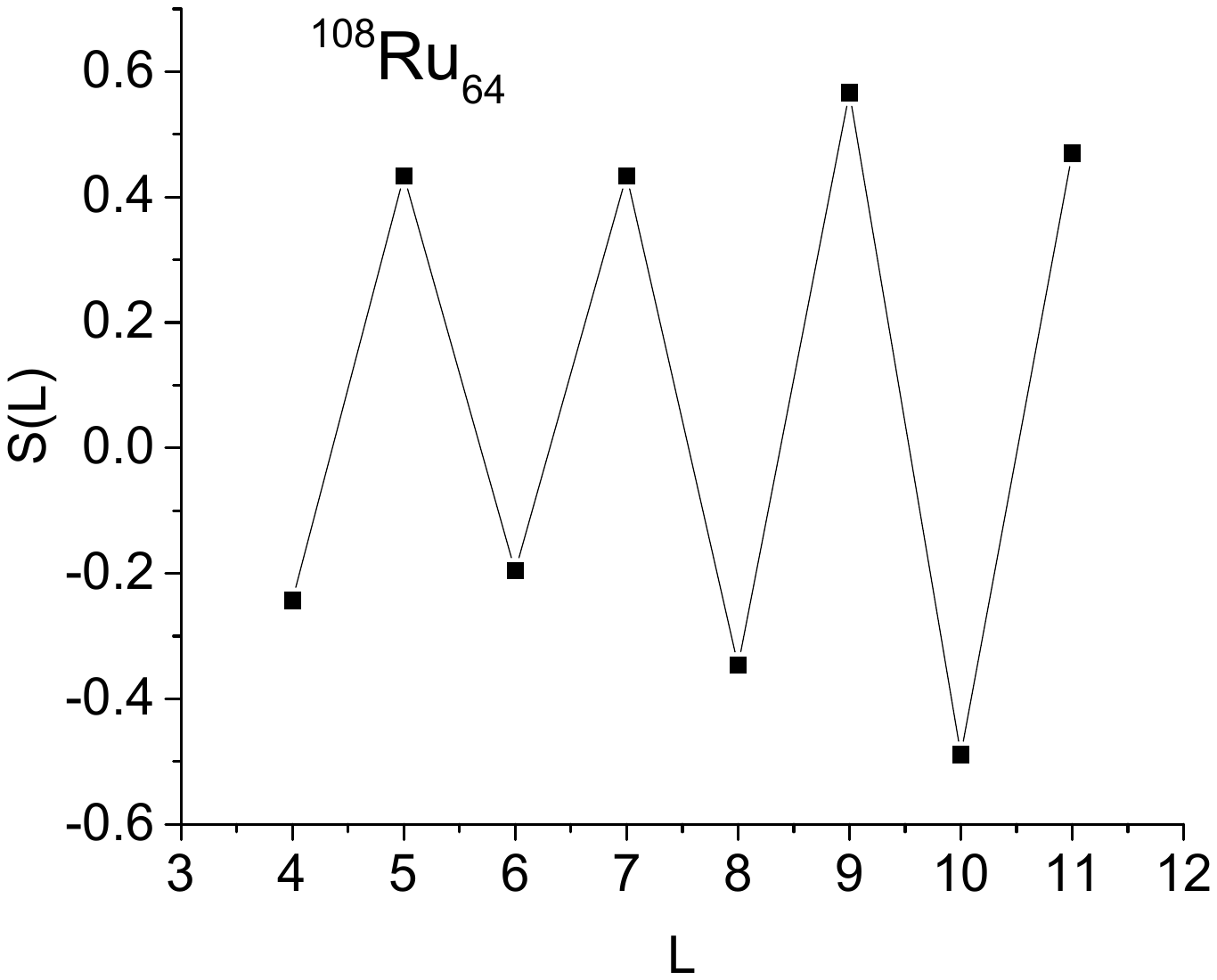}}
      {\includegraphics[width=60mm]{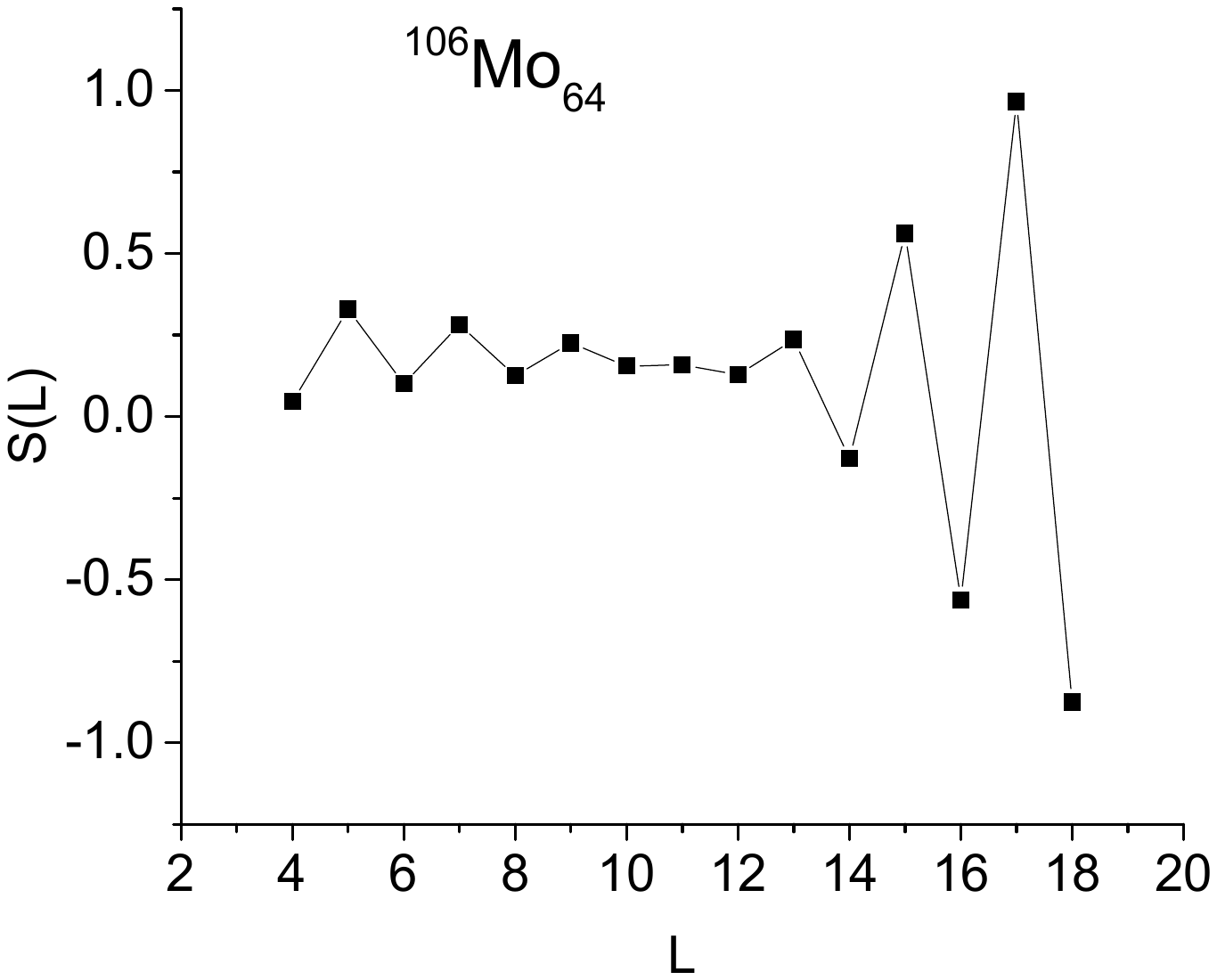}\hspace{5mm}
    \includegraphics[width=60mm]{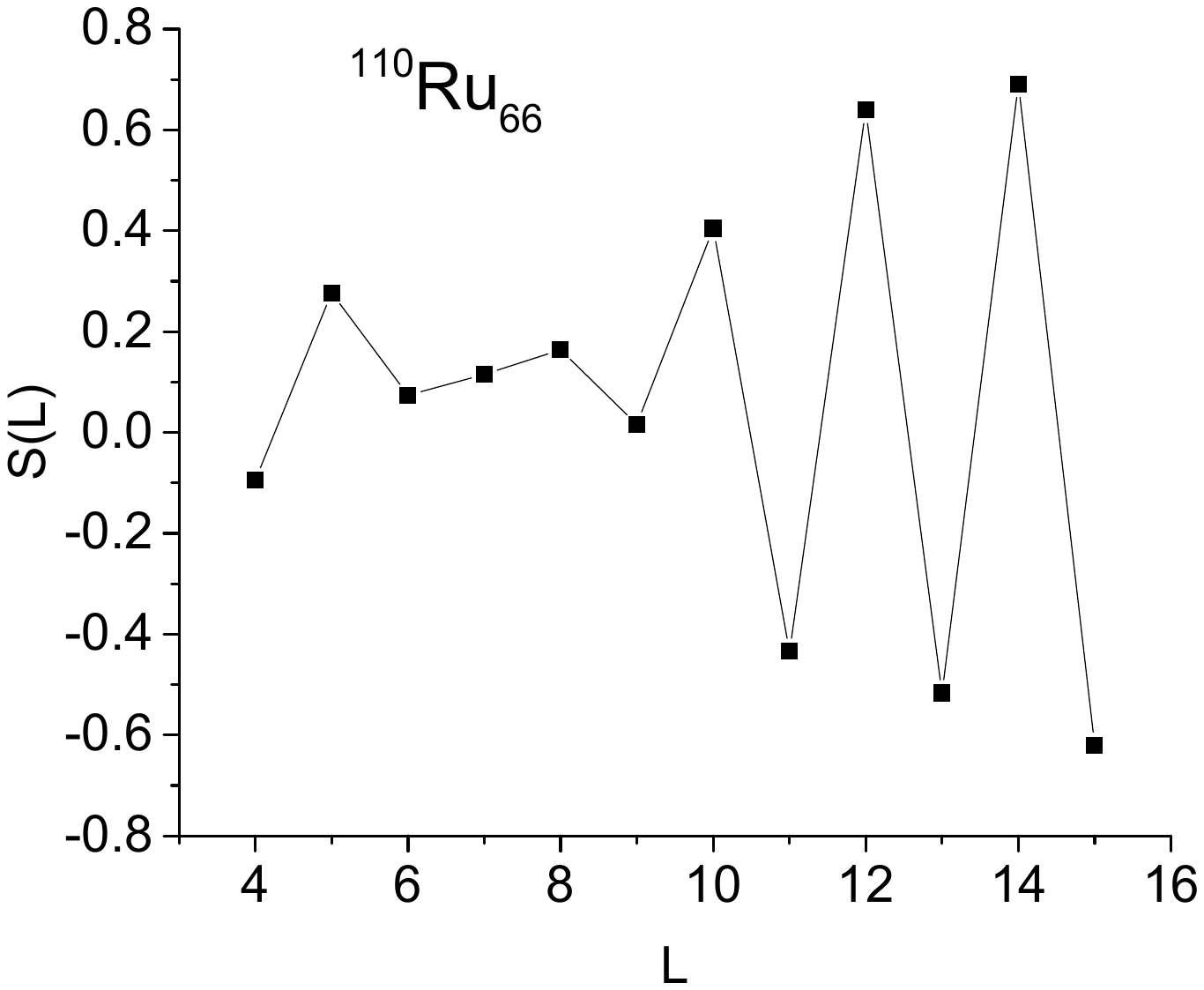}}
      {\includegraphics[width=60mm]{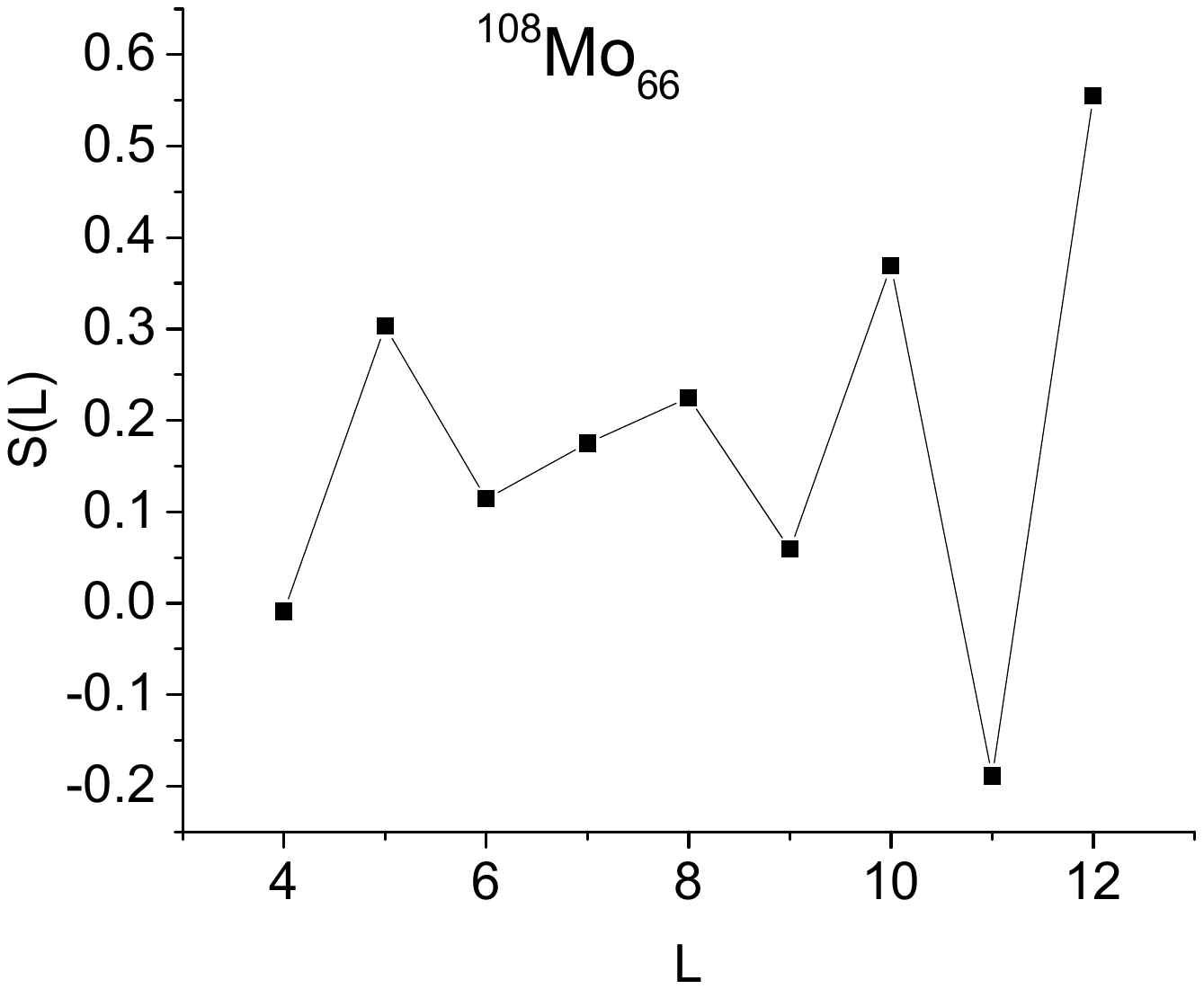}\hspace{5mm}
    \includegraphics[width=60mm]{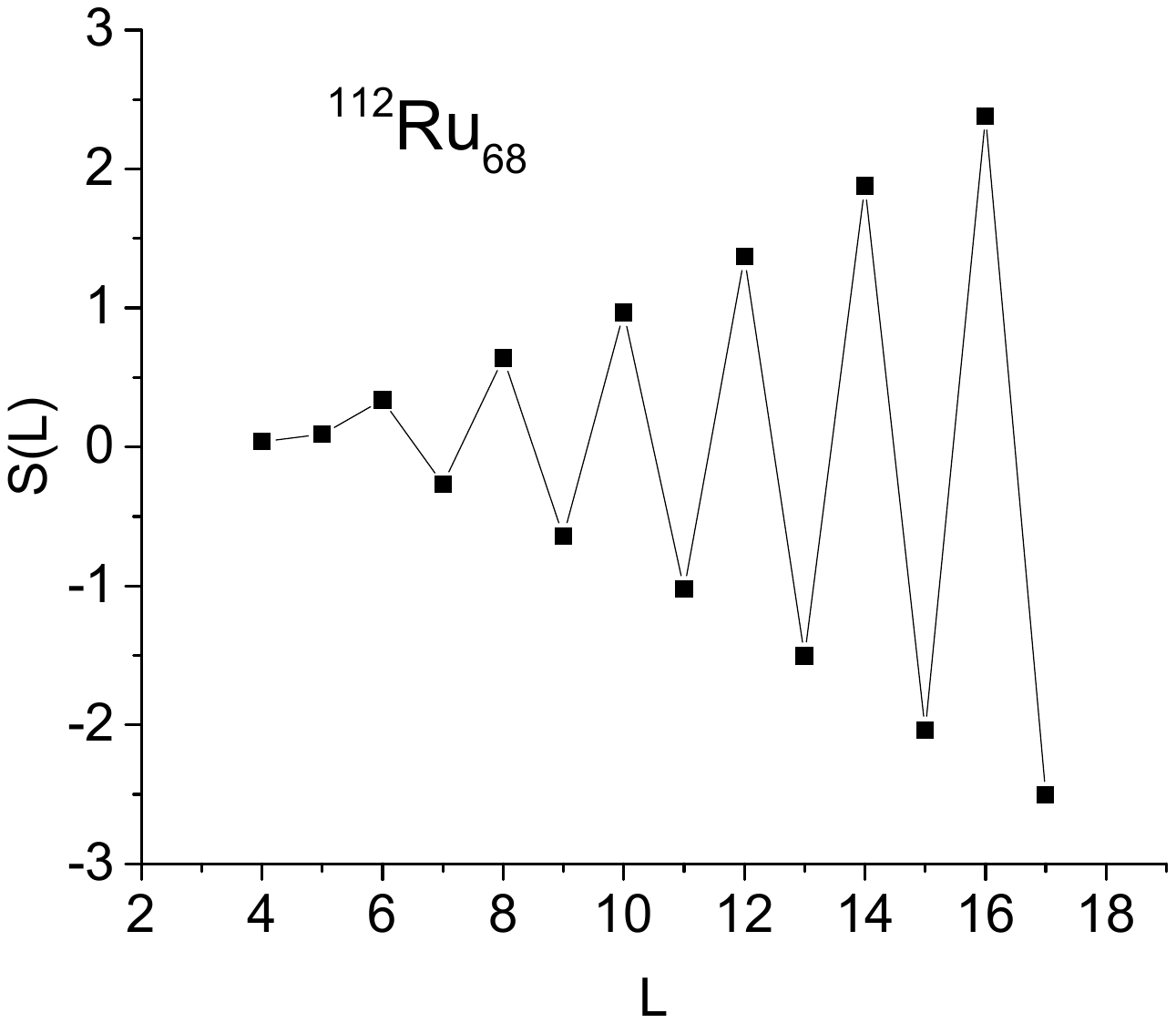}}
   
    \caption{Odd-even staggering in $\gamma$-bands, given by Eq. (\ref{SL}), for various Mo and Ru isotopes. See Sec. \ref{EmpTheor} for further discussion.} 
    \label{Fig4}
\end{figure*}

Experimental odd-even staggering in $\gamma$-bands of Mo and Ru isotopes, for which enough data exist, is shown in Fig.~\ref{Fig4}. We see that $\gamma$-soft staggering with minima at even $L$ appears for both series of isotopes at $N=64$, while at $N=66$ triaxial staggering with minima at odd $L$ appears from $L=9$ onwards. At $N=68$ the triaxial staggering starts earlier, at $L=7$.


These observations indicate a sudden change from $\gamma$-unstable to triaxial behavior as one passes from $N=64$ to $N=66$, i.e., at the neutron mid-shell and beyond. 
A possible interpretation could be given by taking into account the strong $Z=40$ subshell closure. If this subshell closure is taken into account, then Mo and Ru have two and respectively four proton particles above it, while the valence neutrons become holes beyond their $N=66$ midshell. Therefore the occurrence of triaxiality, as expected from  the SU(3)$^*$ limit of the IBM becomes possible. 

The appearance of triaxial staggering in $^{104}$Mo ($N=62$) at $L=15$ and beyond, seen in Fig.~\ref{Fig4}, poses a puzzle, which could be attributed to band-crossing occurring at this higher angular momentum.  

At this point it is instructive to examine the alternative signature of rigid triaxiality discussed in Sec. \ref{sig}, namely $R(L)$, the ratio of energy differences of levels of the $\gamma$ and ground state bands. 

\begin{figure*}[t!]
    {\includegraphics[width=60mm]{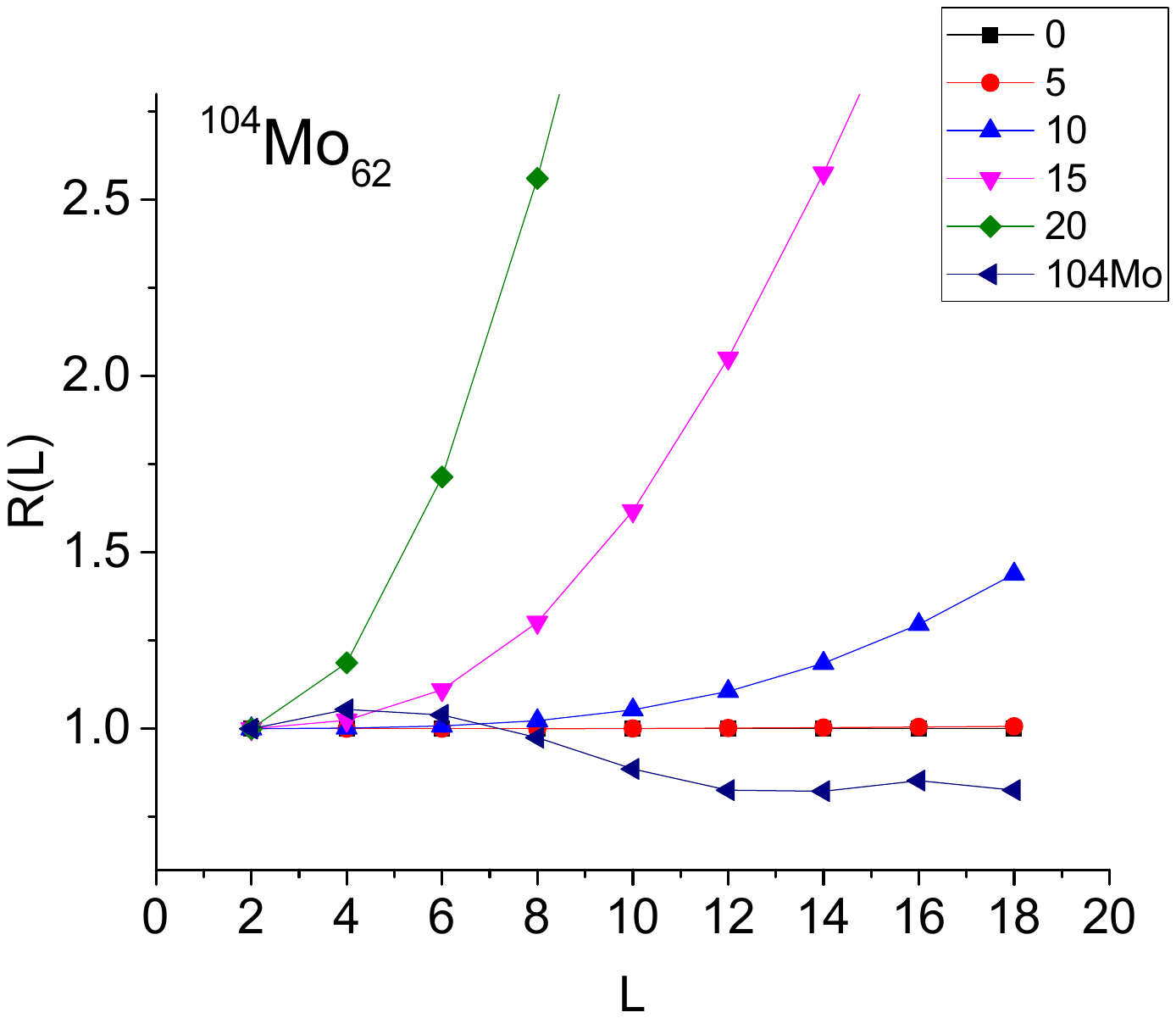}\hspace{5mm}
    \includegraphics[width=60mm]{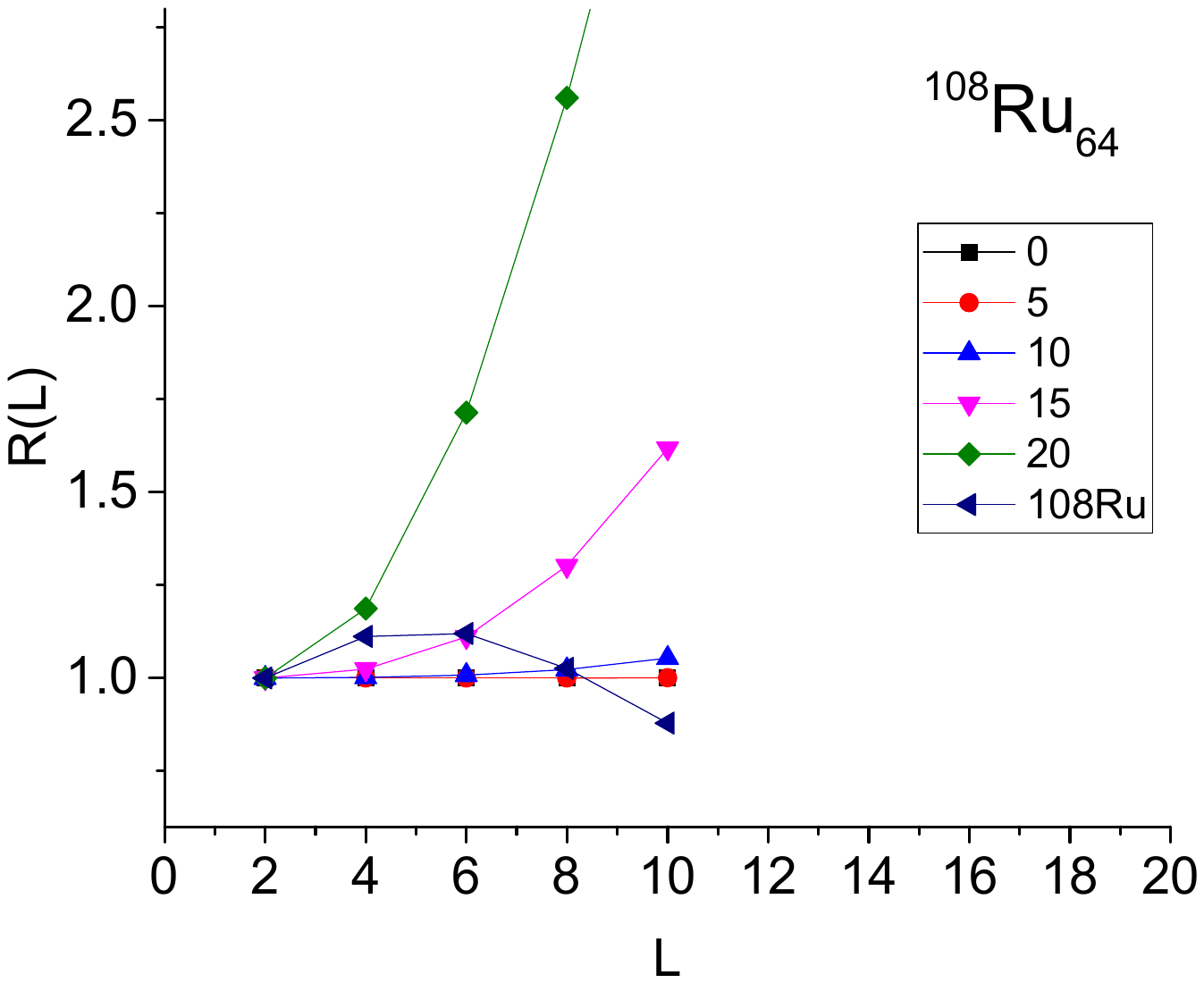}}
      {\includegraphics[width=60mm]{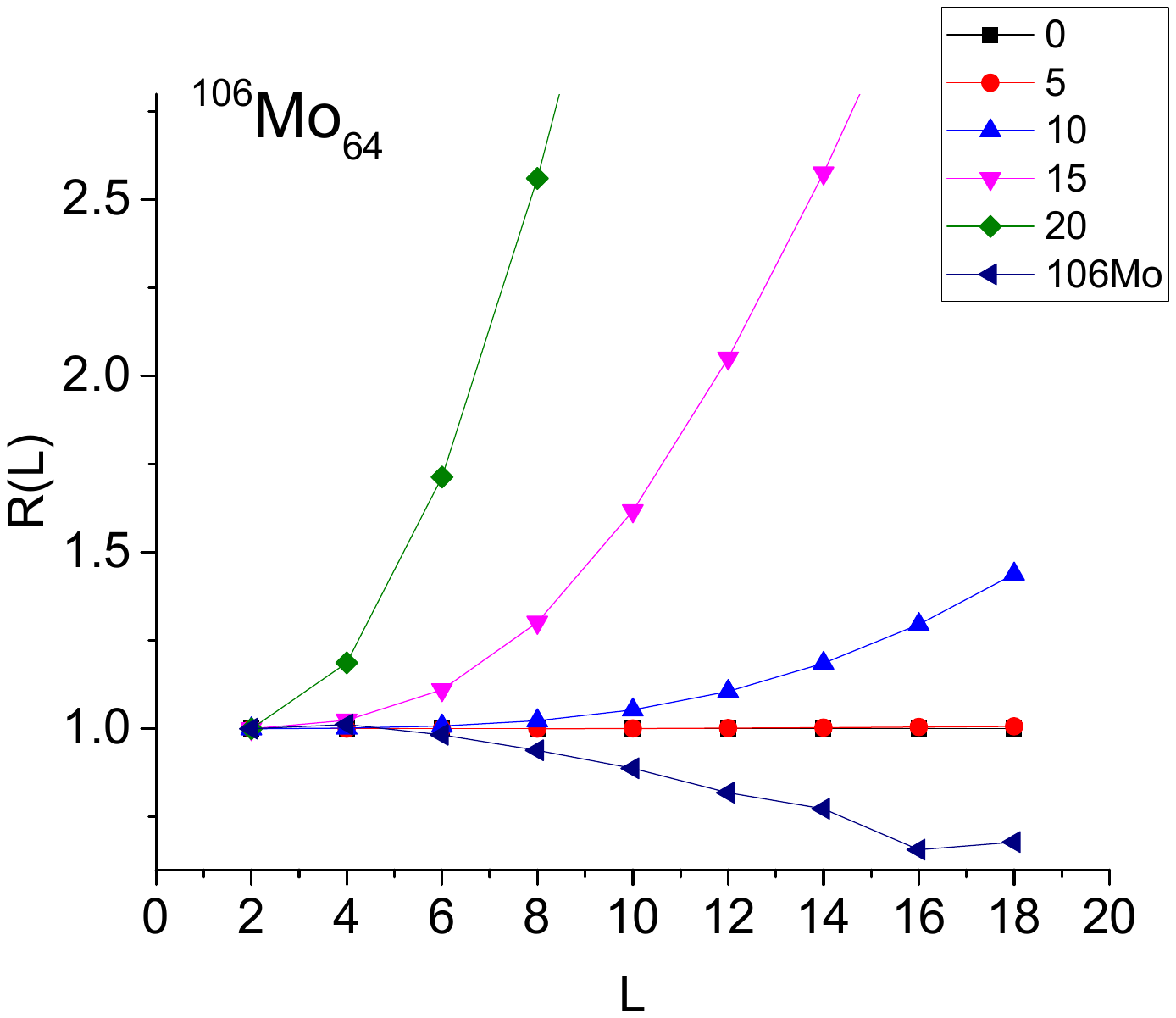}\hspace{5mm}
    \includegraphics[width=60mm]{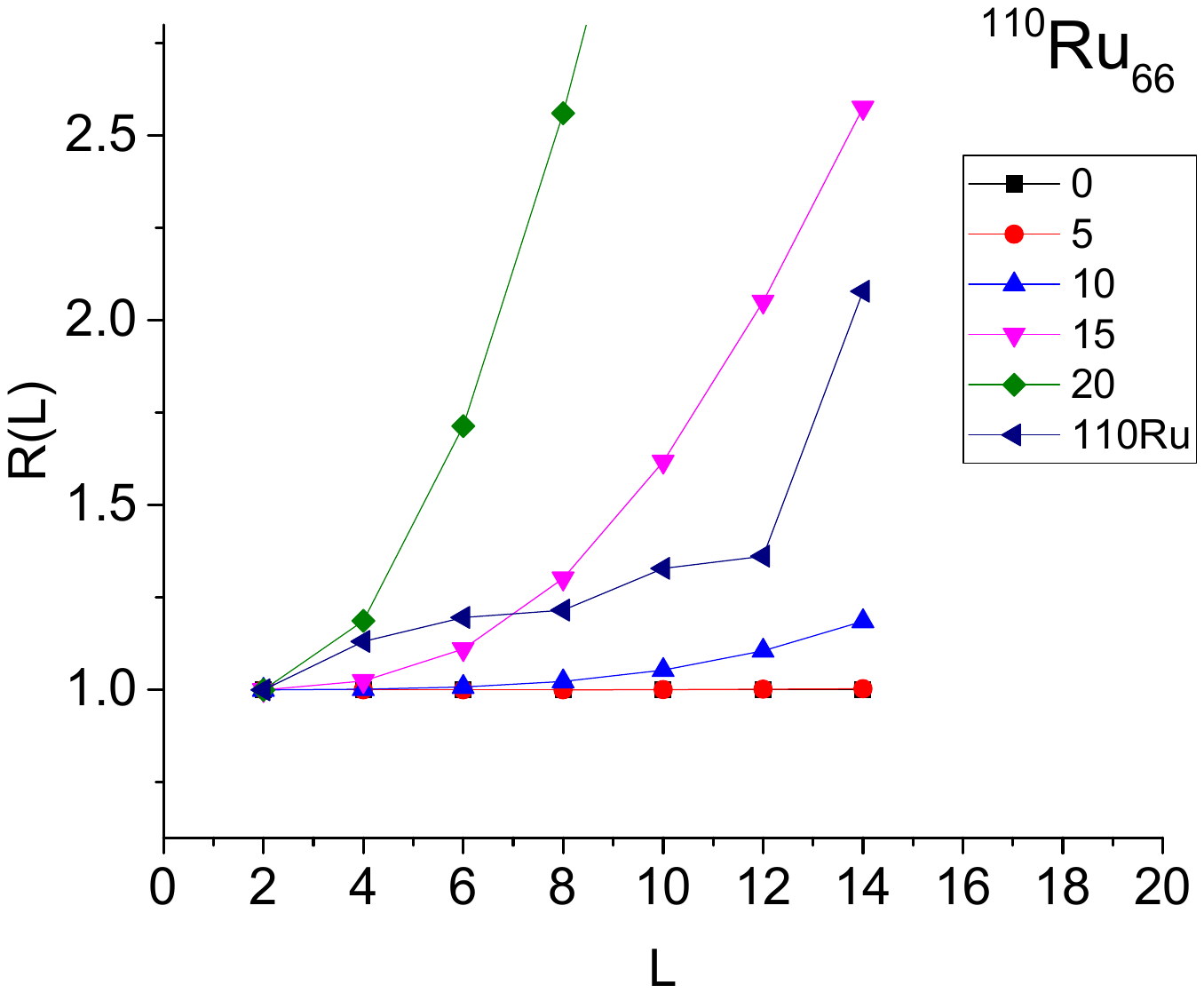}}
      {\includegraphics[width=60mm]{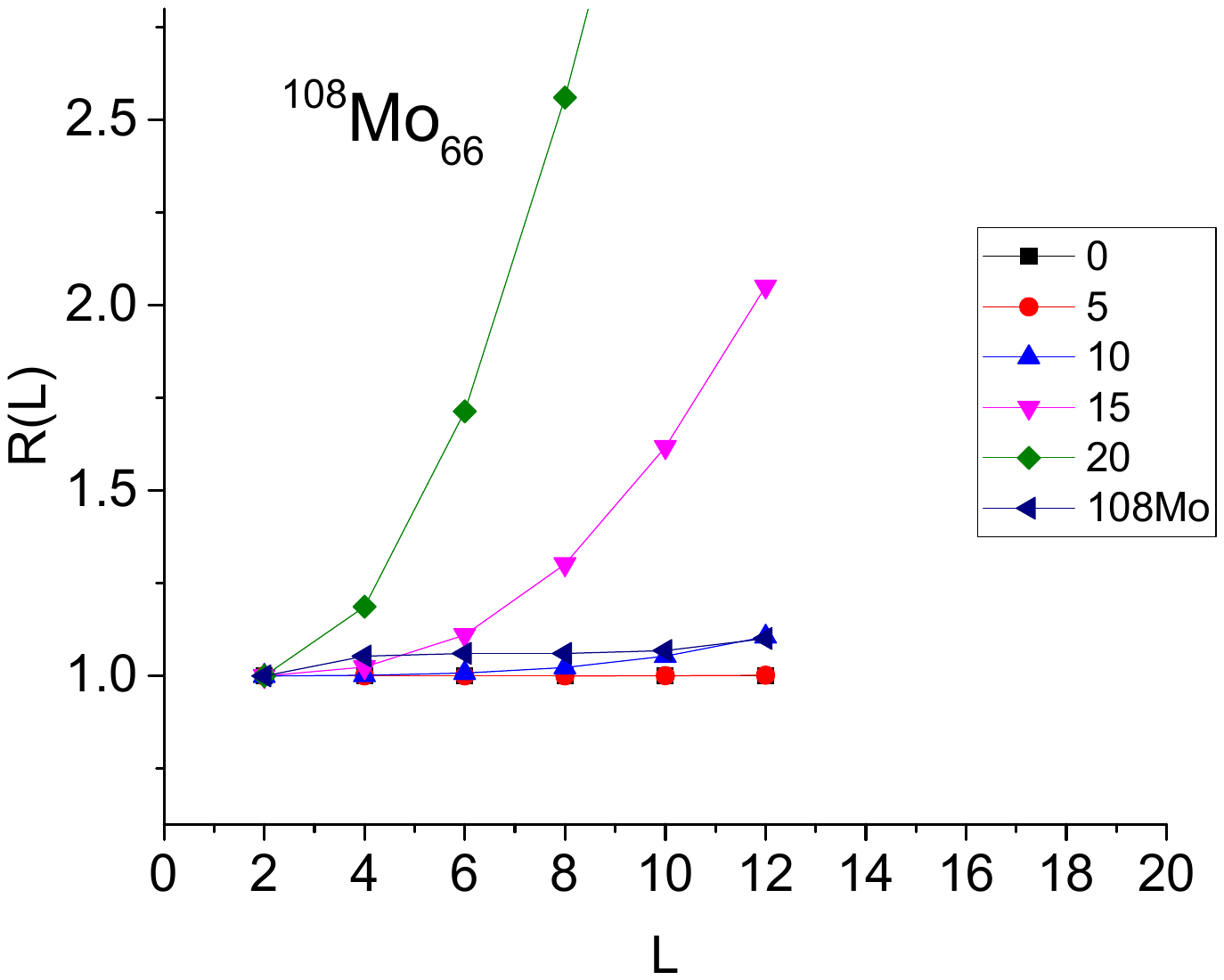}\hspace{5mm}
    \includegraphics[width=60mm]{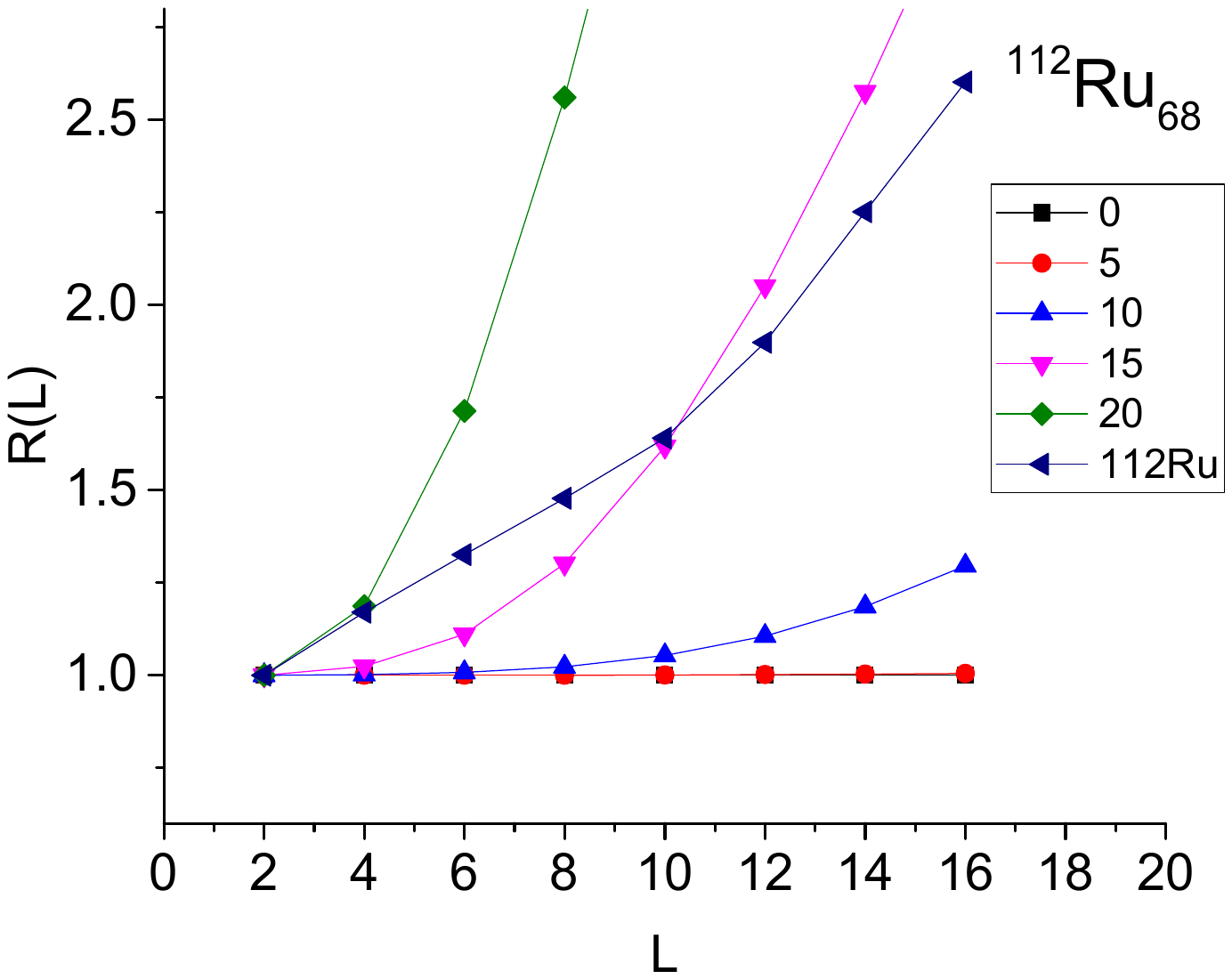}}
        \includegraphics[width=60mm]{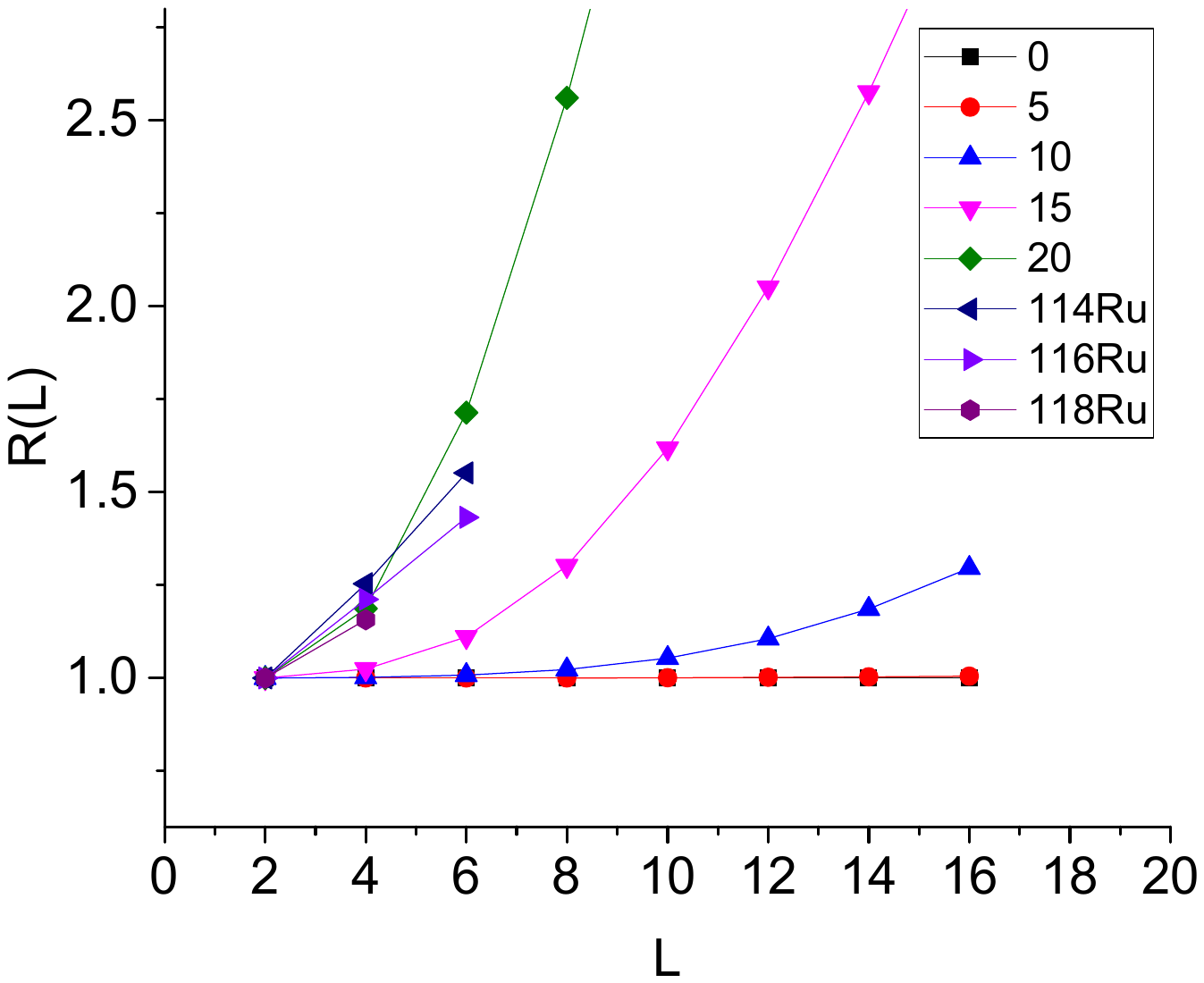}
    \caption{Ratios $R(L)$, given by Eq. (\ref{RL}), for various Mo and Ru isotopes. See sec. \ref{EmpTheor} for further discussion.} 
    \label{Fig7}
\end{figure*}

In Fig.~\ref{Fig7} it is evident that only the $N=66$ isotope of Mo and the $N=66$, 68 isotopes of Ru fall within the region corresponding to rigid triaxiality, while the $N=62$, 64 isotopes of Mo and the $N=64$ isotope of Ru clearly fail to do so.  


We therefore conclude that the ratios $R(L)$ lead to the same conclusions with the staggering quantity $S(L)$, being even better in clarifying questionable cases like $^{104}$Mo. In addition, the ratios $R(L)$ provide some estimate for the value of $\gamma$ for each nucleus. In particular, in $^{108}$Mo $\gamma$ ranges between 
15$^{\rm o}$ and 10$^{\rm o}$, while in $^{110,112}$Ru it acquires values between 20$^{\rm o}$ and 15$^{\rm o}$.


Empirical information and theoretical predictions for the $Z=40$-44, $N=60$-68 nuclei have been gathered in Table~\ref{dat}. 


\begin{table}[t!]

\caption{Empirical information and theoretical predictions for the $Z=40$-44, $N=60$-68 nuclei. The experimental \cite{ensdf} energy ratios  $R_{4/2}$ of Eq. (\ref{R42}) and
 $R$ of Eq. (\ref{R}) are shown, along with the empirical value of $\gamma$ extracted by Eq. (\ref{g}) from the $R$ values.  
 In addition, parameter-free predictions by the proxy-SU(3) model,    
  using the highest weight $(\lambda,\mu)$ irreps (proxy-SU(3)$_{\mathrm{hw}}$) and the irreps carrying the highest eigenvalue of the second order Casimir operator of SU(3)
  (proxy-SU(3)$_{\mathrm{C}}$), 
  derived using Table~I of  Ref.~\cite{Bonatsos2017b}, for $\gamma$ (derived from Eq. (\ref{mu})) and for $\beta$ (derived as described in Ref. \cite{Bonatsos2017b}), as well as predictions derived in the framework of the TPSM \cite{Rouoof2024,Jehangir2021,Zhang2015b}, by the HFB 
approach with the UNEDF0 parametrization of the Skyrme EDF \cite{Zhang2015b},  
as well as through covariant density functional theory involving the DD-ME2 and DD-PC1 parametrizations \cite{Abusara2017a,ElBassem2024} are also shown.  See Sec.~\ref{EmpTheor} for further discussion. 
}
\begin{tabular}{ l  c c c c c c }
\hline
N  & 60 &  62 & 64 & 66 & 68 & Ref. \\
\hline
$_{40}$Zr             &       &       &       &       &     & \\ 
$R_{4/2}$ exp  & 2.656 & 3.151 & 3.246 & 3.132 & 2.992 & \cite{ensdf}\\
R exp          & 5.628 & 6.828 &       &       & 3.466 &  \cite{ensdf} \\
$\gamma$ emp   & 16.5  & 15.0  &       &       & 20.7  &   \\
$\gamma$ proxy-SU(3)$_{\mathrm{hw}}$ &  7.0  & 1.3   & 9.4   & 12.4  & 10.0  &  \cite{Bonatsos2017b}\\
$\gamma$ proxy-SU(3)$_{\mathrm{C}}$ &  21.8  & 15.4   & 23.7   & 45.4  & 53.4  &  \cite{Bonatsos2017b}\\
$\beta$ exp   & 0.356 & 0.387 & 0.460 & 0.360   &      &  \cite{Pritychenko2016} \\
$\beta$ proxy-SU(3)$_{\mathrm{hw}}$ & 0.299 & 0.311 & 0.305 & 0.298 & 0.285 &  \cite{Bonatsos2017b}\\
$\beta$ proxy-SU(3)$_{\mathrm{C}}$ & 0.293 & 0.297 & 0.302 & 0.309 & 0.307 &  \cite{Bonatsos2017b}\\

\hline
$_{42}$Mo             &       &       &       &       &       & \\ 
$R_{4/2}$ exp  & 2.507 & 2.917 & 3.045 & 2.924 & 2.805 & \cite{ensdf}\\
R exp          & 2.859 & 4.227 & 4.142 & 3.040 & 2.312 & \cite{ensdf}\\
$\gamma$  emp  & 22.8  & 18.9  & 19.0  & 22.1  & 25.6  &   \\
$\gamma$ proxy-SU(3)$_{\mathrm{hw}}$ &  16.3 & 10.0  & 18.6  & 21.8  & 19.7  &  \cite{Bonatsos2017b}\\
$\gamma$ proxy-SU(3)$_{\mathrm{C}}$ &  26.5  & 19.7   & 28.4   & 50.6  & 58.7  &  \cite{Bonatsos2017b}\\
$\gamma$ TPSM  &       & 22.1  & 19.5  & 25.4  &       & \cite{Jehangir2021,Rouoof2024} \\
$\gamma$ TPSM  &       &       &  20.  &  25.  &       &  \cite{Zhang2015b} \\
$\gamma$ HFB  &       &       &  16.  &  18.  &       &  \cite{Zhang2015b} \\
$\gamma$ CDFT DD-ME2 & 20. & 17. & 17. &  20.  &       &  \cite{Abusara2017a} \\
$\gamma$ CDFT DD-PC1 & 20. & 17. & 17. &  20.  &       &  \cite{Abusara2017a} \\
$\gamma$ CDFT DD-ME2 & 20. & 15.&  15.  &  60.  &   60. & \cite{ElBassem2024} \\ 
$\gamma$ CDFT DD-PC1 & 20. & 15.&  60.  &  60.  &   60. &  \cite{ElBassem2024} \\ 
$\beta$ exp   & 0.314 & 0.354 & 0.351 & 0.403   &      &  \cite{Pritychenko2016} \\
$\beta$ proxy-SU(3)$_{\mathrm{hw}}$ & 0.281 & 0.289 & 0.290 & 0.285 & 0.271 &  \cite{Bonatsos2017b}\\ 
$\beta$ proxy-SU(3)$_{\mathrm{C}}$ & 0.274 & 0.276 & 0.284 & 0.301 & 0.304 &  \cite{Bonatsos2017b}\\
$\beta$ HFB &             &       &   0.19  &  0.18   &       & \cite{Zhang2015b} \\
$\beta$ CDFT DD-ME2 & 0.33 & 0.37 & 0.39 &  0.37  &       & \cite{Abusara2017a} \\
$\beta$ CDFT DD-PC1 & 0.33 & 0.38 & 0.36 &  0.37  &       & \cite{Abusara2017a} \\
$\beta$ CDFT DD-ME2 & 0.35 & 0.4 &  0.4  &  0.25  &  0.25 & \cite{ElBassem2024} \\ 
$\beta$ CDFT DD-PC1 & 0.3 & 0.4 &  0.25  &  0.25  &  0.25 & \cite{ElBassem2024} \\

\hline

\hline
\end{tabular}
\label{dat}
\end{table}

\begin{table}[t!]
\begin{tabular}{ l  c c c c c c }
\hline
$_{44}$Ru             &       &       &       &       &       &  \\ 
$R_{4/2}$ exp  & 2.482 & 2.646 & 2.746 & 2.756 & 2.725 &  \cite{ensdf}\\
R exp          & 2.495 & 2.934 & 2.922 & 2.546 & 2.212 &  \cite{ensdf}\\
$\gamma$ emp  &  24.5 & 22.5  & 22.5  & 24.2  & 24.1  &  \\
$\gamma$ proxy-SU(3)$_{\mathrm{hw}}$ &  20.9 & 13.9  & 23.1  & 26.5  & 24.5  &  \cite{Bonatsos2017b}\\
$\gamma$ proxy-SU(3)$_{\mathrm{C}}$ &  20.9  & 13.9   & 23.1   & 46.9  & 55.7  &  \cite{Bonatsos2017b}\\
$\gamma$ TPSM  &       &       & 28.2  & 27.3  & 24.2  & \cite{Jehangir2021,Rouoof2024} \\
$\gamma$ TPSM  &       &       &  29.  &  28.  & 25.   &  \cite{Zhang2015b} \\
$\gamma$ HFB  &       &       &  24.  &  25.  & 24.   &  \cite{Zhang2015b} \\
$\gamma$ CDFT DD-ME2& 22. & 25. & 40.   &55.    & 55.   & \cite{Abusara2017a} \\
$\gamma$ CDFT DD-PC1& 22. & 22. & 40.   &55.    & 55.   & \cite{Abusara2017a} \\
$\beta$ exp   & 0.272 & 0.306 & 0.276 & 0.298   & 0.299 & \cite{Pritychenko2016} \\
$\beta$ proxy-SU(3)$_{\mathrm{hw}}$ & 0.260 & 0.266 & 0.270 & 0.267 & 0.252 &  \cite{Bonatsos2017b}\\ 
$\beta$ proxy-SU(3)$_{\mathrm{C}}$ & 0.260 & 0.266 & 0.270 & 0.279 & 0.278 &  \cite{Bonatsos2017b}\\
$\beta$ HFB &             &       & 0.16  &  0.16    &  0.15  &  \cite{Zhang2015b} \\
$\beta$ CDFT DD-ME2 & 0.25 & 0.25 & 0.25 &  0.23  & 0.24  &  \cite{Abusara2017a} \\
$\beta$ CDFT DD-PC1 & 0.27 & 0.27 & 0.25 &  0.23  & 0.24  &  \cite{Abusara2017a} \\

\hline
\end{tabular}
\label{dat}
\end{table}

On the empirical side, the energy ratio 
\begin{equation} \label{R42}
R_{4/2} = {E(4_1^+) \over E(2_1^+)} ,
\end{equation}  
a well known indicator of collectivity, is shown, in addition to the energy ratio $R$ of Eq. (\ref{R}) and the empirical value of $\gamma$ extracted by Eq. (\ref{g}).    

On the theoretical side, predictions for $\gamma$ by several models are shown. Proxy-SU(3) predictions have been derived from Eq. (\ref{mu}), using the 
highest weight $(\lambda,\mu)$ irreps (proxy-SU(3)$_{\mathrm{hw}}$) and the irreps carrying the highest eigenvalue of the second order Casimir operator of SU(3) 
(proxy-SU(3)$_{\mathrm{C}}$), derived using Table~I of  Ref.~\cite{Bonatsos2017b}. In addition, predictions derived in the framework of the 
TPSM \cite{Rouoof2024,Jehangir2021,Zhang2015b}, by the HFB 
approach with the UNEDF0 parametrization of the Skyrme EDF \cite{Zhang2015b},  
as well as through covariant density functional theory involving the DD-ME2 and DD-PC1 parametrizations \cite{Abusara2017a,ElBassem2024} are also shown. 

On the contents of Table~\ref{dat} the following comments apply.

\begin{enumerate}
\item 
The nuclei shown possess $R_{4/2}$ values between the axially deformed limit 10/3=3.333 and the triaxial deformed limit 8/3=2.667
 (corresponding to $\gamma =30^{\mathrm{o}}$), or slightly below it. Equation~(\ref{g}) is based on the rigid triaxial rotor, thus it should be applicable \cite{Varshni1970} for the determination of the empirical value of $\gamma$ in the region of $R_{4/2}$ around 2.667 and not very far from it. 

\item 
The experimental ratio $R$ of Eq. (\ref{R}) remains below 6.85, resulting through Eq. (\ref{g}) in empirical $\gamma$ values 15$^{\rm o}$ or higher, up to maximal triaxiality (30$^{\rm o}$). In most cases, $R$ remains below 3.73, which corresponds to $\gamma$ values above 20$^{\rm o}$. 

\item 
For the Ru and Mo isotopes, there is in general good agreement between the theoretical predictions of the various models and the empirical $\gamma$ values.
The constraints discussed in 1) and 2) are satisfied in this case. 

\item 
In the case of the Zr isotopes, however, a different picture appears. First, there is no extensive literature with theoretical predictions, as for Mo and Ru. Second, the proxy-SU(3) predictions are much lower than the empirical values of $\gamma$. This discrepancy can be due to the fact that in the case of the Zr isotopes the 
constraints of 1) and 2) are only loosely satisfied. In most cases in the Zr isotopes the $R_{4/2}$ ratio is well above the rigid triaxial limit of 2.667, while in addition the $R$ ratio is above the 3.73 value corresponding to 20$^{\rm o}$, and therefore the applicability of Eq.~(\ref{g}) becomes questionable. 

\item 
While the predictions of proxy-SU(3)$_{\mathrm{C}}$ and proxy-SU(3)$_{\mathrm{hw}}$ for the $\beta$ collective variable are very similar in all cases, differences appear in their predictions for the $\gamma$ collective variable beyond midshell, where the proxy-SU(3)$_{\mathrm{C}}$ predictions jump to oblate values above $30^{\rm o}$, while the proxy-SU(3)$_{\mathrm{hw}}$ predictions remain at prolate values below $30^{\rm o}$. This jump is also seen in some of the CDFT calculations reported in Table~\ref{dat},
 suggesting that in those cases the restrictions of the Pauli principle and the short-range nature of the nucleon-nucleon interaction \cite{Martinou2021b} are not sufficiently incorporated in the relevant CDFT models.
\end{enumerate}

 The above observations imply that while triaxiality is clearly present in the Mo and Ru isotopes under consideration, it seems that it is not present to a comparable extent in the Zr isotopes. The origin of this discrepancy calls for further investigation. A reasonable argument is based on the particle/hole nature of the valence nucleons. For $N=60$-66, valence neutrons are particles. In the case of $_{42}$Mo and $_{44}$Ru, valence protons correspond to holes. Therefore, for the $N=60$-66 isotopes of the Mo and Ru chains, the particle-hole environment favoring the appearance of SU(3)$^*$ appears. This is not the case for the $_{40}$Zr isotopes, which lie very close to the $Z=39$ midshell of the 28-50 shell. Therefore triaxial effects are expected to be much weaker in Zr, in comparison to Mo and Ru.  
 
 It should be noticed that in general no data exist \cite{ensdf} for the $B(E2)$ transition rates involved in the branching ratio $R_2$ of Eq. (\ref{R2}) for the Zr, Mo, and Ru isotopes under consideration, while such data do exist for the Mo isotopes with $N=50$-58 and the Ru isotopes with $N=52$-58. The only exception is $^{104}$Ru$_{60}$, for which the data give $R_2 = 19.64$, which is consistent with the empirical value of $\gamma=24.5^{\rm o}$ obtained from its $R=2.495$ ratio, corroborating the early characterization of $^{104}$Ru$_{60}$ as a textbook example of SU(3)$^*$ \cite{Dieperink1982}. Measuring the branching ratio $R_2$ for the $Z=40$-44, $N=60$-68 nuclei would be very useful for understanding triaxiality in this region.

\section{Algebraic collective model}\label{ACMsec}

The ACM, introduced as a computationally tractable version of the BM \cite{BM}  restricted to rotational and quadrupole vibrational degrees of freedom, is characterized by a well defined algebraic structure.
Unlike the conventional $U(5)\supset SO(5)\supset SO(3)$
dynamical subgroup chain used, for example,
 in the Frankfurt program \cite{HessMG81,GneussG71}, the ACM
makes use of the subgroup chain
\begin{equation}
 SU(1,1) \times SO(5) \supset U(1) \times SO(3) \supset SO(2) 
 \label{group-chain}
\end{equation}
to define basis wave functions as products of $\beta$ wave functions and
$SO(5)$ spherical harmonics.
Several advantages result from this choice of dynamical subgroup chain:
(i) with the now available SO(5) Clebsch-Gordan coefficients \cite{CRW,TAW}, and explicit expressions for SO(5) reduced matrix elements,
matrix elements of BM operators can be calculated analytically;  (ii) by appropriate choices of SU(1,1) modified oscillator representations, the $\beta$ basis wave  functions range from those of the $U(5) \supset SO(5)$ harmonic vibrational model to those of the rigid-$\beta$ wave function of the $SO(5)$-invariant Wilets-Jean model; and (iii) with these SU(1,1) representations, collective model calculations converge an order of magnitude more rapidly for deformed nuclei than in $U(5) \supset SO(5)$ bases.
Thus, the ACM  combines the advantages of  the BM and the IBM and
makes collective model calculations a simple routine procedure
\cite{RWC,DJR,RT,Caprio}. A pedagogical treatment of the geometrical and algebraic foundations of the ACM can be found in the book by Rowe and Wood \cite{bookRW}.

In our implementation of the ACM, observables of interest are formed by taking sums of products of several generating operators, that are SO(3) invariant, such as $\beta^{2}$, 
$\beta^{-2}$, $\beta\frac{d}{d\beta}$, $\Delta^{2}$.
The Hamiltonians that can be analyzed are polynomials in these operators with coefficients that may be arbitrary real numbers or functions of the quantum numbers on which they operate.

The ACM enables a wide range of model Hamiltonians to be analyzed. This range includes essentially all Hamiltonians that are rational functions of the model quadrupole moments and contain terms that are at most quadratic functions of the corresponding conjugate momenta. In the case of application to nuclei that exhibit some degree of triaxiality, either soft or rigid, an important concern is the extent to which it is possible to obtain minima at arbitrary values of $\beta$ and $\gamma$ and corresponding arbitrary rigidity. The Hamiltonian below is suitable for this purpose. The advantage is that it is relatively simple and contains a term proportional to the parameter $\kappa$, which can induce triaxiality
\begin{align}
\label{ACM}
\hat H(B,\alpha, \kappa, \chi) 
& = -\frac{1}{2B}\nabla^2+\frac{1}{2}B[(1-2\alpha)\beta^{2}+\alpha\beta^{4}] 
    \nonumber \\
& \quad -\chi\beta^{3}\cos 3\gamma + \kappa \cos^{2}3\gamma ,
\end{align}
where
\begin{equation}
 \nabla^2 = \frac{1}{\beta^4} \frac{\partial}{\partial\beta} \beta^4 \frac{\partial}{\partial\beta} - \frac{1}{\beta^2} \hat\Lambda^2
\end{equation}
is the  Laplacian on the five-dimensional (5D) collective model space  and $\hat\Lambda^2$ is the $SO(5)$ angular momentum operator \cite{bookRW}.

\begin{figure*}[t!]

    {\includegraphics[width=60mm]{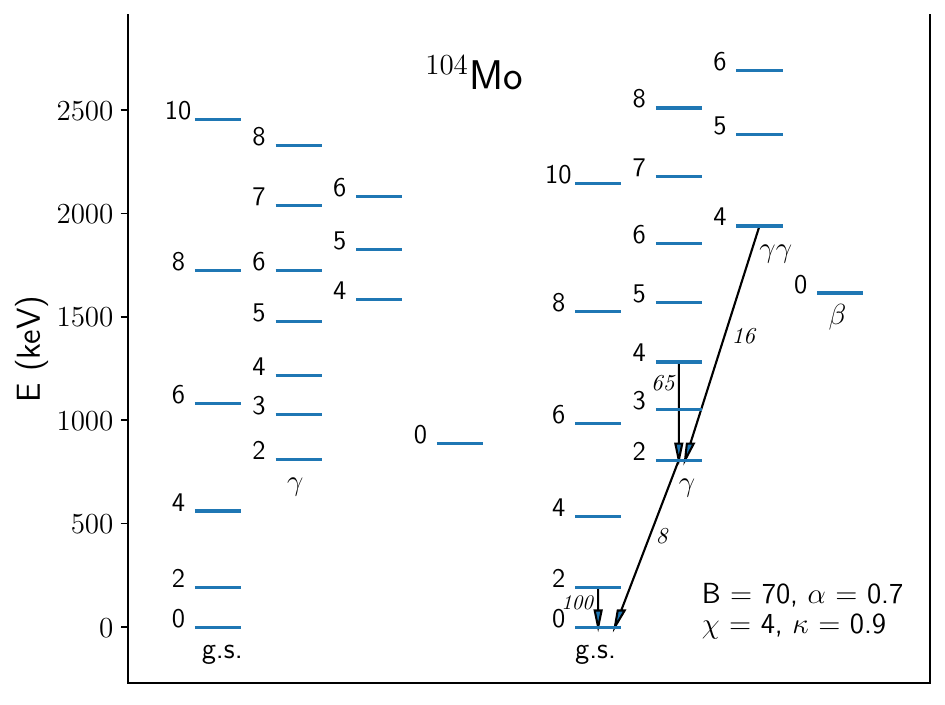}\hspace{5mm}
    \includegraphics[width=60mm]{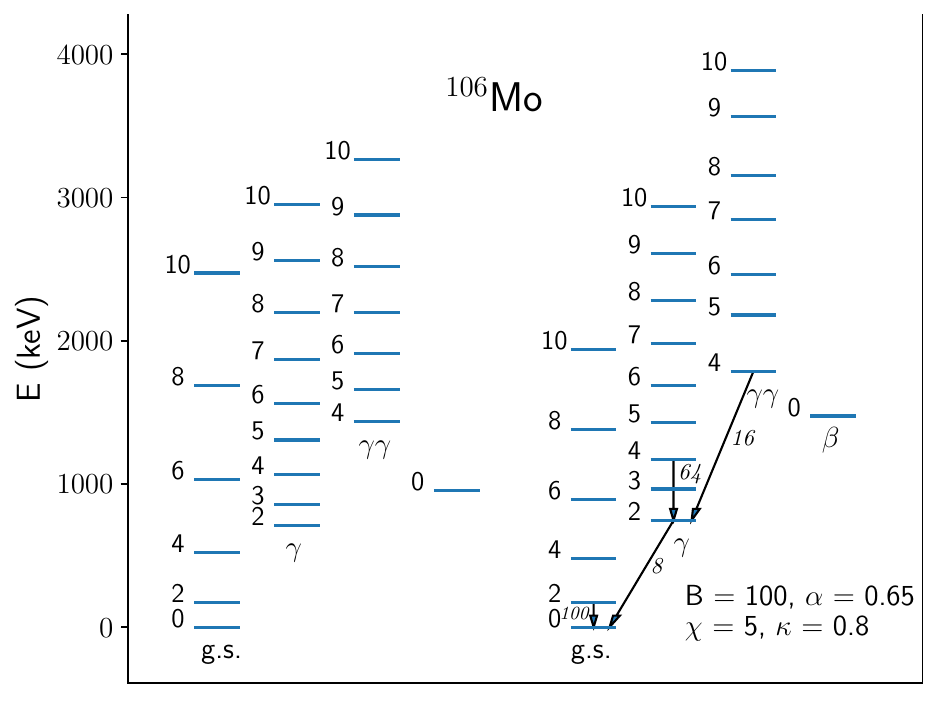}}
      {\includegraphics[width=60mm]{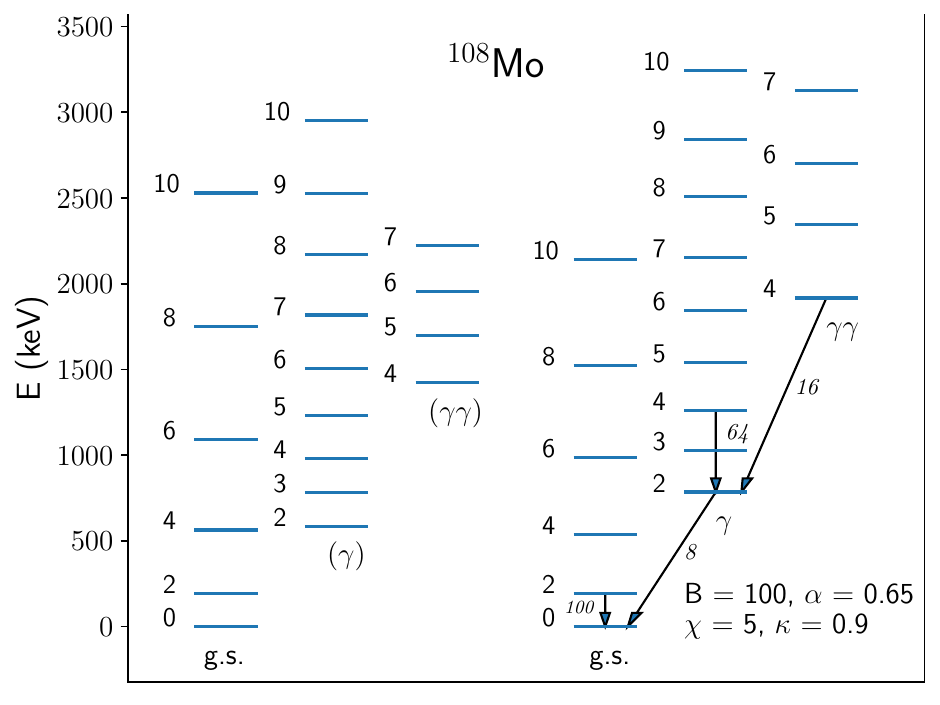}\hspace{5mm}
    \includegraphics[width=60mm]{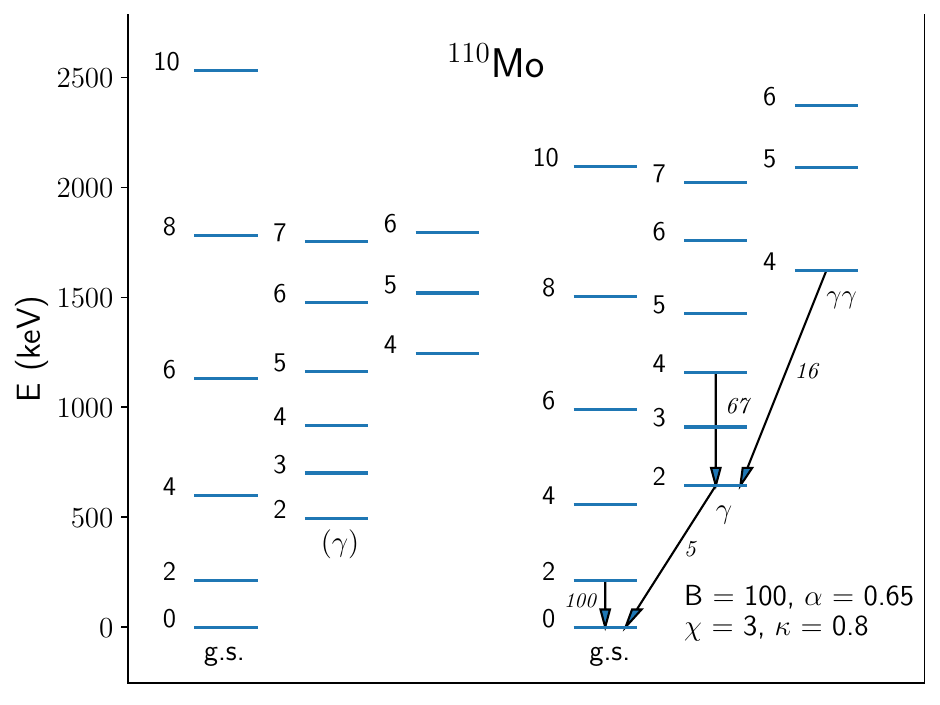}}
   
    \caption{Comparison of experimental (left) and theoretical ACM (right) spectra of $^{104-110}$Mo. 
    Uncertainties in the spin-parity or band assignment of experimental levels are denoted by parentheses around the corresponding bandheads. Only positive parity states
    are shown.} 
    \label{Fig1-ACM}
\end{figure*}

\begin{figure*}[t!]

    \includegraphics[width=140mm]{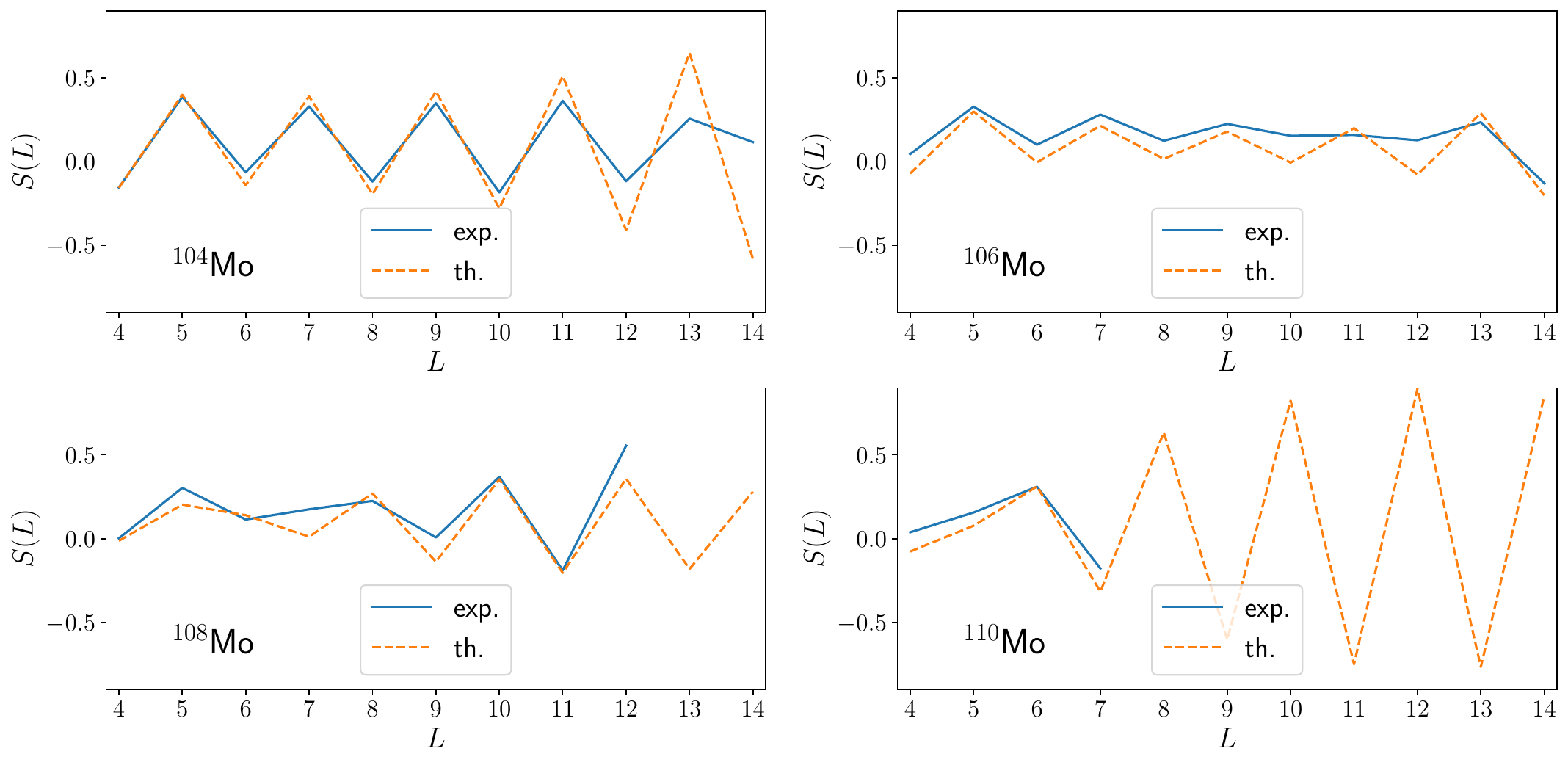}
   
    \caption{Comparison of experimental and theoretical ACM values of the odd-even staggering $S(L)$ in $^{104-110}$Mo.} 
    \label{Fig2-ACM}
\end{figure*}

The transition from axial to triaxial shapes has been studied in the framework of the Bohr Hamiltonian by Caprio \cite{Caprio2009,Caprio2011}, using the potential
\begin{equation}\label{Capr}
V(\gamma)= \chi [ (1-\cos 3\gamma) + \xi \cos^2 3\gamma] .
\end{equation} 
For $\xi=0$  a simple potential $1-\cos3\gamma$ is obtained. Using the Taylor expansion of $\cos 3\gamma$, one sees that this potential is harmonic ($\propto \gamma^2$) 
around $\gamma=0$. At $\xi=1/2$, the potential is still confining but softer, being proportional to $\gamma^4$ around $\gamma=0$. From Ref. \cite{Iachello2003} it is known that this is corresponding to the critical point of a second order quantum phase transition. For $\xi>1/2$, the potential has a minimum at the point $\gamma_0$, with $\cos 3 \gamma_0 = 1/(2\xi)$. In Fig.~1 of Ref. \cite{Caprio2009} one sees that for $\xi=0$ strong odd-even staggering is obtained in the $\gamma$ band, with minima at even $L$, indicating 
$\gamma$-softness, while for $\xi=1/2$ a rather flat staggering is obtained, with minima still at even $L$. The picture is opposite for $\xi>1/2$, for which strong staggering is seen with minima at odd $L$, indicating $\gamma$-rigidity. In short, the potential of Eq. (\ref{Capr}) describes a second order quantum phase transition from $\gamma$-soft to rigid triaxial shapes.  

This potential has been used in the further development of the ACM \cite{RWC,Caprio2011}, in which the onset of rigid triaxiality 
has been explicitly considered \cite{Caprio2011}.
From the physics point of view, Fig.~1 of Ref. \cite{Caprio2009} says that for small $\xi$ one has a $\gamma$-soft ($\gamma$-unstable) behavior, while at high $\xi$ 
$\gamma$-rigid triaxiality is obtained. In the case of the ACM, the $\gamma$-soft behavior is due to the O(5) symmetry involved in the SU(1,1)$\times$SO(5) dynamical group of the ACM. 
In the IBM framework, it is due to the O(5) symmetry underlying both the U(5) and O(6) dynamical symmetries, allowing the use of seniority as a good quantum number 
and leading to the grouping 2, (3,4), (5,6), \dots within the $\gamma$-band, which corresponds to minima at even $L$ for the $S(L)$ staggering quantity. 
It should be remembered that the $\cos 3\gamma$ term appears in the classical limit of IBM-1 \cite{Iachello1987}, involving only one-body and two-body terms, while the term $\cos^2 3\gamma$ arises in IBM-1 when three-body (cubic) terms are added \cite{VanIsacker1981,Heyde1984}. 

At this point, it is appropriate to include a comment concerning the $\beta$ parameter in the ACM. In the ACM, the parameter $\beta$ quantifies the quadrupole deformation of a nucleus -- that is, how much its shape deviates from spherical symmetry. It serves as a coordinate in the five-dimensional collective space, paired with $\gamma$, which captures triaxiality. Unlike the traditional Bohr model (BM), the ACM employs a mathematically  rigorous algebraic structure based on the subgroup chain of Eq.~(\ref{group-chain}) 
to construct nuclear wavefunctions. This formalism enables greater flexibility and precision in modeling nuclear shapes, but it also means that $\beta$ in the ACM is 
model dependent.  Rescaling $\beta$ in the ACM is conceptually straightforward but not trivial in practice. Since $\beta$ in the ACM is defined in oscillator units, its numerical values do not directly correspond to deformation parameters from macroscopic or microscopic models. To compare with, say, HFB or Nilsson model predictions, one would need to apply a scaling factor that accounts for the oscillator length and the specific basis used. Thus, this rescaling is not routinely performed, partly because the scaling factor depends on model details and is not uniquely defined and the ACM  is often used for relative comparisons (e.g. shape evolution, transition patterns), where absolute
$\beta$ values are less critical. For this reason, our comparison between the ACM and HFB approaches focuses primarily on the $\gamma$ parameter, which retains a more consistent geometric interpretation across models and provides a more reliable basis for cross-framework analysis.

In Fig.~\ref{Fig1-ACM} we show a comparison of experimental and theoretical spectra of $^{104-110}$Mo.
As can be seen, the theoretical spectra reasonably coincide with the experimental ones. In all cases we also show a 
double-$\gamma$ $K=4$ band where it is believed that one 
of the possible reasons for its anharmonicity may be an emerging triaxiality of the nucleus. However, these bands lie systematically higher than their experimental 
counterparts and thus their anharmonicity is somewhat overestimated. It is important to note that, in general the ACM tends to overestimate the energies of $\beta$ 
and double-$\gamma$ vibrational bands due to its simplified treatment of collective dynamics and limited coupling to non-collective degrees of freedom. In particular, 
it does not explicitly account for interactions with particle-hole excitations or other microscopic configurations that can lower vibrational energies in real nuclei. 
As spin increases, additional effects such as centrifugal stretching further complicate the spectrum, reducing the model’s accuracy. Consequently, while ACM reproduces 
low-spin spectra reasonably well, it systematically predicts higher vibrational bands at energies above experimental values. Since the ACM provides its most reliable
 predictions in the low- to mid-spin regime, we have limited our calculations to spin values up to about 11. The above arguments are, of course, also valid for the 
 further discussed Ru isotopes.

One can observe that the ACM parameters are more or less stable with the exception of $^{104}$Mo 
where the value of the mass parameter $B$ is smaller than for the other isotopes. This can be attributed to several factors. Positioned before mid-shell ($N=66$), 
$^{104}$Mo exhibits greater $\gamma$-softness and structural flexibility, which naturally leads to a lower mass parameter. In contrast, Ru isotopes ($Z=44$) have two
additional protons, enhancing shell effects and resulting in a more rigid structure. The potential energy surface (PES) of $^{104}$Mo is likely softer and more complex, with multiple competing 
shallow minima. During the ACM fitting process, this softness may cause the model to settle into a configuration with reduced collective inertia, reflected in the 
lower value of the mass parameter. 
The parameter $\chi$ is the smallest in the case of $^{110}$Mo, thus emphasizing the role 
of the parameter $\kappa$ which is quite essential for the description of the triaxial deformation (in the case of $\kappa=0$ the axial symmetry is preserved). In some 
cases there is also a candidate for $\beta$ vibration but its theoretical partner is found typically much higher. In any case the experimental assignment of this 
vibration is problematic and uncertain in most of the cases.

As discussed in Sec.~\ref{sig} the odd-even staggering $S(L)$ is a very sensitive indicator of triaxiality. 
It is expected that in the case of soft triaxiality the values of $S(L)$ should be negative for $L$ even and positive for $L$ odd. In the case of rigid triaxiality the 
opposite is true. The transition between the two staggering patterns is expected for the mid-shell isotope i.e.\ $^{108}$Mo in this case. Experimental staggerings shown in 
Fig.~\ref{Fig2-ACM} are more or less in line with this expected trend. Theoretical staggerings (even if very sensitive to slight parameter changes) are in reasonable 
agreement with the experimental ones. The exception is $^{106}$Mo where it is quite irregular which is difficult to describe in the framework of the ACM. In the case of $^{110}$Mo which is already beyond the mid-shell there is unfortunately not enough data, but the staggering tends to show a pattern expected for more rigid triaxial shape 
with $L=6$. 

\begin{figure*}[t!] 

    {\includegraphics[width=60mm]{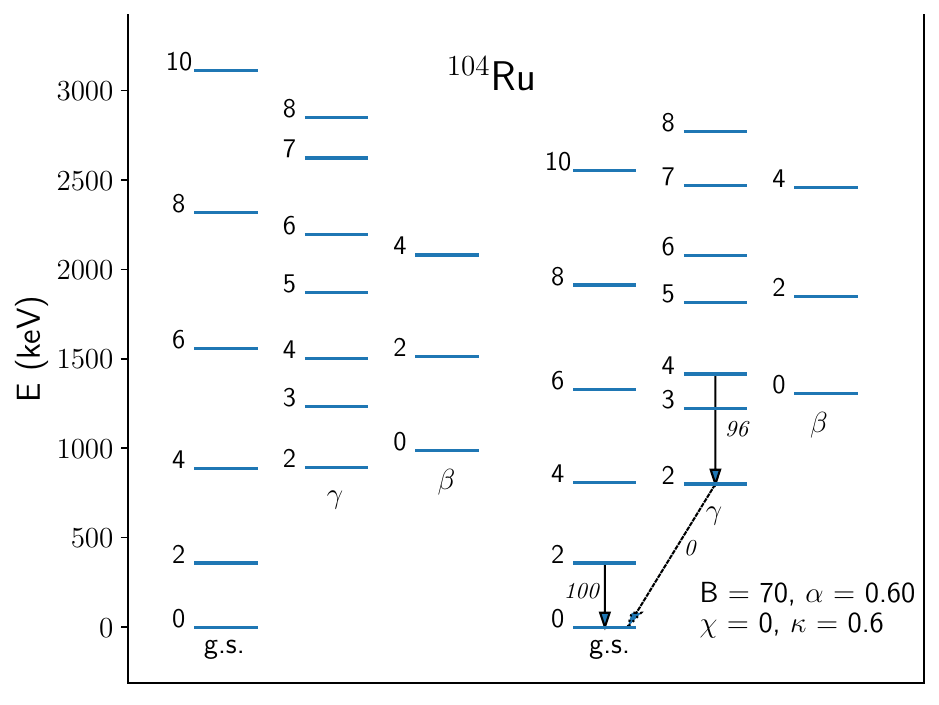}\hspace{5mm}
    \includegraphics[width=60mm]{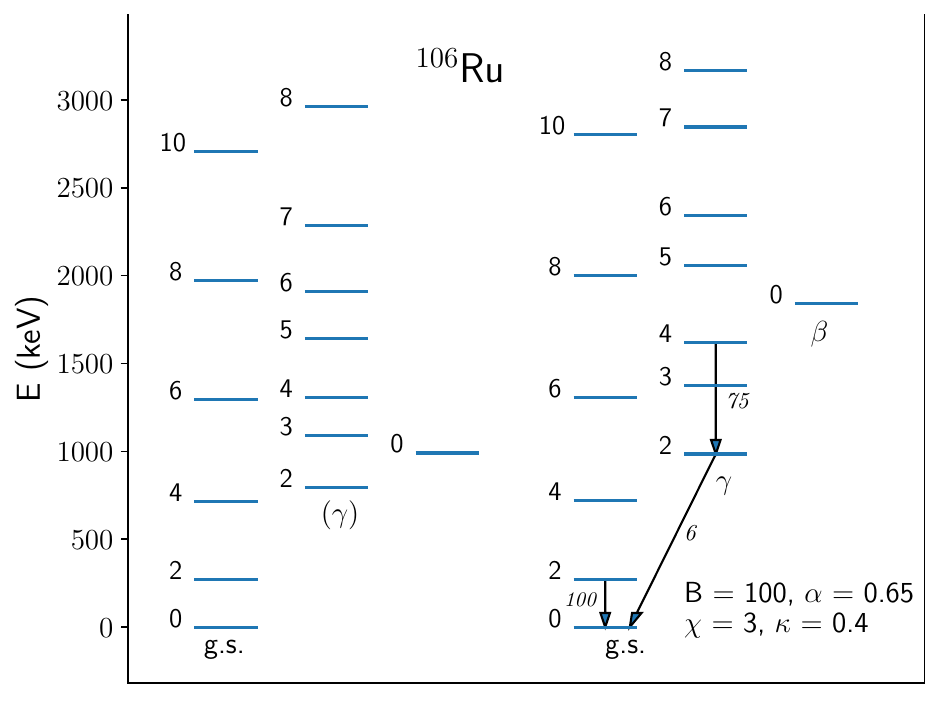}}
      {\includegraphics[width=60mm]{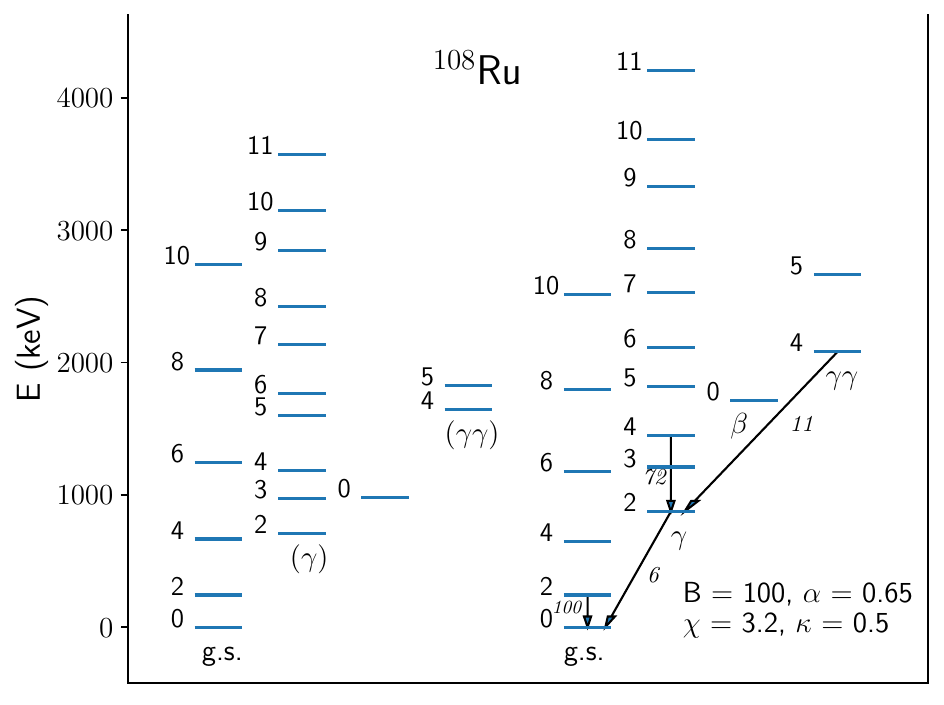}\hspace{5mm}
    \includegraphics[width=60mm]{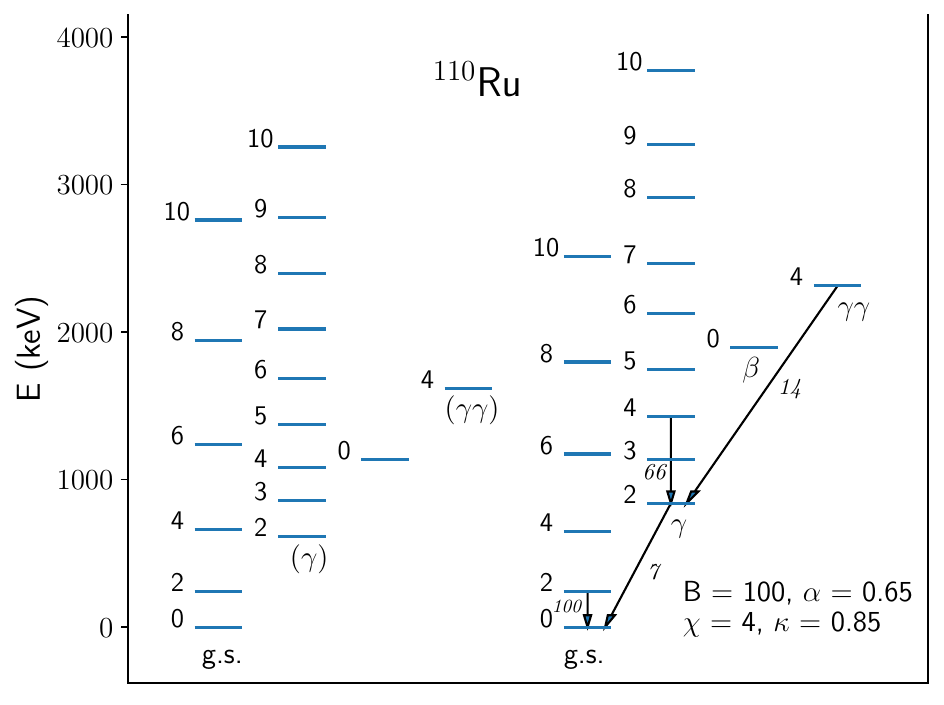}}
    {\includegraphics[width=60mm]{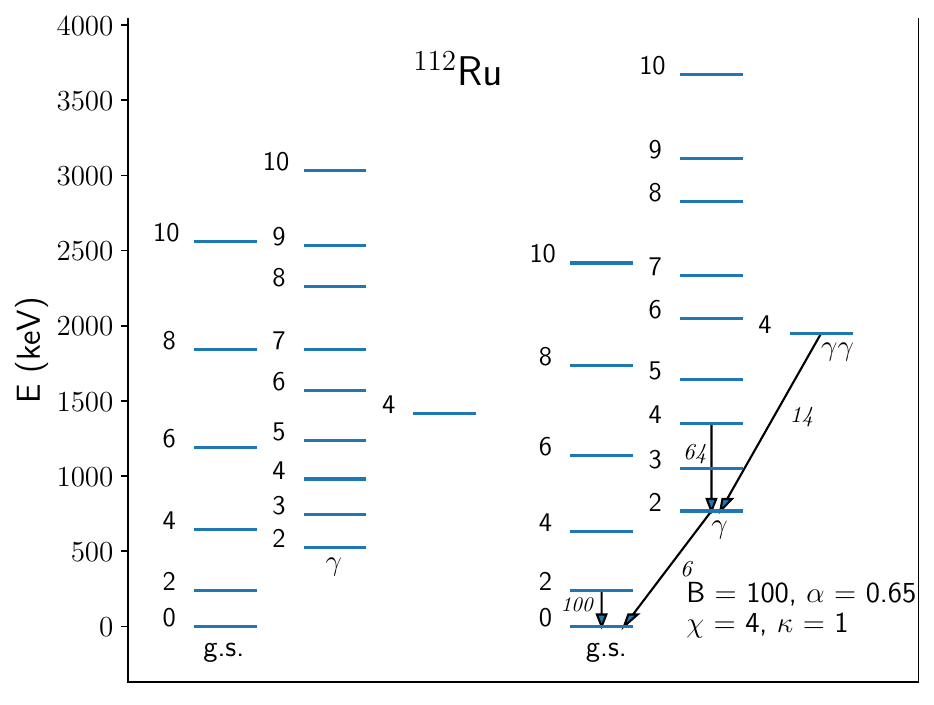}\hspace{5mm}
    \includegraphics[width=60mm]{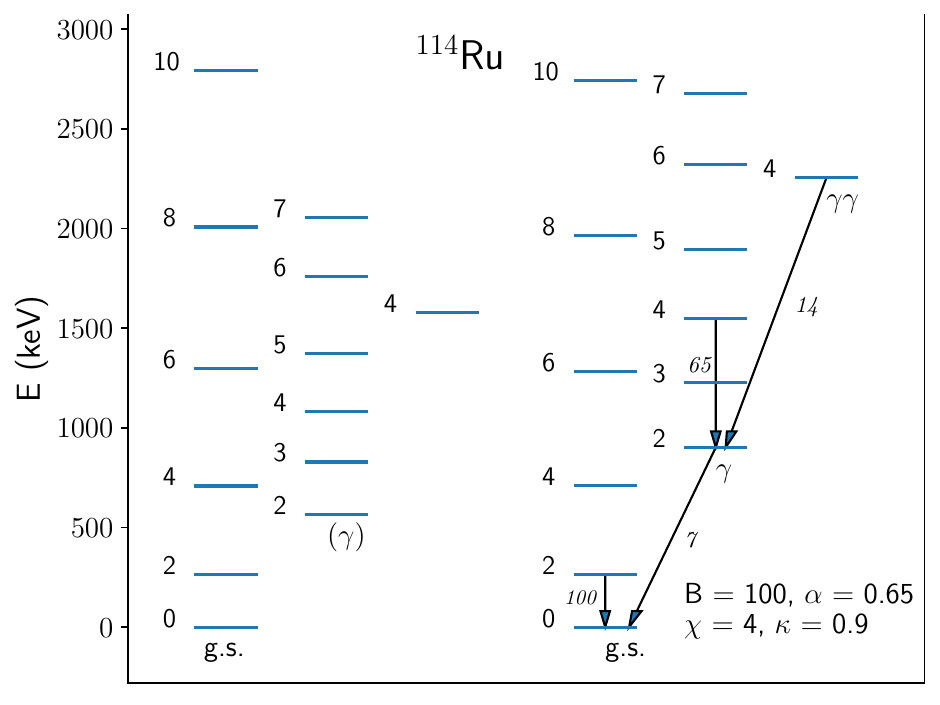}}
      \includegraphics[width=60mm]{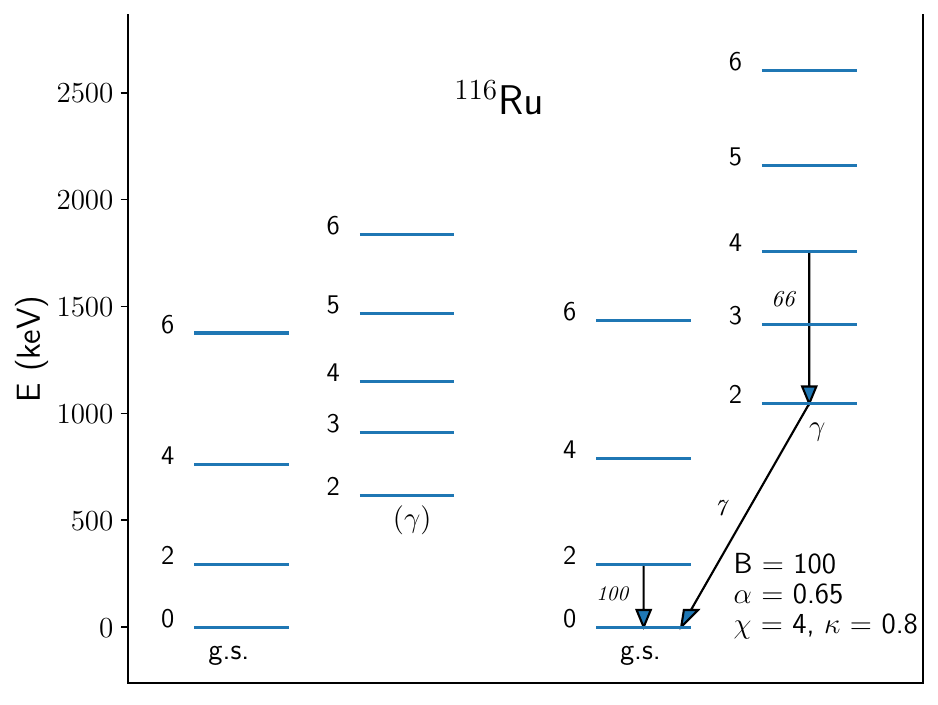}
   
    \caption{Comparison of experimental (left) and theoretical ACM (right) spectra of $^{104-116}$Ru.
    As in Fig.~\ref{Fig1-ACM}, parentheses are used to denote experimental level uncertainties and only positive parity states are shown.} 
    \label{Fig3-ACM}
\end{figure*}

\begin{figure*}[t!] 
    \includegraphics[width=140mm]{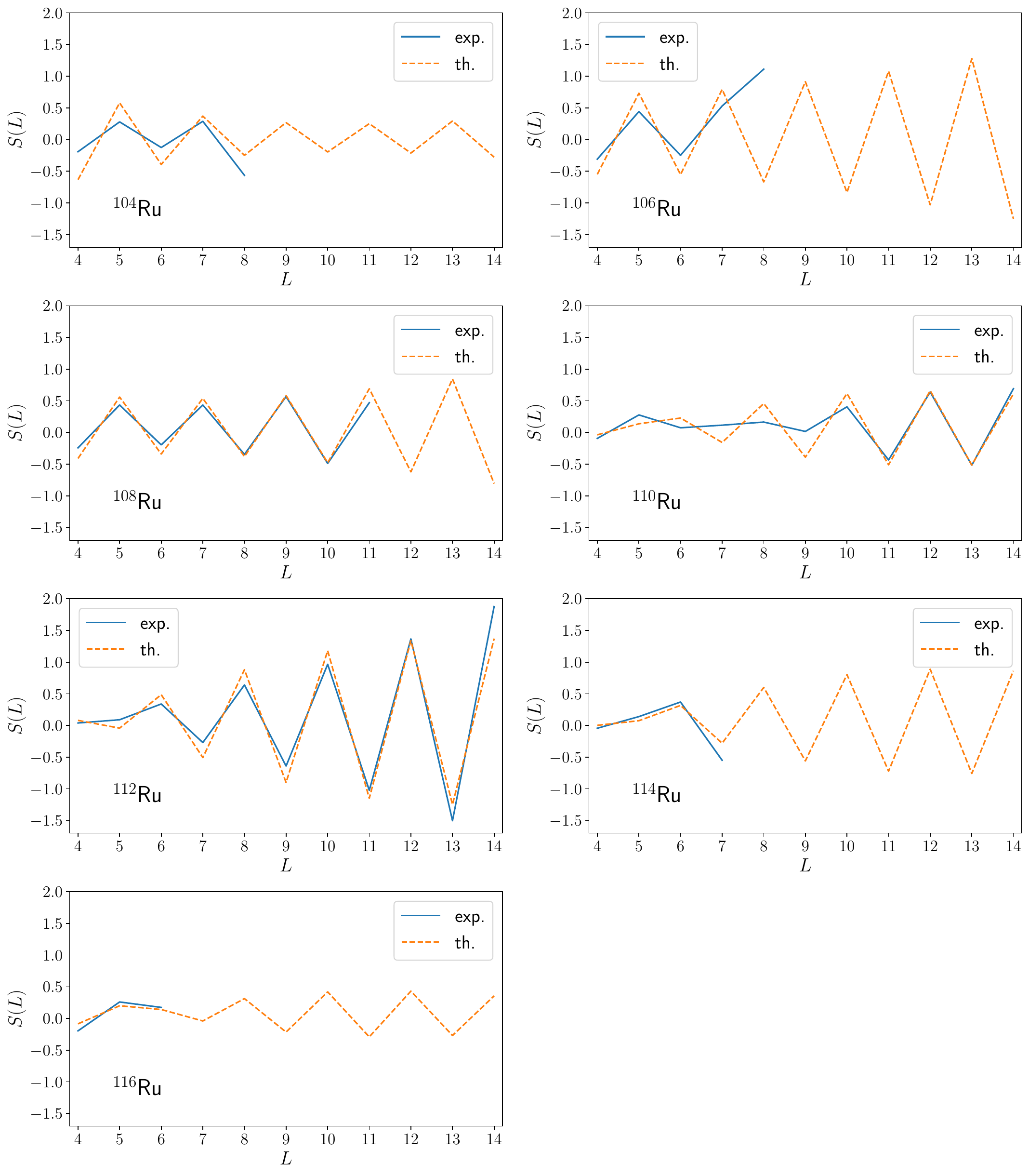}
    \caption{Comparison of experimental and theoretical ACM values of the odd-even staggering $S(L)$ in $^{104-116}$Ru.} 
    \label{Fig4-ACM}
\end{figure*}

In Fig.~\ref{Fig3-ACM} we show the experimental and theoretical spectra for $^{104-116}$Ru. The agreement of the two spectra is reasonable, but again, as in the previous 
case, the double $\gamma$ $K=4$ and $\beta$ vibrational bands are located higher than their experimental counterparts. The value of the parameter $\chi=0$ for $^{104}$Ru
 reflects the fact that the experimental ratio $R_{4/2}=2.48$ i.e. the value expected for a $\gamma$ independent potential.
The mid-shell nucleus is $^{110}$Ru and it is here that we expect a change in staggering pattern. This is indeed confirmed as can be seen in Fig.~\ref{Fig4-ACM}. 
$^{110}$Ru still exhibits a behavior typical for the $\gamma$-soft regime while the staggering pattern of $^{112}$Ru is already switched to the $\gamma$-rigid regime. 
In the case of $^{112,114}$Ru we do not have enough data points and the staggerings show some deviations from the expected behavior but the calculation reproduces it well. 
This evolution of staggering is reflected in the value of the parameter $\kappa$, which increases from $^{104}$Ru to $^{110}$Ru and then slightly decreases.

In Table~\ref{TabI-ACM} we show the theoretical values of ratios $R_{4/2}$, $R$, $R_{2}$ as defined in Secs.~\ref{sig} and \ref{EmpTheor} and the values of $\gamma$ determined from the 
ratio $R_{2}$ using the Eq. \ref{R2gamma}. In the case of Mo isotopes, the values of $R_{4/2}$ are slightly smaller than the experimental ones but generally in good agreement with
them. Moreover, they are close to the triaxial deformed limit of 2.67 and change gradually with increasing number of neutrons, as expected for a second-order phase transition.
 As discussed in Sec.~\ref{sig}, both $R$ and $R_{2}$ are needed to guarantee the appearance of triaxial shapes. These 
 calculated ratios are in the range expected for $\gamma$ ranging from  30$^{\mathrm{o}}$ to 15$^{\mathrm{o}}$. The calculated values of $R$, agree well with the experimental ones for 
 $^{104,106}$Mo but are somewhat bigger for $^{108,110}$Mo. The $\gamma$ values derived from $R_{2}$ are in reasonable agreement with the experimental ones in Table~\ref{dat} and 
 it is particularly interesting that its value increases significantly from $^{108}$Mo to $^{110}$Mo, i.e., when passing from $N=66$ to $N=68$. This is reproduced also by the
  calculation where an increase from 21.5$^{\mathrm{o}}$ to 24.6$^{\mathrm{o}}$ is observed and correlates with a big jump in $R_{2}$ from 7.17 to 17.32. 

In the case of Ru isotopes the experimental values of $R_{4/2}$ range from about 2.5 to 2.7, which is well reproduced by the calculation. The theoretical value is, however, 
quite small for $^{104}$Ru. The $R$ values are somewhat larger than the experimental ones but in both cases one can observe a plateau due to very close values of $R$ in
 $^{106,108}$Ru and a relatively important decrease of $R$ in $^{110}$Ru. This behavior can also be observed in Mo around the mid-shell (a decrease of $R$ from $^{106}$Mo 
 to $^{108}$Mo). This behavior may be once again linked to the second order character of the shape phase transition from a soft to a rigid triaxial regime.

\begin{table}
\caption{Theoretical ratios $R_{4/2}$ , $R$, $R_{2}$ and $\gamma$ obtained from $R_{2}$ using Eq.~(\ref{R2gamma}).}
\label{TabI-ACM}
\begin{tabular}{ccccc}
\hline\noalign{\smallskip}
    & $R_{4/2}$   & $R$  &  $R_{2}$  & $\gamma$ \\
\noalign{\smallskip}\hline\noalign{\smallskip}
$^{104}$Mo & 2.79   & 4.20 &  6.92  &  21.4$^{\mathrm{o}}$  \\
$^{106}$Mo &  2.80  & 4.34 &  6.49  &  21.0$^{\mathrm{o}}$  \\
$^{108}$Mo &  2.78  & 4.07 &  7.17  &  21.5$^{\mathrm{o}}$  \\
$^{110}$Mo &  2.60  & 3.02 &  17.32  &  24.6$^{\mathrm{o}}$  \\
\noalign{\smallskip}\hline\noalign{\smallskip}
$^{104}$Ru & 2.26   & 2.23 & $\infty$   & 30.0$^{\mathrm{o}}$   \\
$^{106}$Ru & 2.67   & 3.65  & 10.72   &  23.1$^{\mathrm{o}}$  \\
$^{108}$Ru & 2.67   & 3.60 &  10.68  & 23.0$^{\mathrm{o}}$   \\
$^{110}$Ru & 2.69   & 3.48 &  10.41  &  23.0$^{\mathrm{o}}$  \\
$^{112}$Ru & 2.67   & 3.24 & 12.33   &  23.5$^{\mathrm{o}}$  \\
$^{114}$Ru &  2.68  & 3.40 &  11.01  &  23.2$^{\mathrm{o}}$  \\
$^{116}$Ru &  2.70  & 3.58 &  9.85  &  22.8$^{\mathrm{o}}$  \\
\noalign{\smallskip}\hline
\end{tabular}
\end{table}

\begin{figure*}[t!]

    \includegraphics[width=150mm]{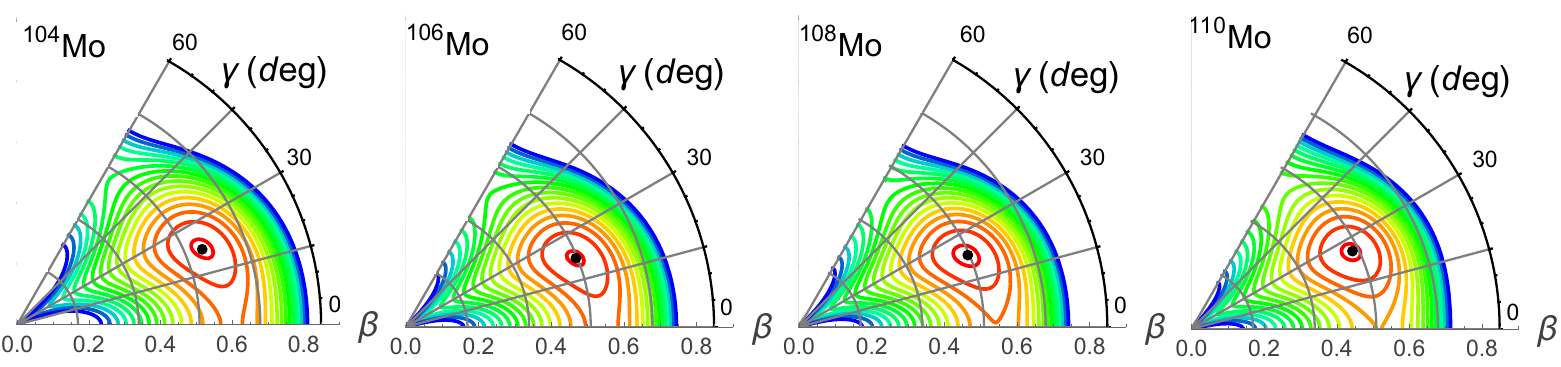}
   
    \caption{ACM potential energy surfaces for $^{104-110}$Mo.} 
    \label{PES-ACM-Mo}
\end{figure*}

\begin{figure*}[t!] 

    \includegraphics[width=150mm]{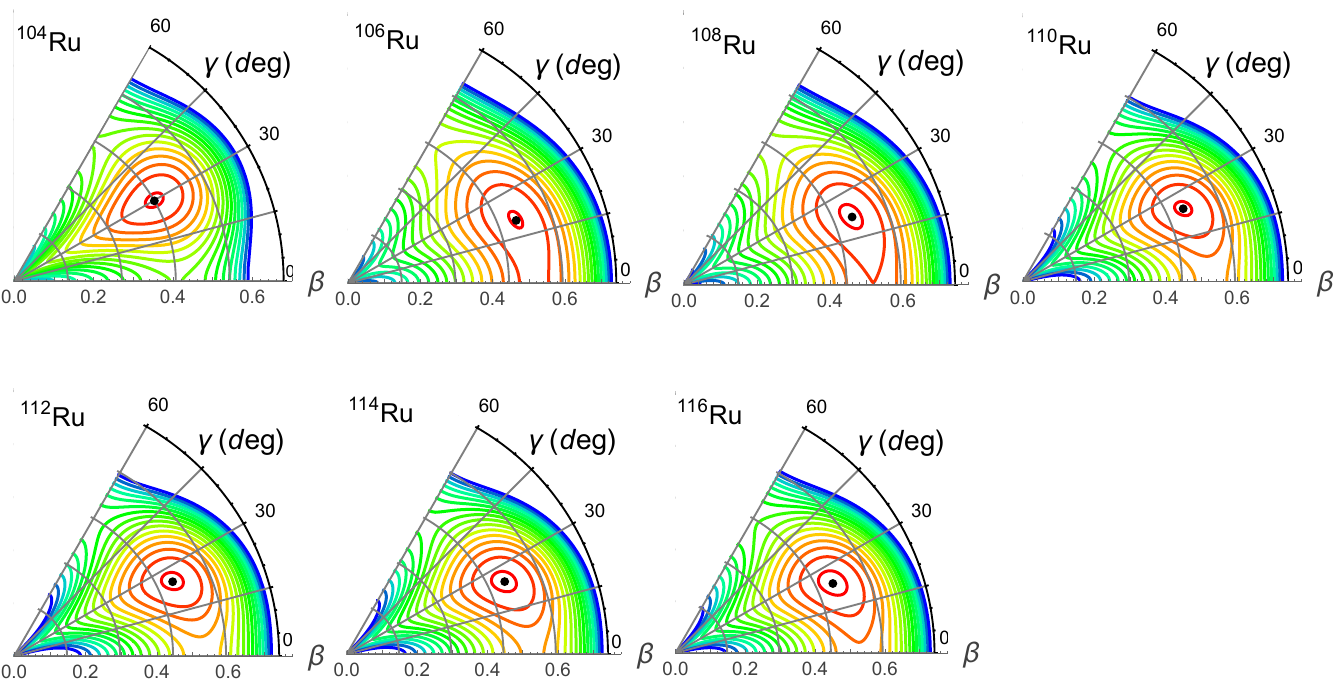}
   
    \caption{ACM potential energy surfaces for $^{104-116}$Ru.} 
    \label{PES-ACM-Ru}
\end{figure*}

ACM PESs for the same ACM parameters used in calculations of the spectra for both Mo and Ru isotopes are shown in Figs.~\ref{PES-ACM-Mo} and
\ref{PES-ACM-Ru}. They reveal a gradual evolution in nuclear shape rigidity as neutron number increases.

In the evolution of Mo isotopes from $^{104}$Mo to $^{108}$Mo, the morphology of the PES contours computed within the ACM reveals a transition from $\gamma$ softness to 
triaxial rigidity. In $^{104}$Mo and $^{106}$Mo, the secondary energy contours surrounding the minimum at $\gamma$ remain open toward $\gamma =0^{\mathrm{o}}$, indicating 
a relatively flat and soft potential in that direction. However, in $^{108}$Mo, the contours become fully enclosed and form a drop-like structure with a pointed extension 
toward $\gamma =0^{\mathrm{o}}$, marking a significant stabilization of the triaxial shape. This pronounced closure, accompanied by a sharpening of the energy gradient, 
signals enhanced structural rigidity and reduced susceptibility to axial shape fluctuations near mid-shell neutron number.

A comparable trend is observed in the Ru isotopic chain. The triaxial minimum in the PESs remains relatively stable in position, with $\gamma$ migrating only slightly across 
the isotopes. However, a notable change occurs at $^{108}$Ru, where the second energy contour around the minimum becomes fully enclosed—indicating a shift toward greater shape
localization. This structural consolidation intensifies around the mid-shell nucleus $^{110}$Ru, which serves as a turning point in the chain. Beyond mid-shell, the PES contours
become increasingly dense and circular, reflecting enhanced rigidity of the triaxial shape and reduced $\gamma$ softness. These contour transformations across the Ru isotopes
point to a gradual transition from shape fluctuation to stabilization, consistent with expectations based on neutron number and shell effects. Note that for $^{104}$Ru, 
the vanishing $\chi$ parameter in the $\cos{3\gamma}$ term results in a potential energy surface that is inherently symmetric about $\gamma =30^{\mathrm{o}}$ reflecting a
natural gamma softness.

\section{Skyrme HFB results}\label{SkyrmeInt}

Now we will investigate the possible phase transition from $\gamma$-soft to rigid triaxial nuclei within a more fundamental microscopic approach.
We performed microscopic HFB calculations for the even-even $^{102-116}$Mo and $^{104-118}$Ru isotopes
with several representative parametrizations of the effective Skyrme interaction. For each of these nuclei, we construct the deformation energy surface as a function 
of the total quadrupole deformation $\beta_2$ and the triaxiality angle $\gamma$ that 
are related to dimensionless multipole deformations $\beta_{2 0}$ and $\beta_{2 2}$: 
\begin{align}
\label{eq:beta2}
\beta_2
& = \sqrt{\beta_{20}^2 + 2\beta_{22}^2} \, ,
    \\
\label{eq:gamma}
\gamma
& = \mathrm{arctan} \bigg( \sqrt{2} \, \frac{\beta_{22}}{\beta_{20}} \bigg)  \, ,
\end{align}
that in turn are related to spherical multipole moments $\langle \hat{Q}_{\ell m} \rangle$ 
of the mass density $ \rho(\vec{\bf{r}})$ through
\begin{equation}
\beta_{\ell m} 
 = \frac{4 \pi } {  3 R_0^{\ell} A }  \langle \hat{Q}_{\ell m} \rangle 
 = \frac{4 \pi } {  3 R_0^{\ell} A }  
    \int \! \mathrm{d}^3 r \, r^\ell \, Y_{\ell m} (\vartheta, \varphi) \, \rho(\vec{\bf{r}}) \, .
\label{eq:betalm}
\end{equation}
We use the same convention for the reference radius $R_0 = 1.2 \, A^{1/3} \, \mathrm{fm}$ 
\cite{Ryssens19b,Maass25x} as all widely-used tabulations of transition quadrupole moments.
Note that the volume multipole moments $\beta_{\ell m}$ cannot be directly compared with 
the expansion coefficients of the multipole expansion of the nuclear surface as used
in microscopic-macroscopic models \cite{Ryssens19b} or the TPSM, nor does $\beta_2$ represent 
the same deformation as the $\beta$ of the ACM.

The calculations are performed with the MOCCa \cite{MOCCa} code that solves the HFB equations
with the two-basis method \cite{Ryssens19a} in a Cartesian 3D coordinate-space representation.
The code is based on the same principles as the earlier \textsc{Ev8} \cite{ev8} code, with 
the main differences being that it solves the full HFB equations instead of the HF+BCS approximation, 
calculates the derivatives always with the very accurate Lagrange-mesh technique \cite{Ryssens15a}, 
and uses more advanced algorithms to solve various aspects of the self-consistency problem 
\cite{Ryssens19a}.

The calculations reported here were run with symmetry restrictions, imposing that the single-particle 
states are eigenstates of parity, $z$ signature, and $y$ time simplex, and also pairwise connected by
time-reversal. These choices are sufficient to describe triaxial shapes of even-even nuclei, i.e.\
configurations with three plane reflection symmetries as also assumed in the Davydov model 
\cite{Davydov1958,Davydov1959} and the ACM \cite{RWC,Caprio2011}. Imposing these symmetries, the numerical solution
of the HFB equations can be reduced to one eighth of the full numerical box \cite{ev8}.
This choice implies that all $\beta_{\ell m}$ are real with $\beta_{\ell m} = \beta_{\ell -m}$, 
and that all deformations $\beta_{\ell m}$ with either odd $\ell$ or odd $m$ are zero \cite{MOCCa}.
These are the symmetry settings that are hard coded in the \textsc{Ev8} code, see
Ref.~\cite{ev8} for a detailed discussion, whereas for the MOCCa code they are a specific choice among 
others that can be selected to accelerate and stabilize convergence when calculating configurations 
for which phenomenology and previous theoretical studies indicate the absence of certain shape 
degrees of freedom. 

The energy surfaces are mapped with constraints on the values of $\beta_{2 0}$ and $\beta_{2 2}$
that fall on a regular mesh of the quantities $q_1$ and $q_2$ defined in Ref.~\cite{ev8}.
We recall that in a self-consistent mean-field model, higher deformations compatible 
with the imposed symmetries in general adopt non-zero values when minimizing the total  energy.

\begin{figure*}[t!]

    {\includegraphics[width=35mm]{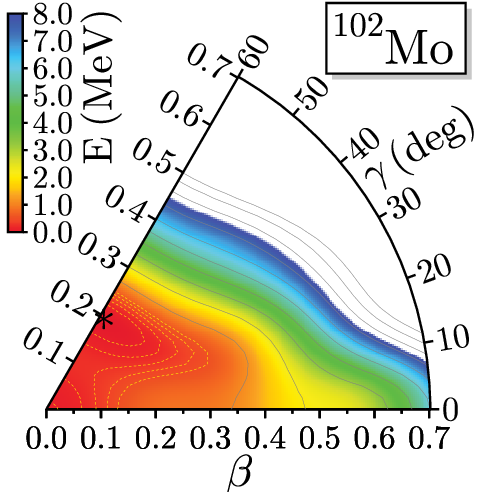}\hspace{2mm}
    \includegraphics[width=35mm]{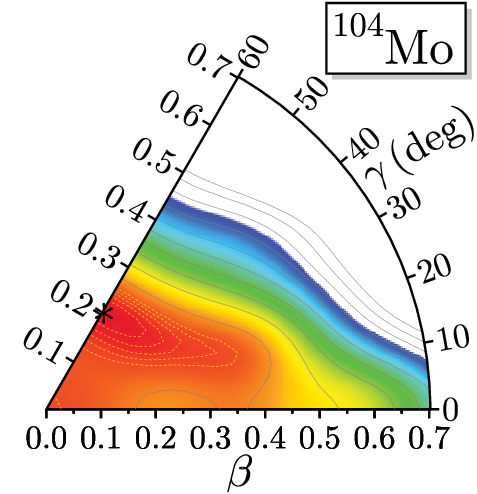}\hspace{2mm}
    \includegraphics[width=35mm]{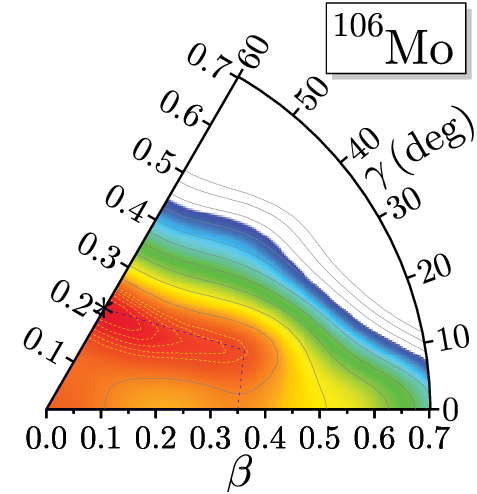}\hspace{2mm}
    \includegraphics[width=35mm]{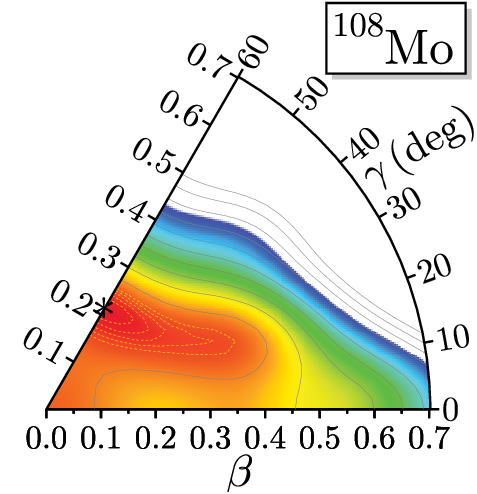}\hspace{2mm}}
    {\includegraphics[width=35mm]{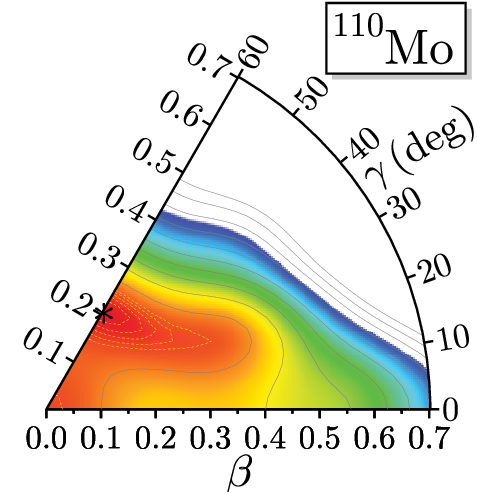}\hspace{2mm}
    \includegraphics[width=35mm]{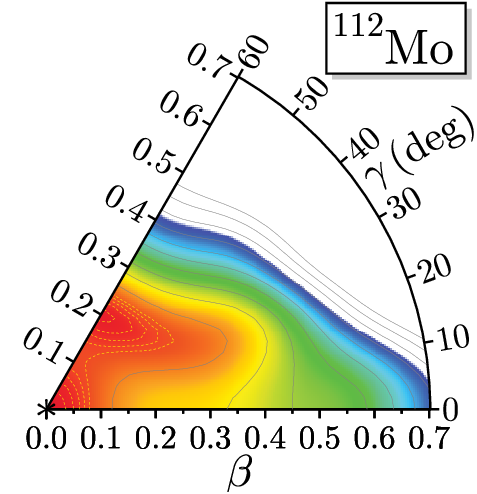}\hspace{2mm}
    \includegraphics[width=35mm]{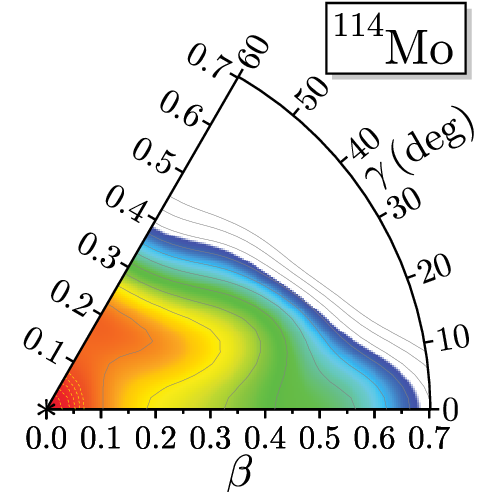}\hspace{2mm}
    \includegraphics[width=35mm]{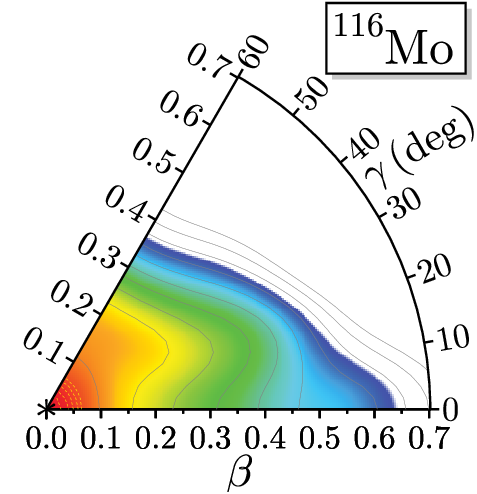}\hspace{2mm}}
    \caption{Skyrme-HFB PESs obtained with SV-bas for $^{102-116}$Mo.} 
    \label{FigMo-SV-bas}
\end{figure*}

\begin{figure*}[t!]

    {\includegraphics[width=35mm]{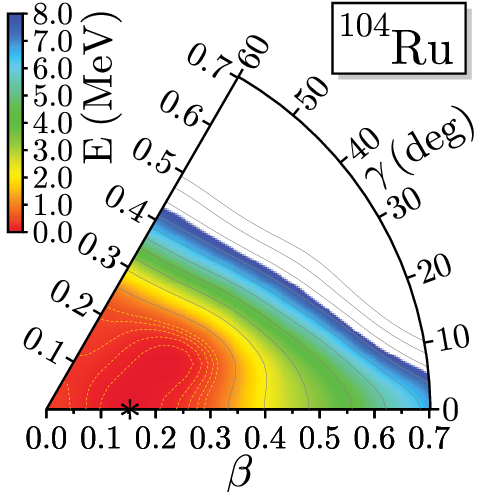}\hspace{2mm}
    \includegraphics[width=35mm]{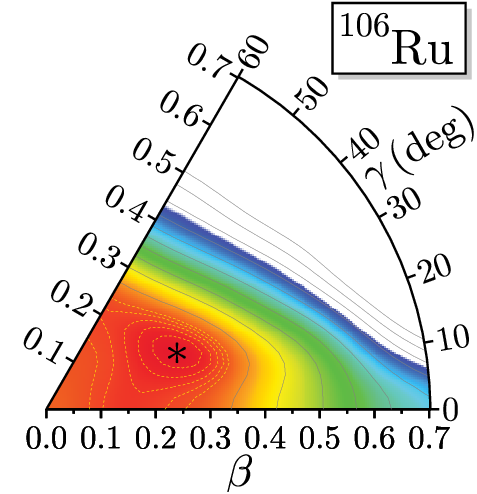}\hspace{2mm}
    \includegraphics[width=35mm]{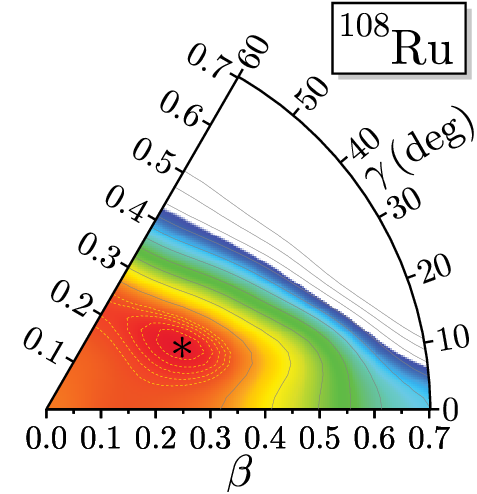}\hspace{2mm}
    \includegraphics[width=35mm]{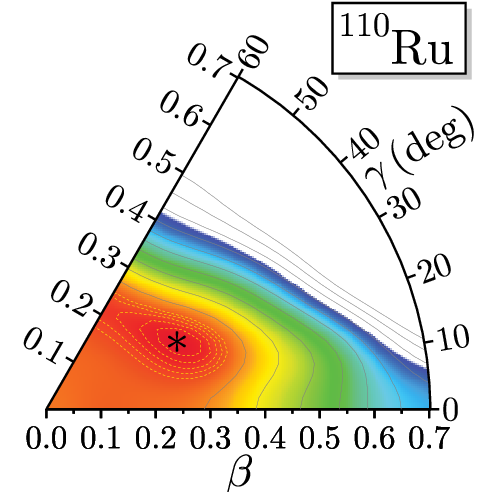}\hspace{2mm}}
    {\includegraphics[width=35mm]{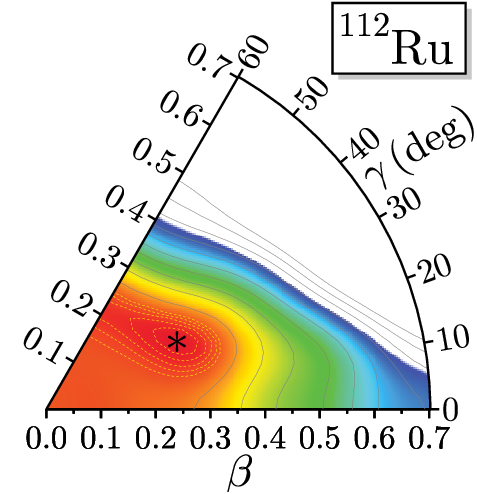}\hspace{2mm}
    \includegraphics[width=35mm]{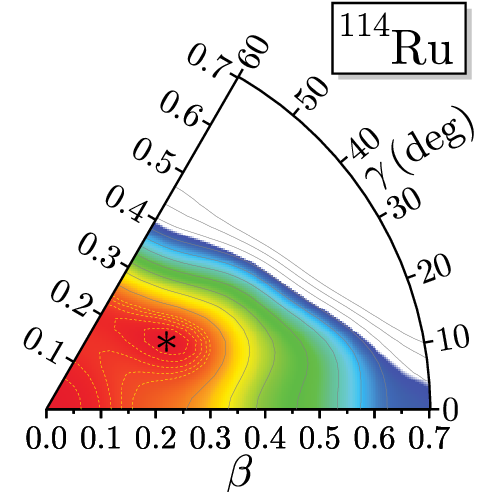}\hspace{2mm}
    \includegraphics[width=35mm]{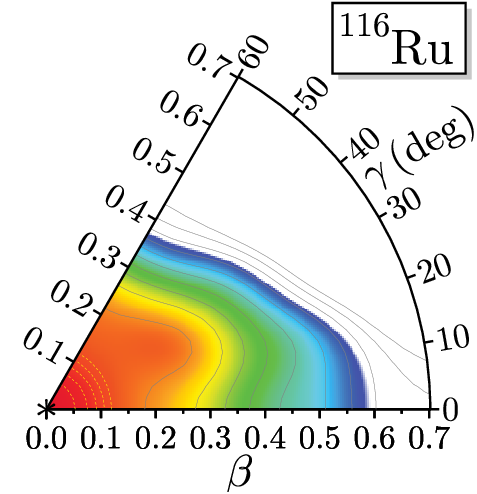}\hspace{2mm}
    \includegraphics[width=35mm]{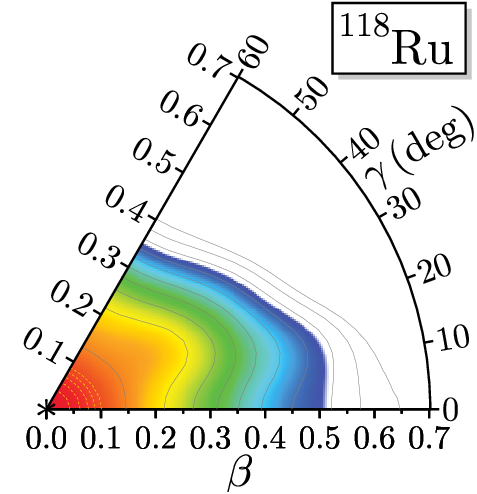}\hspace{2mm}}
   
    \caption{Skyrme-HFB PESs obtained with SV-bas for $^{104-118}$Ru.} 
    \label{FigRu-SV-bas}
\end{figure*}

\begin{figure*}[t!]

    {\includegraphics[width=35mm]{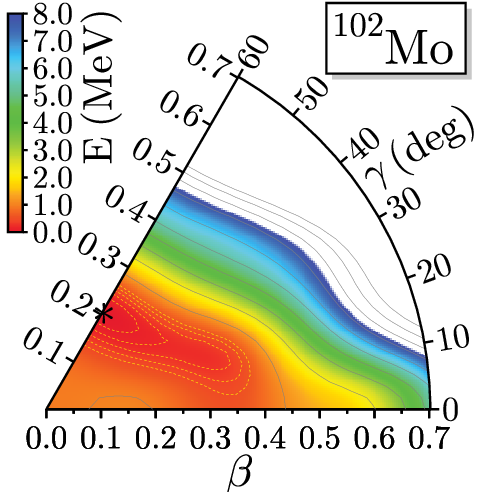}\hspace{2mm}
    \includegraphics[width=35mm]{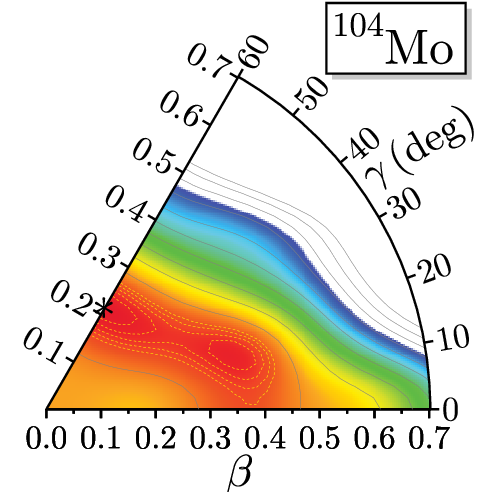}\hspace{2mm}
    \includegraphics[width=35mm]{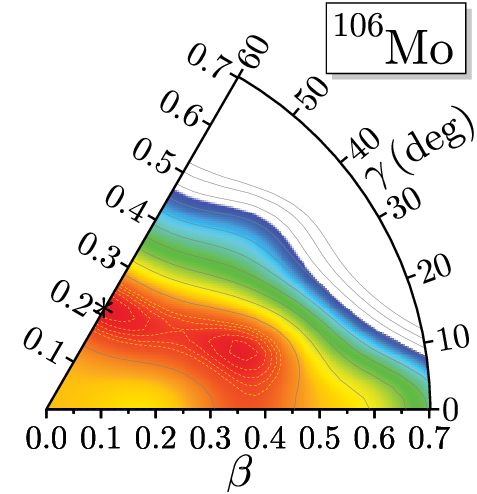}\hspace{2mm}
    \includegraphics[width=35mm]{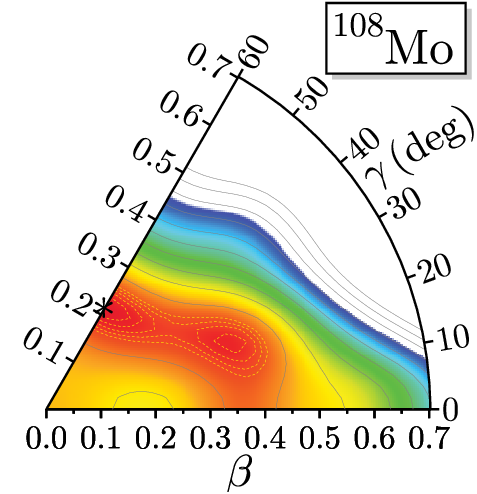}\hspace{2mm}}
    {\includegraphics[width=35mm]{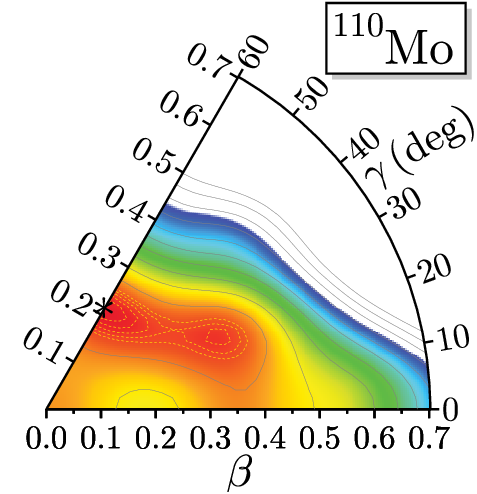}\hspace{2mm}
    \includegraphics[width=35mm]{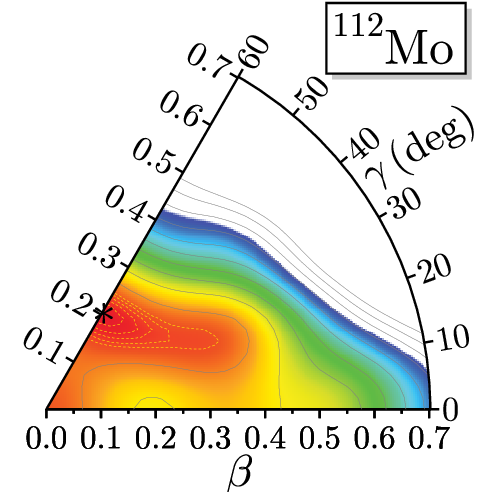}\hspace{2mm}
    \includegraphics[width=35mm]{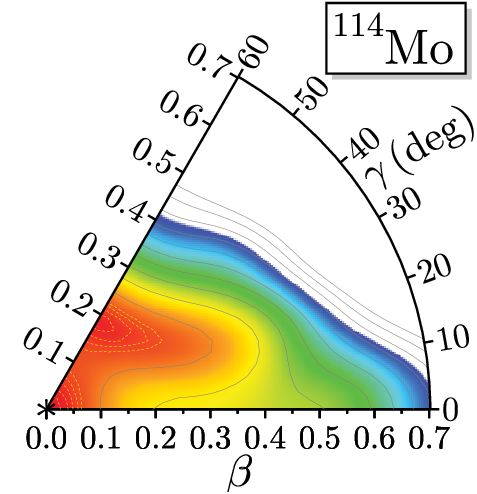}\hspace{2mm}
    \includegraphics[width=35mm]{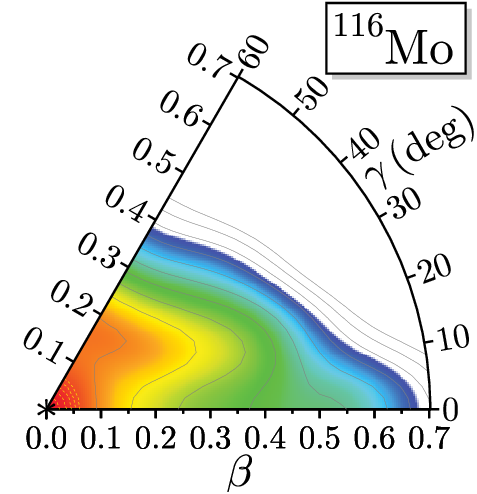}\hspace{2mm}}
   
    \caption{Skyrme-HFB PESs obtained with SLy7$^*$ for $^{102-116}$Mo.} 
    \label{FigMo-1t2t80}
\end{figure*}

\begin{figure*}[t!]

    {\includegraphics[width=35mm]{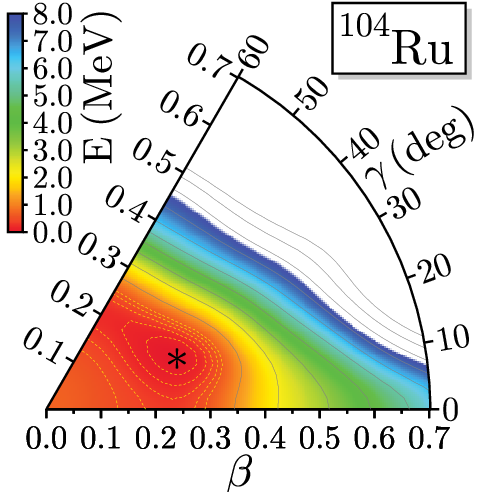}\hspace{2mm}
    \includegraphics[width=35mm]{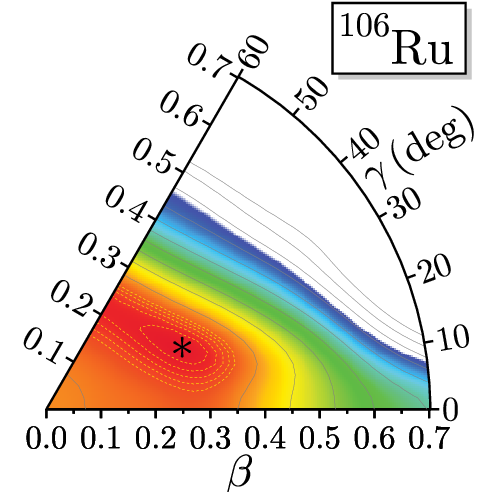}\hspace{2mm}
    \includegraphics[width=35mm]{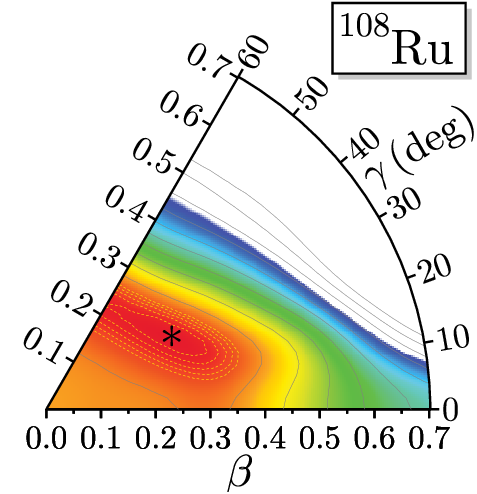}\hspace{2mm}
    \includegraphics[width=35mm]{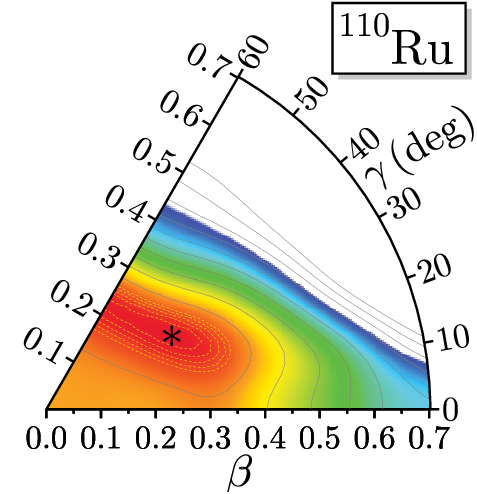}\hspace{2mm}}
    {\includegraphics[width=35mm]{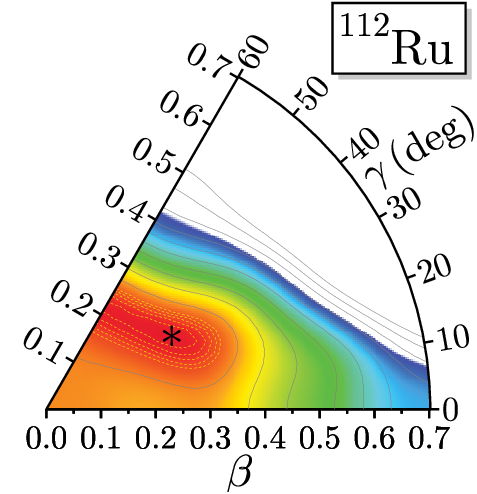}\hspace{2mm}
    \includegraphics[width=35mm]{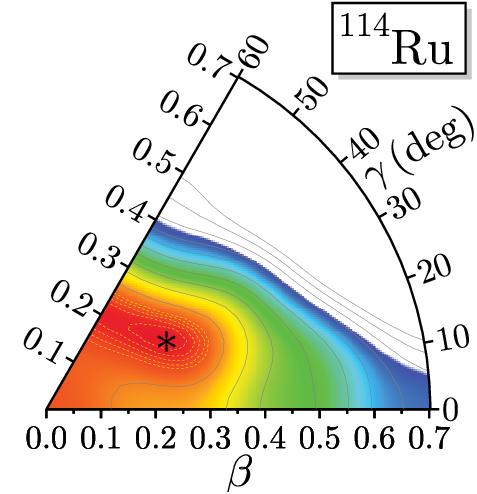}\hspace{2mm}
    \includegraphics[width=35mm]{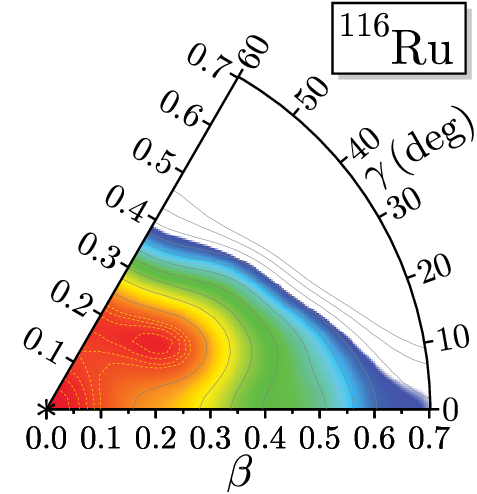}\hspace{2mm}
    \includegraphics[width=35mm]{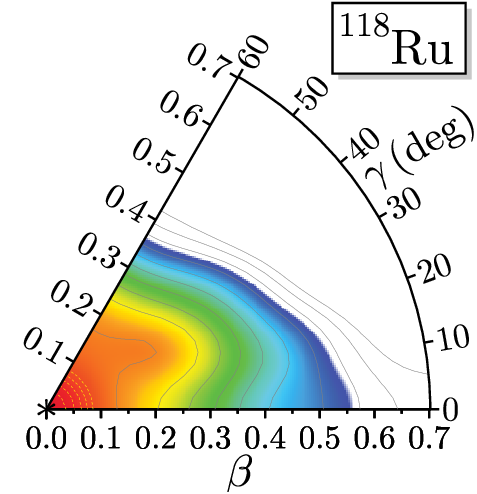}\hspace{2mm}}
   
    \caption{Skyrme-HFB PESs obtained with SLy7$^*$ for $^{104-118}$Ru.} 
    \label{FigRu-1t2t80}
\end{figure*}

\begin{figure*}[t!]

    {\includegraphics[width=35mm]{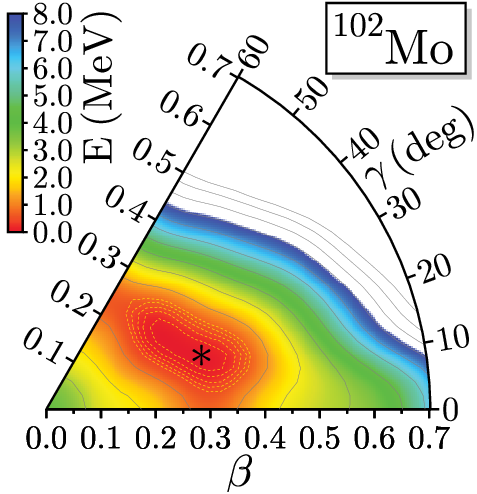}\hspace{2mm}
    \includegraphics[width=35mm]{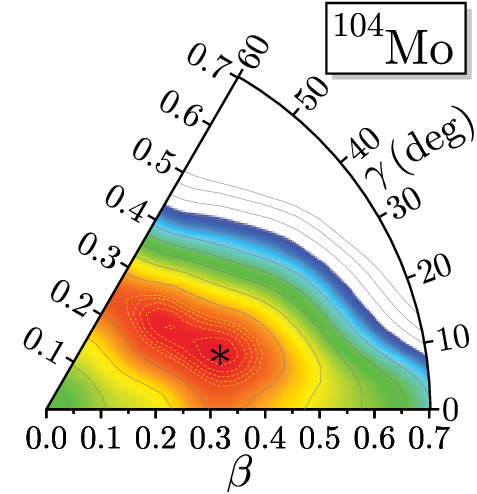}\hspace{2mm}
    \includegraphics[width=35mm]{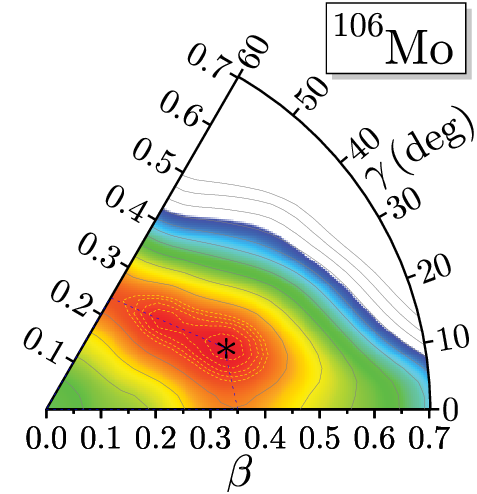}\hspace{2mm}
    \includegraphics[width=35mm]{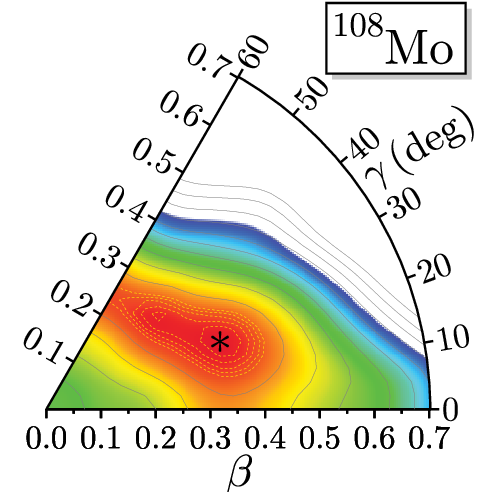}\hspace{2mm}}
    {\includegraphics[width=35mm]{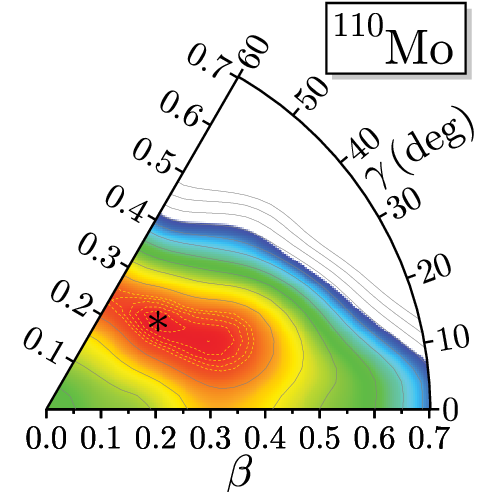}\hspace{2mm}
    \includegraphics[width=35mm]{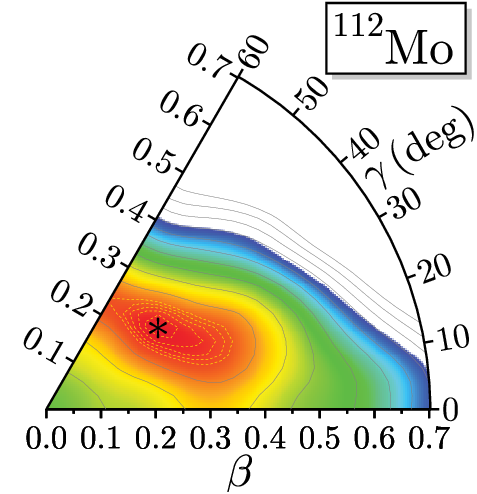}\hspace{2mm}
    \includegraphics[width=35mm]{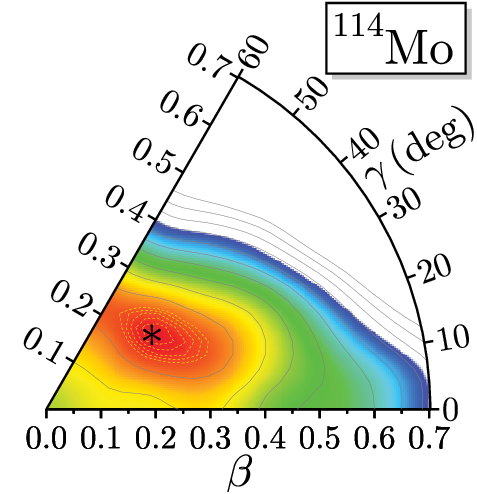}\hspace{2mm}
    \includegraphics[width=35mm]{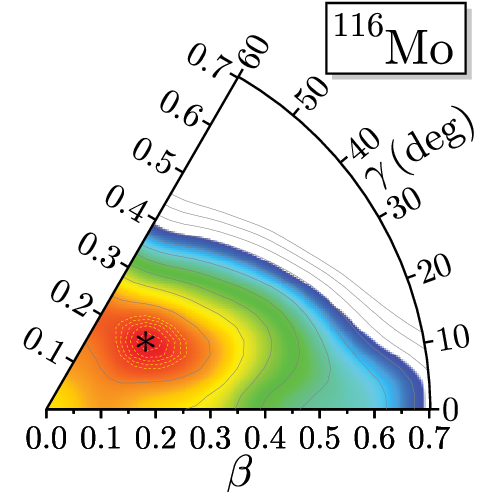}\hspace{2mm}}
   
    \caption{Skyrme-HFB PESs obtained with BSkG2 for $^{102-116}$Mo.} 
    \label{FigMo-bskg2}
\end{figure*}

\begin{figure*}[t!]

    {\includegraphics[width=35mm]{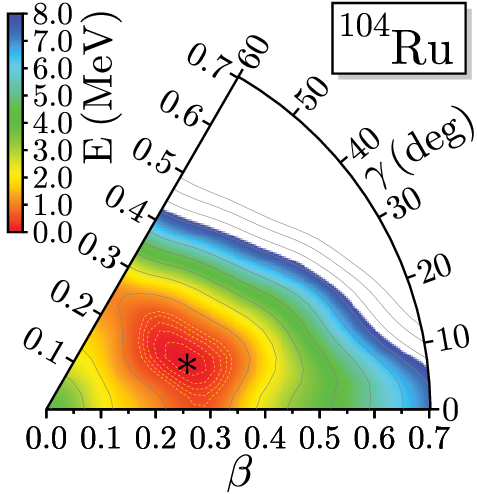}\hspace{2mm}
    \includegraphics[width=35mm]{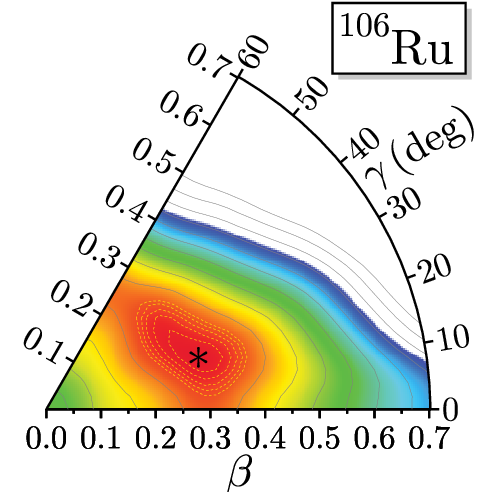}\hspace{2mm}
    \includegraphics[width=35mm]{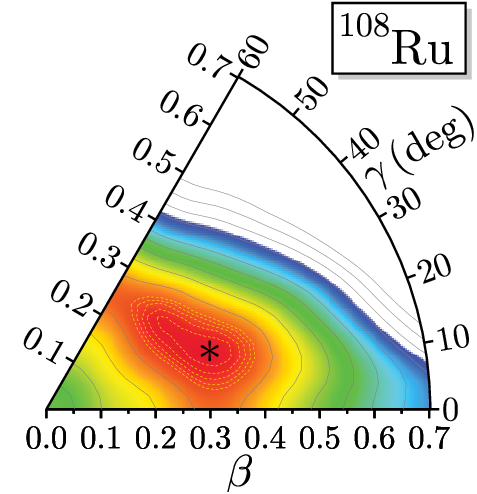}\hspace{2mm}
    \includegraphics[width=35mm]{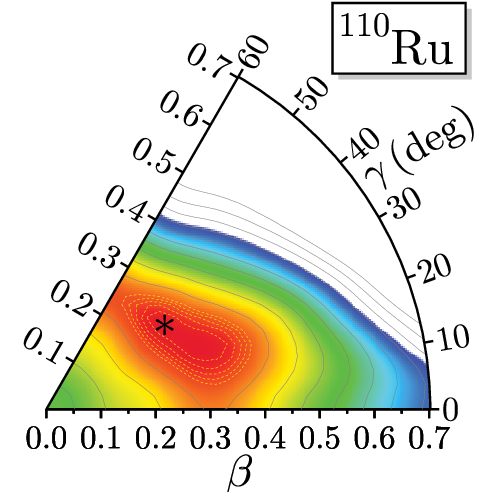}\hspace{2mm}}
    {\includegraphics[width=35mm]{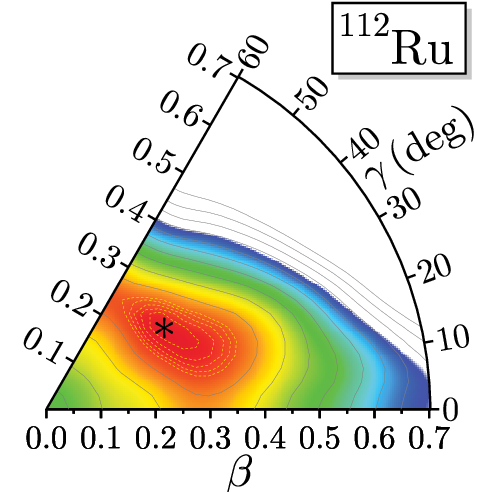}\hspace{2mm}
    \includegraphics[width=35mm]{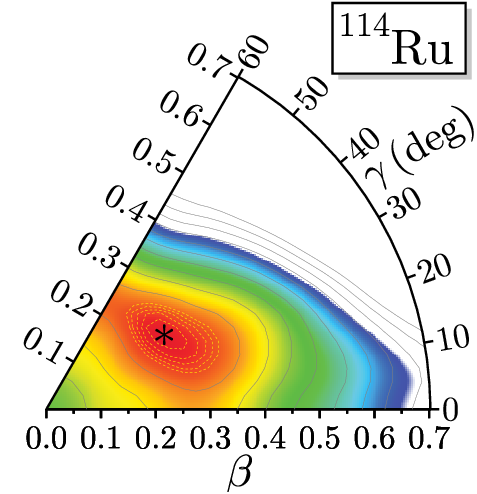}\hspace{2mm}
    \includegraphics[width=35mm]{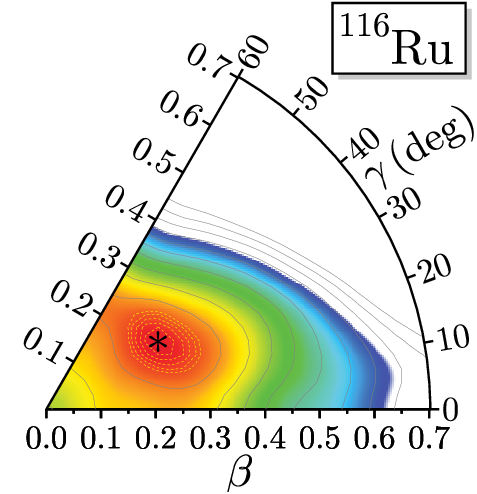}\hspace{2mm}
    \includegraphics[width=35mm]{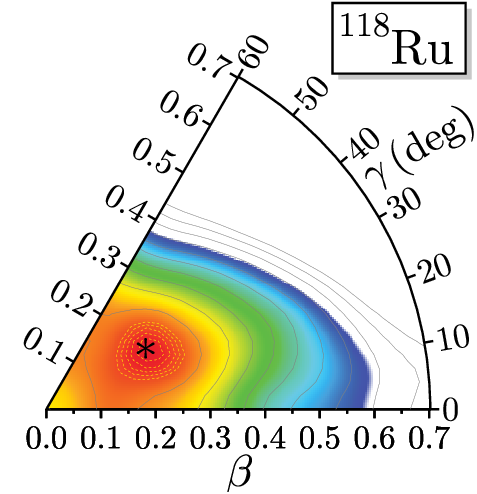}\hspace{2mm}}
   
    \caption{Skyrme-HFB PESs obtained with BSkG2 for $^{104-118}$Ru.} 
    \label{FigRu-bskg2}
\end{figure*}

Results displayed in Figs.~\ref{FigMo-SV-bas}--\ref{FigRu-bskg2} were obtained with
SV-bas \cite{Klupfel}, 
SLy7* \cite{SLy7s} (also called 1T2T(0.80) in that reference), and 
BSkG2 \cite{BSkG2a,BSkG2b}.
We also ran calculations with the SV-tls \cite{Klupfel}, 
BSkG1 \cite{BSkG1} and BSkG3 \cite{BSkG3} parametrizations,
but will not report results for energy surfaces
because those obtained with SV-tls do not differ 
significantly from the ones obtained with SV-bas, nor do results obtained with
BSkG1 or BSkG3 differ significantly from the ones obtained with BSkG2. 
We will come back to what can be deduced from this observation below. 

The parameters of SV-bas and SV-tls are adjusted to binding energies 
and information on the charge distribution of a set of selected 
spherical nuclei \cite{Klupfel}, whereas the parameters of SLy7* \cite{SLy7s} 
are adjusted with an updated version of the Saclay-Lyon protocol to 
binding energies of doubly-magic spherical nuclei and selected properties
of infinite nuclear matter. By contrast, the parameters of the BSkG$x$ models 
\cite{BSkG2a,BSkG2b,BSkG1,BSkG3} are adjusted to all binding 
energies of spherical and deformed nuclei with $Z \geq 8$ tabulated in AME2020, 
properties of nuclear matter relevant for astrophysics and, beginning with BSkG2, also
fission barriers and excitation energies of fission isomers. The 
BSkG2 model achieves a root-mean-square (rms) deviation of 0.678~MeV on 
2457 known masses, 0.027~fm on 884 measured charge radii, 
0.44 and 0.47~MeV, respectively, on 45 reference values for primary and 
secondary fission barriers of actinide nuclei, and 0.49~MeV on the excitation 
energy of 28 fission isomers.
All BSkG$x$ models yield a triaxial ground state of $^{104}$Ru
with $\beta_2$ and $\gamma$ close to their values determined from the
Kumar invariants extracted via Coulomb excitation \cite{BSkG1,Maass25x}.

There are some noteworthy features of and differences between these parametrizations
that are relevant for our discussion.
SLy7* and SV-tls keep the terms bilinear in the spin-current tensor density $J_{\mu\nu}(\vec{\bf{r}})$ 
with coupling constants determined from the HF expectation value of the central two-body 
Skyrme interaction, whereas these terms are dropped from SV-bas and all BSkG$x$ as
done for the majority of published parameter sets. BSkG1 and BSkG2 employ the same 
standard form of the Skyrme EDF as SV-bas, whereas BSkG3 has  
additional density dependencies of the derivative terms, and a pairing EDF tailored 
to reproduce microscopic calculations for the pairing gap in infinite matter.
The latter extensions are relevant for nuclear astrophysics, but make no significant
difference for the ground states of Mo and Ru isotopes discussed here \cite{Maass25x}.

All parametrizations used here include the full one- and two-body terms when calculating
the correction for spurious center-of-mass motion.
As discussed in Refs.~\cite{Bender00a,SLy7s}, doing so during the parameter adjustment leads 
to significantly smaller values of the surface energy coefficient than the more widely used
scheme that neglects the numerically more demanding two-body part of the correction. This,
in turn makes deformation energy surfaces more susceptible to higher-order 
deformation modes such as non-axiality, which in general then leads to a larger 
number of nuclei actually adopting a triaxial shape for their ground state, 
see also the discussion in Ref.~\cite{Ryssens19b}. 

For the BSkG$x$ models 
some further corrections for zero-point motion 
are added, most importantly for spurious rotational and vibrational motion. 
Unlike the center-of-mass correction that itself does only weakly depend on deformation, 
the combined rotational and vibrational corrections in general increases with 
deformation, which often pushes minima to larger deformation and also tends 
to enhance the likelihood of finding triaxial ground states.

\begin{table}[t!]
\caption{\label{tab:3d:fits:sly4}
Effective isoscalar mass at the saturation point of 
infinite matter and parameters of the pairing EDF used 
with SLy7*, SV-bas, and SV-tls (see text).
The original SLy4+ULB values are given for comparison.
}
\begin{tabular}{lllll}
\hline\noalign{\smallskip}
       & \multicolumn{1}{c}{$m^*_0/m$} 
       & \multicolumn{1}{c}{$V_1$} 
       & \multicolumn{1}{c}{$\eta$}   
       & \multicolumn{1}{c}{$\sigma$} \\
\noalign{\smallskip}\hline\noalign{\smallskip}
SLy4+ULB  & 0.6947 & $-1250.0$    & 1.000   & 1.000    \\
SLy7*     & 0.8000 & $-1267.0$    & 1.011   & 0.896    \\   
SV-bas    & 0.8995 & $-1284.0$    & 1.022   & 0.828    \\
SV-tls    & 0.8996 & $-1285.0$    & 1.020   & 0.824    \\
\noalign{\smallskip}\hline
\end{tabular}
\end{table}

Calculations with SV-bas, SV-tls and SLy7* are run with 25 points 
of distance $dx = 0.8 \, \mathrm{fm}$ in each direction in the octant of the
full box in which the HFB equations are numerically solved.
The parameters of the BSkG$x$ models \cite{BSkG1,BSkG2a,BSkG2b,BSkG3}
were adjusted using the MOCCa code, and the calculations with these parameter 
sets were run in each case with the same settings for box size and number of
single-particle states as used during their parameter adjustment, including 
the choice for the pairing interaction and cutoff. 

For SV-bas, SV-tls and SLy7*, we employ a density-dependent pairing EDF complemented 
by the stabilization proposed in Ref.~\cite{Erler08a} with 
$E_{\mathrm{cutp}} = 0.3 \, \mathrm{MeV}$, and by a smooth cutoff at 
$\Delta \epsilon_q = \pm 5 \, \mathrm{MeV}$ above and below the chemical potential 
of width $\mu_q = 0.5 \, \mathrm{MeV}$ as defined in Ref.~\cite{ev8}.
The parameters of the pairing EDF for its overall strength $V_1$, 
surface-to-volume mixing $\eta$ and power of the density dependence $\sigma$ were 
in each case adjusted to reproduce the density-dependence of the pairing gap 
obtained in SLy4+ULB pairing calculations in infinite symmetric matter as explained 
in Ref.~\cite{Gaps}, resulting in the values listed in Table~\ref{tab:3d:fits:sly4}.

\begin{table}[t!]
\caption{\label{tab:hfb:minima}
Ground-state quadrupole deformations $\beta$ and $\gamma$ found in HFB calculations with 
the Skyrme parametrizations as indicated in the same format as in 
Table~\ref{dat}.
Values for the BSkG$x$ are taken from the supplementary material of the original
publication that lists the ground-state properties of systems throughout the chart of nuclei.}
\begin{tabular}{llllllllll}
\hline\noalign{\smallskip}
         &        & 60    & 62    & 64    & 66    & 68    & 70    & 72    & 74     \\ 
\noalign{\smallskip}\hline\noalign{\smallskip}
{${}_{42}$Mo} \\
$\beta$  & SV-bas & 0.20  & 0.21  & 0.21  & 0.21  & 0.21  & 0.00 & 0.00 & 0.00  \\ 
$\gamma$ & SV-bas & 60    & 60    & 60    & 60    & 60    &  0    &  0    &  0     \\ 
$\beta$  & SV-tls & 0.20  & 0.21  & 0.21  & 0.21  & 0.21  & 0.19  & 0.00 & 0.00  \\ 
$\gamma$ & SV-tls & 55    & 60    & 60    & 60    & 60    & 60    &  0    &  0     \\ 
$\beta$  & SLy7*  & 0.21  & 0.21  & 0.21  & 0.21  & 0.21  & 0.21  & 0.00 & 0.00  \\ 
$\gamma$ & SLy7*  & 60    & 60    & 60    & 60    & 60    & 60    &  0    &  0     \\ 
$\beta$  & BSkG1  & 0.29  & 0.35  & 0.36  & 0.35  & 0.32  & 0.28  & 0.26  & 0.22   \\ 
$\gamma$ & BSkG1  & 22    & 19    & 18    & 21    & 26    & 29    & 29    & 30     \\ 
$\beta$  & BSkG2  & 0.30  & 0.33  & 0.35  & 0.34  & 0.26  & 0.25  & 0.24  & 0.22   \\ 
$\gamma$ & BSkG2  & 19    & 17    & 19    & 21    & 38    & 36    & 35    & 34     \\ 
$\beta$  & BSkG3 & 0.31   & 0.33  & 0.35  & 0.34  & 0.29  & 0.28  & 0.25  & 0.22   \\ 
$\gamma$ & BSkG3 & 19     & 17    & 16    & 19    & 28    & 30    & 30    & 30     \\ 
\noalign{\smallskip}\hline\noalign{\smallskip}
{${}_{44}$Ru} \\
$\beta$  & SV-bas & 0.15  & 0.26  & 0.27  & 0.27  & 0.27  & 0.25  & 0.00  & 0.00   \\ 
$\gamma$ & SV-bas &  0    & 23    & 25    & 27    & 27    & 29    &  0    &  0     \\ 
$\beta$  & SV-tls & 0.24  & 0.25  & 0.25  & 0.25  & 0.23  & 0.00  & 0.00  & 0.00   \\ 
$\gamma$ & SV-tls & 23    & 27    & 29    & 33    & 39    &  0    &  0    &  0     \\ 
$\beta$  & SLy7*  & 0.26  & 0.27  & 0.27  & 0.27  & 0.27  & 0.25  & 0.00  & 0.00   \\ 
$\gamma$ & SLy7*  & 21    & 25    & 30    & 30    & 30    & 29    &  0    &  0     \\ 
$\beta$  & BSkG1  & 0.27  & 0.29  & 0.31  & 0.31  & 0.29  & 0.27  & 0.25  & 0.22   \\ 
$\gamma$ & BSkG1  & 23    & 24    & 22    & 25    & 28    & 30    & 30    & 27     \\ 
$\beta$  & BSkG2  & 0.27  & 0.29  & 0.32  & 0.27  & 0.26  & 0.26  & 0.24  & 0.21   \\
$\gamma$ & BSkG2  & 18    & 19    & 20    & 36    & 35    & 32    & 31    & 32     \\ 
$\beta$  & BSkG3  & 0.27  & 0.30  & 0.31  & 0.29  & 0.28  & 0.26  & 0.25  & 0.21   \\ 
$\gamma$ & BSkG3  & 23    & 21    & 20    & 24    & 28    & 30    & 29    & 30     \\ 
\noalign{\smallskip}\hline
\end{tabular}
\end{table}

All deformation energy surfaces are normalized to the energy at the respective global minimum
that in each case is indicated by a black asterisk. Solid iso-energy lines in gray are 1~MeV apart. 
For a better resolution of the fine structure of the energy surfaces around the minima,
dotted yellow isolines indicate an energy of 0.1, 0.2, 0.3, 0.4 and 0.5 MeV above 
the ground state.

Triaxial ground states are predicted by SV-bas only for $^{106-114}$Ru, and by 
SLy7* for $^{104-114}$Ru. In both cases, the heavier $^{116-118}$Ru isotopes are predicted 
to have a spherical ground state, as do $^{112-116}$Mo when calculated with SV-min
and $^{114-116}$Mo when calculated with SLy7*, respectively. For the lighter molybdenum
isotopes, both parameter sets yield an oblate ground state.  
By contrast, with BSkG2 a triaxial ground state is found for all $^{102-116}$Mo 
and $^{104-118}$Ru isotopes studied here. In addition, BSkG2 predicts quite simple 
deformation energy surfaces with a single quite pronounced minimum that with 
increasing neutron number in both isotopic chains slowly shifts to larger $\gamma$.
This noticeable difference in predictions can be mainly attributed to the inclusion of 
a rotational and vibrational correction in the BSkG$x$ models \cite{BSkG1}: Its contribution 
significantly deepens the deformed minima that are generated by shell effects, 
and this slightly more so for triaxial shapes. Indeed, compared to the spherical
configuration, the energy gain of the deformed ground states found with SV-min and SLy7*
does not exceed 1 and 1.5~MeV, respectively, often being smaller. Conversely,
with the exception of the two heaviest isotopes in each chain, the deformation 
energy found with BSkG2 is typically between 3.5 and 4.5~MeV. 

Beyond these observations, however, for many nuclei the deformation energy surfaces
obtained with SV-bas and SLy7* exhibit quite complicated substructures.

On the energy surfaces from SV-bas, for $^{102-114}$Mo an often gamma-soft oblate 
minimum coexists with a spherical one. These minima are separated by a small barrier
that for $^{102}$Mo and $^{112}$Mo is less than 500 keV high. Going to $^{114}$Mo,
there is a shape transition where the spherical configuration becomes more bound 
than the oblate one that with increasing $N$ also becomes more rigid in $\gamma$
direction.
Coexistence of spherical and oblate minima is also found when using SLy7* instead.  
For $^{104-110}$Mo, the SLy7* parametrization predicts an additional shallow excited 
triaxial minimum that is nearly degenerate with the oblate one and separated from 
it by a barrier that does not exceed 400 keV. From the topography of these energy 
surfaces, however, it cannot be directly concluded that, when fed into a collective
Hamiltonian of some kind, they would produce collective states that can be attributed 
to each of the minima.

For the $^{104-124}$Ru isotopes, SV-bas actually predicts that the energy surface is
very flat for $\beta_2$ values up to about 0.3 and varies by at most 1 MeV, irrespective 
of $\gamma$. Beginning with $^{116}$Ru, the energy surfaces become slightly more rigid, 
and the minimum is found for a spherical configuration. SLy7* predicts in general 
slightly more pronounced minima for all Ru isotopes, in particular with respect to 
variations of $\beta_2$, but these differences remain minor compared to what is obtained
with BSkG2.

We note in passing that SV-tls yields deformation energy surfaces that are 
very similar to those obtained with SV-bas, implying that including or not the 
spin-current tensor terms from the central Skyrme interaction in the EDF does not
make a significant difference for these nuclei, which is also corroborated by 
the similarity to results obtained with SLy7* for which these terms are kept.
This does, however, not rule out that parameter sets with genuinely independent
tensor interactions \cite{Les07a} might produce significantly different results.

The quadrupole deformations $\beta$ and $\gamma$ of the ground states found in these
calculations are summarized in Table~\ref{tab:hfb:minima}.

Note that for some nuclei the BSkG$x$ values listed in the table slightly differ from
those given in the supplemental material to the original papers \cite{BSkG2a,BSkG1,BSkG3}.
The reason is that because of the rotational correction the minimum has to 
be constructed from the interpolation of the energy surface around the minimum. 
To better resolve details of the very soft energy surfaces of the Mo and Ru isotopes,
where in some cases the energy changes only by a few keV over a wide range of $\gamma$,
we used more discretization points than the original publication.

For nuclei for which
triaxial minima are found, the $\gamma$ values are similar in size to those listed
in Table~\ref{dat}, although they do not closely
follow all details of their evolution. There might be many reasons for these
differences, most importantly the assumptions that on the one hand the 
observed data are perfectly described by a rigid triaxial rotor and that 
on the other hand the state at the minimum of the microscopic energy surfaces 
is representative for the description of these nuclei.  
To study fluctuations in $\gamma$ would require a beyond-mean-field approach,
which is beyond the scope of our study.
Still, it is gratifying to see that even when making all of these assumptions, 
the appearance of minima at large triaxiality is well reproduced by the 
BSkG$x$ models.

For $^{104}$Ru, empirical values of $\beta_2 = 0.272(4)$ and $\gamma = 23(3)^{\circ}$ 
have also been deduced as Kumar shape invariants from Coulomb excitation \cite{Srebrny06a}. 
Both SLy7* and BSkG2 predict a minimum that is close to these values, whereas 
SV-bas predicts this nucleus to be $\gamma$ soft with an overall prolate minimum.

We also note that for these Mo and Ru isotopes, results obtained with the widely-used Gogny 
interaction D1S reported in the AMADEE data base \cite{Hilaire07a,Amadee} 
are qualitatively very similar to what is found for SV-bas and SLy7*; see 
also Ref.~\cite{Deloncle00a}. The energy surfaces from HFB calculations with
the Skyrme interaction UNEDF0 \cite{Zhang2015b} are also very similar to
these.

\begin{figure}[t!]
\centerline{\includegraphics[width=8.2cm]{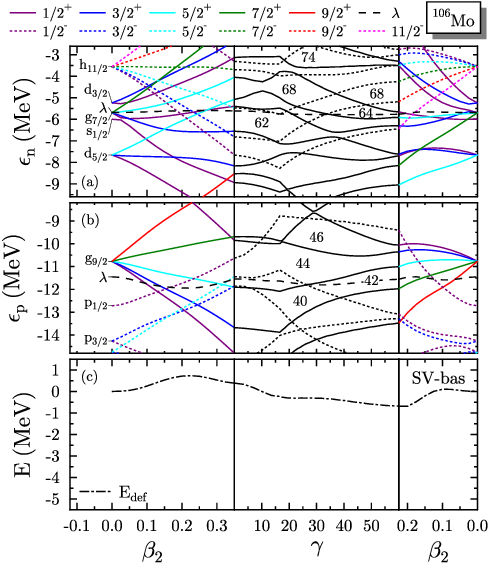}}
\caption{\label{fig:Nilsson:mo106:sv-bas}
Nilsson diagrams of neutrons (a) and protons (b) together with the deformation energy 
(c) along a closed path in the $\beta$-$\gamma$ plane obtained for $^{106}$Mo ($Z=42$, $N=64$) with SV-bas.
On the left, the path starts at spherical shape, passes through axial shapes with increasing $\beta_2$ 
at $\gamma = 0^\circ$ to the prolate saddle point, then through triaxial shapes with increasing $\gamma$ but slowly
decreasing $\beta_2$ to the oblate minimum, and then returns through axial shapes with decreasing $\beta_2$ 
at $\gamma = 60^\circ$ to the spherical point. In the two axial parts of the Nilsson diagram,
colors represent the projection of angular momentum on the symmetry axis $K$, whereas in the triaxial 
part the levels indicated in black have mixed in $K$. In the Nilsson diagrams, solid lines indicate levels 
with positive parity, whereas dotted lines indicate levels of negative parity. The dashed curves 
represent the chemical potential $\lambda$ of the respective nucleon species.
The deformation energy in panel (c) is plotted relative to the spherical configuration.
}
\end{figure}

\begin{figure}[t!]
\centerline{\includegraphics[width=8.2cm]{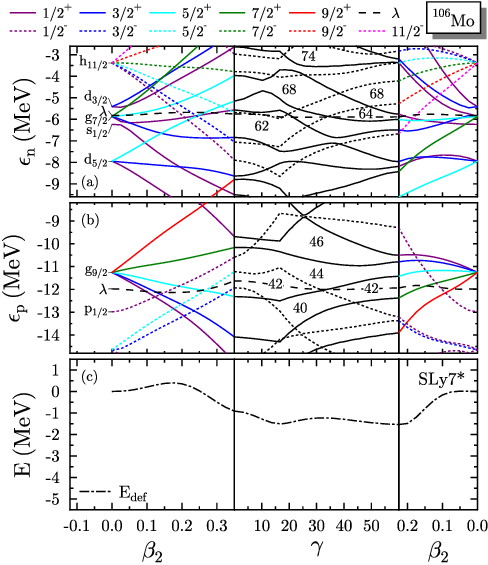}}
\caption{\label{fig:Nilsson:mo106:sly7s}
Nilsson diagrams of protons and neutrons as well as the mean-field deformation energy along a closed path 
in the $\beta$-$\gamma$ plane obtained with SLy7* for $^{106}$Mo represented 
in the same way as in Fig.~\ref{fig:Nilsson:mo106:sv-bas}.
}
\end{figure}

\begin{figure}[t!]
\centerline{\includegraphics[width=8.2cm]{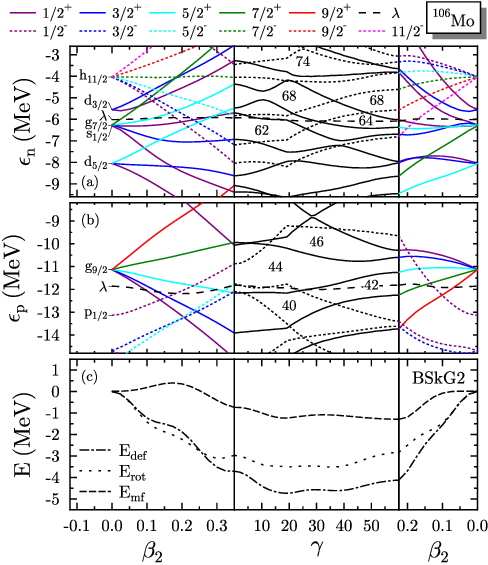}}
\caption{\label{fig:Nilsson:mo106:bskg2}
Nilsson diagrams of protons and neutrons as well as the mean-field deformation energy along a closed path 
in the $\beta$-$\gamma$ plane obtained with BSkG2 for $^{106}$Mo represented 
in the same way as in Fig.~\ref{fig:Nilsson:mo106:sv-bas}.
In addition to the total deformation energy, panel (c) also displays its decomposition
$E_{\text{def}} = E_{\text{mf}} + E_{\text{rot}}$ into the mean-field deformation energy 
$E_{\text{mf}}$ and the rotational correction $E_{\text{rot}}$ considered for this 
parameter set.
}
\end{figure}

In self-consistent mean-field models, the presence of deformed minima in the PESs
is a consequence of shell effects related to the variation of the single-particle
spectra with deformation. Without such shell effects, all nuclei up to about 
$Z \simeq 100$ had spherical ground states, and would instantaneously fission for 
larger $Z$.
According to Strutinsky's theorem, the variation of the bunching of single-particle 
levels of protons and neutrons around the respective chemical potential $\lambda$ with 
deformation gives a contribution to the binding energy that generates the
complex structure of PESs with local minima and barriers that evolve quickly
with nucleon number: Configurations with a large bunching of levels are less bound, 
whereas configurations with smaller then average level density gain additional
binding \cite{Nilsson95a,Brack75a,Brack81a}. 

The energy from shell effects is in competition with the surface and surface 
symmetry energy of nuclear matter whose sum grows with the geometrical size 
of the nuclear surface and thereby characterizes the average stiffness of 
nuclei against deformation.
Like the precise position of levels in the single-particle spectrum, the surface 
energy depends sensitively on the parameters of the nuclear EDF 
\cite{SLy7s,Ryssens19b,Jodon16a}. When the correction for rotational zero-point 
motion is considered, as done for some parameter sets like the BSkGx, it 
partially counteracts the contribution from the surface energy.

The role of the single-particle spectrum for the deformation energy is 
illustrated by Figs.~\ref{fig:Nilsson:mo106:sv-bas}, \ref{fig:Nilsson:mo106:sly7s}, 
and \ref{fig:Nilsson:mo106:bskg2} that show the Nilsson diagrams of protons and 
neutrons as well as the deformation energy along a closed path in the $\beta$-$\gamma$ 
plane obtained for $^{106}$Mo with SV-bas, SLy7*, and BSkG2. The respective path
for each parameter set is indicated by a dashed blue line in the PESs of 
 $^{106}$Mo obtained with these parameter sets in Figs.~\ref{FigMo-SV-bas}, 
 \ref{FigMo-1t2t80}, and \ref{FigMo-bskg2}. Starting at the spherical 
configuration, the path for the Nilsson diagrams passes first through axially symmetric 
prolate shapes up to the prolate saddle point, then roughly follows the valley 
in $\gamma$ direction to the oblate minimum or saddle, and finally returns 
through axial oblate shapes to the spherical configuration. 
For axial shapes, the single-particle states are eigenstates of the angular-momentum
projection $K$ on the symmetry axis as indicated by the colors of the levels in 
the Nilsson diagrams. For triaxial shapes, however, the angular momenta of the 
single-particle states are mixed, leading to many avoided level crossings.
The Nilsson diagrams predicted by the three parameter sets are very similar, with only
small differences that can be traced back to small shifts between the shells at sphericity. 

The proton $g_{9/2}$ and neutron $h_{11/2}$ shells play a particular role
for the triaxial states of Mo and Ru isotopes. Because of the unique parity 
of these intruder shells, for the entire range of deformations relevant for 
ground states, their magnetic substates split with axial quadrupole 
deformation almost without any interaction with other 
levels of same parity, the only exception being the downsloping $K=1/2$ 
level that originates from the lowest spherical shell in the next major shell.
As a consequence, the splitting of the magnetic substates becomes quite large
for the $\beta$ values at which the ground state of Mo and Ru isotopes is found.
In addition, at fixed $\beta$, these levels are only mildly dependent 
on triaxiality, such that they keep their large splitting and remain quite parallel 
when changing $\gamma$. This results in large gaps between most of these
intruder levels at all triaxial shapes. However, as the Nilsson diagrams 
follow a path along which $\beta$ becomes smaller with increasing $\gamma$, 
the gaps between the intruder levels shrink when going towards oblate shapes
on Figs.~\ref{fig:Nilsson:mo106:sly7s}, \ref{fig:Nilsson:mo106:sv-bas}, 
and \ref{fig:Nilsson:mo106:bskg2}.  

By contrast, most of the normal parity proton levels strongly depend on 
$\gamma$ and have many avoided level crossings when their $K$
quantum number gets mixed, and many of them cross the intruder levels at 
steep angles. Combined with the more equally distributed background of 
intruder levels, this generates a multitude of gaps in the Nilsson 
diagram for many combinations of proton numbers between 40 and 46
and neutron numbers between 60 and 74 at some triaxial shape. And even
where there are no actual gaps, the average level density around the chemical
potential of protons and neutrons is visibly and consistently smaller 
for triaxial shapes then axial ones, explaining the $\gamma$ softness 
of nuclei in this region of the nuclear chart. The situation is quite 
different at axial shape, where the level density around the chemical potential
in general remains quite large, with only a few gaps like the ones at 
$Z=42$ and $N=66$ and 68 that open for oblate shapes.

There is a noteworthy qualitative difference between the proton and neutron 
spectra. At spherical shape, the proton $g_{9/2}$ intruder situated below 
the major shell closure at $Z=50$ is also well separated from 
the normal-parity levels below by the $Z=40$ gap. Therefore, when the magnetic 
substates of the proton $g_{9/2}$ intruder level split with deformation, 
they spread without crossing many of the normal-parity proton levels 
from the $fp$ shell below. By contrast, the spherical neutron $h_{11/2}$ shell 
is nearly degenerate with the normal-parity levels, which leads to an 
overall much larger level density for neutrons than for protons. This
has the consequence that in this mass region the shell effects of protons 
dominate over those of neutrons, with the triaxial deformation being mainly 
driven by the gaps at $Z=42$, $Z=44$, and $Z=46$ visible in 
Figs.~\ref{fig:Nilsson:mo106:sly7s}--\ref{fig:Nilsson:mo106:bskg2}. The
various neutron gaps between 62 and 74 reinforce this effect, which together
explains the outline of this region in the chart of nuclei for which 
triaxial states are found.

Comparing the Nilsson diagram obtained with BSkG2 with the two others, 
it is clear that the systematic large difference between the PESs predicted 
by BSkG2 and those predicted by SLy7* and SV-bas is not the consequence
of a different shell structure. And indeed, the mean-field contribution 
$E_{\text{mf}}$ to the deformation energy obtained with BSkG2 indicated
in Fig.~\ref{fig:Nilsson:mo106:bskg2}(c) is very similar
to the deformation energy obtained with SLy7* and SV-bas. As already 
underlined before, what makes the difference is the rotational correction 
$E_{\text{rot}}$ that is taken into account for the BSkG$x$, but not 
for SLy7* and SV-bas.

\section{Comparison between the models}

It appears that the predictions from the BSkG$x$ parameter sets are closest to
the available empirical data for Mo and Ru isotopes. Taking BSkG2 as a representative
example, the microscopic PESs for Mo isotopes reveal a systematic shift in the location of the
triaxial minimum in $\gamma$. The PESs of $^{104-108}$Mo exhibit minima slightly 
below $\gamma = 30^{\circ}$, thereby indicating a mild triaxial preference on the
prolate side, whereas from $^{110}$Mo onward the isotopes show minima above $30^{\circ}$,
trending toward the oblate side. A striking feature of the PESs is the pronounced 
elongation of the contour lines around the minima: For the prolate-trending minima 
the energy valley extends deep into the oblate region, whereas for the oblate-trending 
minima the valleys reach far toward the prolate side. This bilateral elongation indicates
persistent $\gamma$ softness, where the PES is shallow across a wide range of 
$\gamma$ values, which can be expected to result in large-amplitude shape fluctuations
of low-lying states in these nuclei.
Only in the heavier isotopes, $^{114}$Mo and $^{116}$Mo, do we begin to see tightly 
localized, circular contour lines, suggesting that the nucleus is settling into a more 
rigid and stable configuration.

The Skyrme-HFB PESs of the Ru isotopes exhibit trends that mirror those observed for 
the Mo isotopes. Up to the mid-shell nucleus $^{110}$Ru, the triaxial minima reside 
slightly below $\gamma = 30^{\circ}$, suggesting a mild triaxial tendency on the 
prolate side. At $^{110}$Ru, a distinct shift occurs: The minima jump above $30^\circ$, 
indicating a transition toward oblate-triaxial character. This evolution is 
accompanied by pronounced elongation of the energy contours across the isotopic chain, 
reflecting persistent $\gamma$ softness. From $^{114}$Ru onward, the contours around 
the minima become more densely packed and circular, pointing towards increased shape 
localization and structural rigidity. These transitions underscore a gradual 
consolidation of triaxial configurations beyond mid-shell and reinforce the conclusions 
drawn from the Mo series: Both isotopic chains reveal a transition from softness to 
rigidity, captured vividly through the evolution of the morphology of the contour lines 
and the $\gamma$ position of the minima in the Skyrme-HFB framework.

This transition mirrors what is obtained for the PESs with the ACM, where increasing 
neutron number leads to enhanced triaxial rigidity. However, the $\gamma$ asymmetry 
and migration of the minimum, visible in HFB but absent in ACM, results from 
the minimization of a total energy that is intimately connected to shell structure. 
By contrast, the ACM produces symmetric PESs centered around fixed $\gamma$ values 
(often near $25^{\circ}$ $-$ $30^{\circ}$), governed by collective symmetry principles rather 
than single-particle dynamics. Taken together, these observations point to a 
convergent narrative: Both models predict increasing rigidity when going across the 
Mo and Ru isotopic chains.

Moreover, the transition from prolate to oblate shapes seen in the HFB calculations 
with BSkG2 around $N=66$ appears to be consistent with the proxy-SU(3) picture, 
if the irreps carrying the highest eigenvalue of the second-order Casimir operator
of SU(3) are used, instead of the highest weight irreps. Indeed, using 
Table I of Ref. [2] and Eq. (5) one finds the values of $\gamma$ listed in Table~\ref{Proxy-SU3}.
These results show a clear jump from prolate to oblate shapes between $N=64$ and 66.

\begin{table}[t!]
\caption{Estimates of proxy-SU(3) for the value of the $\gamma$ parameter (in degrees) assuming the use of the SU(3) irreps carrying the highest eigenvalue of the second-order Casimir operator.}
\label{Proxy-SU3}
\begin{tabular}{cccccc}
\hline
N   &   60  &   62   &  64  &   66  &   68\\
\hline
Zr &   21.8  & 15.4 &  23.7  & 45.4 &  53.4  \\   
Mo   & 26.5 &  19.7 &  28.4 &  50.6  & 58.7   \\   
Ru  &  20.9  & 13.9  & 23.1 &  46.9 &  55.7  \\
\hline
\end{tabular}
\end{table}

\section{Conclusions}\label{Concl}

Triaxiality in the region of the Mo and Ru isotopes appears to be of special interest, since empirical systematics suggest that it might represent a case of shape transition 
from $\gamma$-soft to rigid triaxial shapes. Indeed, both the odd-even staggering within the $\gamma$ bands, as well as the energy differences between the energy levels of the $\gamma$ band and the ground-state band for given value of the angular momentum $L$ (normalized to the difference occurring for $L=2$) indicate a sudden change from $\gamma$-unstable to rigid triaxial behavior at the neutron mid shell of $N=66$.

Within a shell-model picture, the subshell closure at $Z=40$ implies that the Mo and Ru isotopes 
are characterized by two and four valence proton particles respectively. It
appears that the shape transition occurs when the neutrons cross the $N=66$ midshell, having as a consequence that isotopes below (above) $N=66$ are characterized by neutron particles (holes). It appears therefore that $\gamma$-soft behavior is observed when both valence protons and valence neutrons correspond to particles, while rigid triaxiality is seen when the valence protons correspond to particles but the valence neutrons correspond to holes, a situation first suggested within the SU(3)$^*$ limiting symmetry of the IBM-2. 

In the present work we considered this region in detail, using two different approaches, the phenomenological ACM and the microscopic Skyrme-HFB approach. 
The ACM is highly suitable for describing various collective excitation modes. When applying this approach to nuclei that show some level of triaxiality, 
whether soft or rigid, a key consideration is how effectively minima can be obtained at various values of $\beta$ and $\gamma$, along with the associated degree of rigidity. 
The strength of the Hamiltonian used in our analysis lies in its relative simplicity and the inclusion of a term proportional to the parameter $\kappa$, which has the 
potential to induce triaxiality.

For Mo isotopes, we observe that the ACM parameters change smoothly and, in principle, only slightly, allowing for a 
reasonable description of the spectra. The energy of the
first excited $\gamma$ band is well reproduced, while higher bands ($\beta$ and double $\gamma$) are less accurately described. However, given the simplicity of the collective model,
further refinement is not expected. Importantly, the staggering pattern, which serves as a strong criterion for distinguishing the type of triaxiality (soft or rigid), is
described quite reasonably. Its pronounced sensitivity to even the slightest parameter variations, particularly those associated with the $\cos{(3\gamma)}$ terms, is observed.

Experimental spectra of $^{104-116}$Ru are also reasonably described, although, as in the case of Mo isotopes, the double $\gamma$ $K = 4$ and $\beta$-vibrational bands 
appear at higher energies than their experimental counterparts. For $^{104}$Ru, the parameter $\chi = 0$ corresponds to an experimental ratio $R_{4/2} = 2.48$, which aligns 
with expectations for a $\gamma$-independent potential. The nucleus at mid-shell, $^{110}$Ru, marks a point where a change in the staggering pattern is anticipated. 
This transition is indeed verified by the calculation: While $^{110}$Ru still exhibits characteristics of the $\gamma$-soft regime, $^{112}$Ru already demonstrates
 a staggering pattern consistent with the $\gamma$-rigid regime. However, for $^{112,114}$Ru, the available data points are limited, preventing us from making more 
 conclusive findings.
Energy ratios $R_{4/2}$ and $R$ and the ratio of transition rates $R_2$, along with the parameter $\gamma$ of the minimum derived from $R_2$, are in reasonable agreement with
 experimental values as well as predictions from other models. Given the stability of ACM parameters, the ACM PESs remain relatively stable, with a minimum at $\gamma$ 
 approximately 22$^{\mathrm{o}}$. From a structural perspective, microscopic models provide more insightful information in this regard, as their PES exhibits greater variability along the isotopic chain.

Within self-consistent mean-field models, the deformed minima of a PES
result from the competition between shell effects that are connected to the 
bunching of single-particle levels around the Fermi energy in the respective Nilsson 
diagram of protons and neutrons, and the average deformability that is mainly
connected to the surface energy, but for some parameter sets
may be enhanced by adding a rotational correction.
Finding triaxial ground states requires that the shell effects 
are more pronounced for triaxial configurations than for axial ones, 
and that the smooth average deformation energy remains sufficiently soft
over a wide range of $\beta$ and $\gamma$ values.

Predictions of triaxial ground states in the Skyrme-HFB approach are  therefore
parametrization dependent: 
They are only found for $^{106-114}$Ru with SV-bas and $^{104-114}$Ru with SLy7$^*$, whereas
the BSkG$x$ family 
predicts triaxial shape for all investigated Mo and Ru nuclei.
Besides differences in the location of the minimum that depend on the parameter 
set, there is a systematic difference between the results from the ACM, 
Figs.~\ref{PES-ACM-Mo} and \ref{PES-ACM-Ru}, and the 
energy surfaces from all Skyrme parameter sets: The ACM yields energy surfaces 
that are softer towards small $\gamma$, whereas the Skyrme results are
all softer towards oblate shapes.

The microscopic HFB PES behavior mirrors what is seen in the ACM, where increasing neutron number leads to enhanced triaxial rigidity. However, the $\gamma$ asymmetry and
 migration of the minimum -- visible in HFB but absent in ACM -- highlights the microscopic sensitivity of the Skyrme approach to shell structure and detailed energy balance. 
 In contrast, ACM produces symmetric PESs centered around fixed $\gamma$ values (often near $25–30^{\mathrm{o}}$), governed by collective symmetry principles rather than 
 single-particle dynamics.
Taken together, these observations point to a convergent narrative: Both models predict increasing rigidity across the Mo and Ru isotopic chain, but the Skyrme-HFB adds 
depth by capturing shape competition and $\gamma$ evolution driven by microscopic configurations -- an aspect that complements the idealized collective trends of the ACM.

Since the relevant experimental data ultimately decide which theoretical model corresponds more to reality, it is necessary to concentrate experimental efforts on the 
investigation of experimental signatures of triaxiality such as $B(E2; 2_2^+ \to 2_1^+)$ transition rates involved in the $B(E2)$ branching ratio $R_2$ or Kumar-Cline shape invariants currently known only for $^{104}$Ru as well as to odd-even staggering in $\gamma$ bands with 
insufficient or missing data for $N > 66$ for Mo isotopes and for $N > 68$ for Ru isotopes. 

\acknowledgments
This work was supported by the Project SP2026/022 financed by the Czech Ministry of Education, Youth 
and Sports and by the Czech Science Foundation grants GA22-14497S and 25-18335S. The Skyrme-HFB calculations were performed using HPC resources from the CC-IN2P3 of the CNRS.

\end{document}